%% file: PRD_master.tex
\RequirePackage{lineno}
\documentclass[prd,twocolumn,superscriptaddress,preprintnumbers,showpacs,amsmath,amssymb,floatfix]{revtex4}

\usepackage{float}
\usepackage{graphicx}
\usepackage{dcolumn}
\usepackage{bm}
\usepackage{slashbox}
\usepackage{setspace}
\usepackage{color}
\usepackage{multirow}
\usepackage{lineno}

\def\ra{\rightarrow}
\newcommand{\GeV}{\ensuremath{\mathrm{Ge\kern -0.1em V}}}
\newcommand{\MeV}{\ensuremath{\mathrm{Me\kern -0.1em V}}}

\newcommand{\pp}{\ensuremath{p\overline{p}}}

\newcommand{\fb}{\ensuremath{\mathrm{fb}^{-1}}}
\newcommand{\bs}{\ensuremath{B_s^0}}
\newcommand{\bd}{\ensuremath{B^0}}
\newcommand{\bsd}{\ensuremath{B_s^0(B^0)}}
\newcommand{\bu}{\ensuremath{B^{+}}}
\newcommand{\jp}{\ensuremath{J/\psi}}
\newcommand{\mm}{\ensuremath{\mu^{+}\mu^{-}}}
\newcommand{\jpmm}{\ensuremath{J/\psi\to\mm}}
\newcommand{\mmk}{\ensuremath{\mu^{+}\mu^{-}K^{+}}}

\newcommand{\hh}{\ensuremath{h^{+}h^{\prime -}}}
\newcommand{\bsmm}{\ensuremath{\bs\ra\mm}}
\newcommand{\bdmm}{\ensuremath{\bd\ra\mm}}
\newcommand{\bsdmm}{\ensuremath{\bsd\ra\mm}}
\newcommand{\bjk}{\ensuremath{\bu\ra J/\psi~K^{+}}}
\newcommand{\bjkmm}{\ensuremath{\bu\ra J/\psi~(\to\mm)~K^{+}}}

\newcommand{\bsjp}{\ensuremath{\bs\ra J/\psi~\phi}}

\newcommand{\brbsmm}{\ensuremath{\mathcal{B}(\bsmm)}}
\newcommand{\brbdmm}{\ensuremath{\mathcal{B}(\bdmm)}}
\newcommand{\brbsdmm}{\ensuremath{\mathcal{B}(B_s^0(B^0)\ra\mm)}}
\newcommand{\Mmm}{\ensuremath{M_{\mm}}}

\newcommand{\pting}{\ensuremath{\Delta\Omega}}
\newcommand{\iso}{\ensuremath{\mathit{I}}}
\newcommand{\nn}{\ensuremath{\nu_{N}}}

\newcommand{\cdf}{CDF~II}

\newcommand{\dedx}{\ensuremath{dE/dx}}
\newcommand{\pval}{\ensuremath{p}-value}
\newcommand{\bhh}{\ensuremath{B\to h^+h^{\prime -}}}
\newcommand{\brbdmmeNF}{\ensuremath{4.2\times 10^{-9}}}

\newcommand{\brbsmmeNF}{\ensuremath{1.3\times 10^{-8}}}

\newcommand{\brbsmmoN}{\ensuremath{2.7\times 10^{-8}}}
\newcommand{\brbsmmoNF}{\ensuremath{3.1\times 10^{-8}}}
\newcommand{\brbdmmoN}{\ensuremath{3.8\times 10^{-9}}}
\newcommand{\brbdmmoNF}{\ensuremath{4.6\times 10^{-9}}}
\begin{document}


\title{\bf{ Search for $\bsmm$ and $\bdmm$ decays with the full CDF Run II data set}} 

\input{November2012_Authors-withSperka}
\date{\today}

\begin{abstract}
We report on a search for \bsmm\ and \bdmm\ decays using proton-antiproton collision data at $\sqrt{s}=1.96$~TeV corresponding to $10~\fb$ of integrated luminosity collected by the \cdf\ detector at the Fermilab Tevatron collider.  The observed number of \bd\ candidates is consistent with background-only expectations and yields an upper limit on the branching fraction of 
$\brbdmm < \brbdmmoNF$ at 95\% confidence level.  We observe an excess of \bs\ candidates.  The probability that the background processes alone could
produce such an excess or larger is $0.94\%$.  The probability that the combination of background and the expected standard model rate of \bsmm\ could produce such an excess or larger is $6.8\%$. These data are used to determine a branching fraction 
$\brbsmm = (1.3^{+0.9}_{-0.7}) \times 10^{-8} $ and provide
an upper limit of $\brbsmm < \brbsmmoNF$ at 95\% confidence level.
\end{abstract}


\maketitle

\section{Introduction}
\label{sec:intro}
The study of decays of the \bs\ meson (with a quark content of 
$\overline{b}s$) and the \bd\ meson ($\overline{b}d$) into a dimuon pair (\mm ) has long been of great interest as a test of the standard model (SM) of particle physics. 
These flavor-changing neutral-current (FCNC) decays occur in the SM only through weak-interaction-mediated loop processes whose amplitudes are suppressed via the Glashow-Iliopoulos-Maiani mechanism~\cite{GIM} and by helicity conservation. In the SM, the branching fractions for \bsmm\ and \bdmm\ are predicted to be $(3.2\pm0.3)\times10^{-9}$ and $(1.1\pm0.1)\times10^{-10}$, 
respectively~\cite{smbr}. 
Note that the inclusion of charge conjugate modes is implied throughout this paper.
Non-SM particles in the loop processes or non-SM coupling mechanisms can significantly alter the rate of these decays so that measurements of their branching fractions serve 
as powerful tools to probe for the effects of new physics beyond the SM. For example, in the minimal supersymmetric standard model the \bsmm\ decay rate is proportional to 
($\tan\beta)^6$~\cite{tanbeta1,tanbeta2,tanbeta3}, where
$\tan\beta$ is the ratio of the vacuum expectation values of the two Higgs fields.  The decay rate can be enhanced relative to the SM by over two orders of magnitude at large $\tan\beta$ values. The search is also sensitive to supersymmetry (SUSY) in cosmologically consistent 
scenarios~\cite{cosConst1,cosConst2,cosConst3,cosConst4}. Other models, such as R-parity violating SUSY~\cite{cosConst1}, littlest Higgs model with T-parity~\cite{littlestHiggs}, or models with extra dimensions~\cite{extraDim1,extraDim2}, predict large effects independent of the value of 
$\tan\beta$.  Substantial negative interference effects can suppress the 
\bsmm\ branching fraction by as much as a factor of three in portions of the SUSY parameter space~\cite{Beskidt:2011qf}.
%
%
In the absence of an observation, limits on \brbsmm\ are complementary to limits provided by direct searches in constraining the new-physics parameter-space.
The status of constraints on new physics in a variety of different models
and in a model independent treatment are discussed in Ref.~\cite{recentReview}.

Recent results on \brbsdmm\ include limits from the ATLAS~\cite{ATLAS}, CMS~\cite{CMS},  D0~\cite{D0}, and CDF~\cite{CDFPRL2011} experiments. The most sensitive result is from the LHCb experiment~\cite{LHCb}, which reported an excess of \bsmm\ events and measured $\brbsmm = 3.2^{+1.5}_{-1.2}\times10^{-9}$ and
an upper limit for the \brbdmm\ within a factor of about nine of the SM \bdmm\ rate.
The previous CDF result reported $\brbsmm = 1.8^{+1.1}_{-0.9}\times10^{-8}$ and sets a two-sided interval at 90\% C.L. of $4.6\times10^{-9} < \brbsmm < 3.9\times10^{-8}$ and an upper limit of $\brbdmm < 3.5\times10^{-8}$ at 90\% C.L. The \bs\ and \bd\ results from all experiments are compatible and indicate that there is no strong enhancement in the \bsmm\ decay rate.
Further measurements of \brbsmm\ are likely to constrain strongly new 
physics models predicting significant deviations from the SM predictions.


We report on a search for \bsmm\ and \bdmm\ decays using the complete Run~II data set of \pp\ collisions at $\sqrt{s}=1.96$ TeV collected by the upgraded Collider Detector at Fermilab (CDF~II) and corresponding to an integrated luminosity of $10\:~\fb$.  Because the previous CDF 
analysis~\cite{CDFPRL2011}, using 7~\fb\ of integrated luminosity, reported an excess of \bsmm\ signal events, the same analysis methodology is applied to the full available data set.  The sensitivity of the analysis reported here
is improved with respect to that reported in Ref.~\cite{CDFPRL2011} due to the 24\% increase in event-sample size. All other aspects of the analysis have remained the same.  

\section{Experimental setup: the CDF II detector}
\begin{figure}
  \label{cdfLayout}
  \centering
  \includegraphics[width=3.5in]{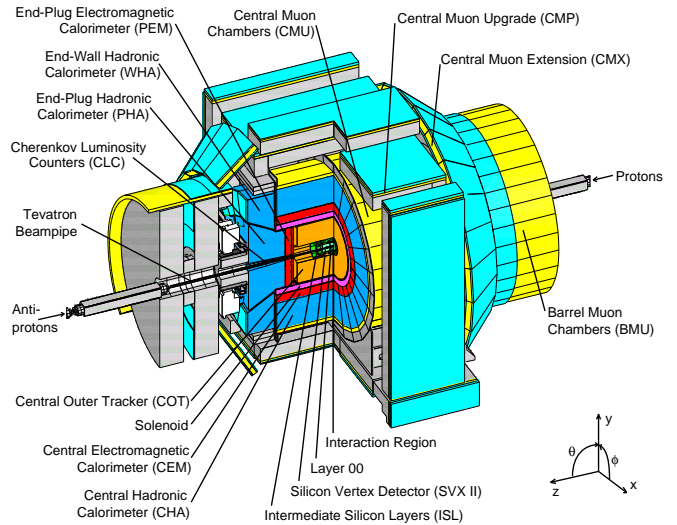}
  \caption{Cutaway isometric view of the CDF II detector.}
  \label{fig:cdfLayout}
\end{figure}

The CDF~II detector is a general-purpose detector~\cite{cdfTDR,cdfDet1,cdfDet2} with cylindrical symmetry (Fig.~\ref{fig:cdfLayout}) designed to detect products of \pp\ collisions at a center-of-mass energy of $\sqrt{s}=1.96$ $\textnormal{TeV}$. 
A cylindrical coordinate system is used to describe particle trajectories. The $z$ axis is defined as the direction of the proton beam. 
Besides the azimuthal angle $\phi$, radius relative to the beam line $r$, and polar angle
$\theta$, we define a pseudorapidity $\eta=-\ln\left(\tan(\theta/2)\right)$. 
The transverse momentum, $p_{T}$, represents the component of a particle's momentum in the plane
perpendicular to the beam axis, $p_{T}= p \sin\theta$.

The most important subdetectors for this analysis are briefly described below and include
the tracking system and the muon system. Additional subsystems such as the calorimeters and luminosity detector system also play a role in the analysis. The calorimeters are used in part of the particle identification process, while information from the luminosity detector system is used in some of the background estimations. A more detailed description of the CDF~II detector can be found in Ref.~\cite{cdfTDR}.

\subsection{Tracking system}
The tracking system consists of silicon microstrip detectors, a multi-wire open-cell drift chamber, and a solenoidal superconducting magnet. 
The innermost tracking system, L00, is a single-sided silicon microstrip system mounted on the beam pipe~\cite{l00Det}. Outside L00 is the SVXII 
detector, with five layers of double-sided silicon microstrip sensors~\cite{svxDet}.  One side of each sensor provides azimuthal ($r\phi$) information while the opposite side provides longitudinal information ($rz$). 
The SVXII hit resolution is 11 $\mu$m, while the impact parameter resolution for charged particles with $p_T>2.0$ GeV/$c$ is about 40 $\mu$m, which includes a 35 $\mu$m contribution due to the size of the \pp\ luminous region.  The association of SVXII $rz$ hits allows measurement
of the $z$ coordinate of charged-particle trajectories (tracks) at the \pp\ interaction point with 70 $\mu$m
resolution.  The combination of excellent $r\phi$ and $z$
resolution allows precise determination of the three-dimensional spacepoint 
defined by the \bsdmm\ decay vertex and the rejection of background from pairs of random muon-candidates that accidentally meet the selection requirements (combinatorics). 

Outside the silicon subsystems is the COT~\cite{cotDet1}, an open-cell multi-wire drift 
chamber divided into eight concentric superlayers. 
The superlayers
themselves are divided in $\phi$ into supercells, each containing 12 sense 
wires. 
In addition to charged-particle trajectories, the COT also measures the ionization \dedx\ per unit path-length for 
particle identification. 
For this analysis the \dedx\ information is mainly used to help reject kaons.

Surrounding the COT is a superconducting solenoidal magnet producing a 1.4 T magnetic 
field parallel to the beam axis. 
The $p_T$ resolution of the COT,
$\sigma_{p_{T}}/p_{T}^2 \approx 0.15\%$ $\textnormal{(GeV}/c)^{-1}$~\cite{cotDet1}, is determined by comparing the curvature of inward- and outward-going tracks of cosmic-ray events. 
The absolute momentum scale is determined using $J/\psi$, $\Upsilon$, and $Z$-boson resonances, where
the resonances decay into two muons~\cite{momentumScale}.

\subsection{Muon system}

\begin{figure}
  \centering
  \includegraphics[width=3in]{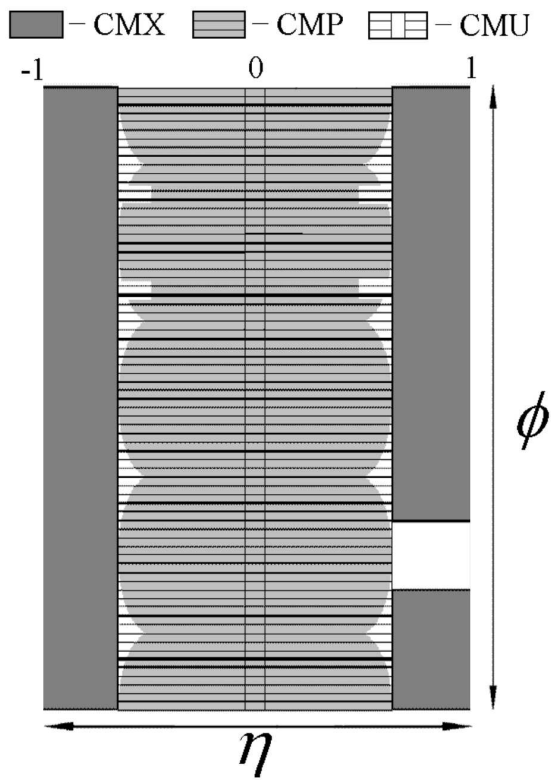}
  \caption{Muon system $\phi$ and $\eta$ coverage.}
  \label{fig:muonCoverage}
\end{figure}

Outside the solenoidal magnet are electromagnetic and hadronic calorimeters, which in turn are surrounded by the muon systems consisting of multi-layer single-wire drift chambers and scintillators. The drift chambers are used to reconstruct muon-track segments (stubs) while the scintillators are used for timing information to match muon candidates to the correct \pp\
collision crossing. The $\eta$ and $\phi$ coverage of the muon subdetectors used in this analysis is shown in Fig.~\ref{fig:muonCoverage}.

In the central region, the cylindrical central muon chambers (CMU)~\cite{cmuDet} provide coverage up to $|\eta|<0.6$. Because material corresponding to 5.5 interaction lengths lies between the \pp\ luminous region and the CMU, a muon must have a minimum $p_T$ of 1.4 GeV/$c$ (range-out threshold) to reach the CMU. The CMU is subdivided in 24 wedges in $\phi$, each with four layers of drift chambers, and lies immediately outside the central hadronic calorimeter.

Beyond the CMU are additional central chambers with nearly the same $\eta$ coverage 
known as the CMP. Additional steel absorber with a thickness of 2.3 interaction lengths is placed between the CMU and CMP yielding a range-out threshold $p_T$ 
of 2.2~GeV/$c$.  The CMP forms a box around the cylindrical CMU and is comprised of four layers of drift chambers, and a layer of scintillator.

The CMX~\cite{cdfTDR} detector extends the muon system coverage to higher pseudorapidity, $0.6<|\eta|<1.0$. The CMX consists of two arches at each end of the detector, with additional upper and lower sets of chambers. The CMX consists of eight layers of drift chambers arranged in conic sections;
the geometry is such that particles can traverse six of the eight layers on average. The range-out threshold $p_T$ for the CMX is about 2.0~GeV/$c$.

In this analysis muons are required to have either a CMU or CMX stub and have the stub matched to an extrapolated track from the tracking system. Further information, such as $z$ position and stub angle, is used in a multivariate likelihood discriminant for muon identification. Information from the CMP detector is used, if available,
to identify high-purity muons. The muons are paired into either a CMU-CMU (CC) or a CMU-CMX (CF) channel. The selection criteria for these two  
channels are discussed in the next section.

\subsection{Muon trigger}
\label{sec:muonTrig}

At the Tevatron \pp\ crossings occurred every 396 ns (2.5 MHz) and peak instantaneous luminosities of $4\times 10^{32}$ cm$^{-2}$s$^{-1}$ were achieved, creating an event rate of 1.7\,MHz. An online system of
custom hardware boards and software algorithms (the ``trigger'' system) was employed to reduce the data rate to about 100 events per second, which were recorded to tape for later analysis. The CDF trigger system was divided into three consecutive levels with increasing granularity, sophistication, and precision.  Trigger level one (L1) was evaluated for every \pp\ crossing and used coarse track, calorimeter, and muon-stub information to identify potentially 
interesting events that were then passed to trigger level two.  Trigger level two (L2) used more precise calorimeter and muon-stub information to 
eliminate events poorly reconstructed in trigger L1.  Accepted events were 
passed to trigger level three (L3), a CPU farm performing full event reconstruction and identifying the most interesting events to record to tape for later analysis.  The data sets used in the present analysis were collected with a set of L1, L2, and L3 triggers that required a pair of muon candidates.

In L1, muons were identified by matching a track reconstructed in the COT to a muon stub reconstructed in one of the muon systems.  The track reconstruction was performed by a custom-built system~\cite{xft} that achieved 
a $p_T$ resolution of $\sigma_{p_{T}}/p^2_{T}=1.7\%/$({\rm GeV}/$c$). Custom electronic boards identified muon stubs in each of the CMU, CMX, and CMP systems and performed a coarse matching to the L1 tracks~\cite{xtrp}.  Events were required to have two separate CMU-track matches or one CMU-track match and one CMX-track match.  The particles associated to the tracks were required to have opposite electric charge. In L2 the track
muon-stub matches were confirmed using more sophisticated algorithms and improved resolutions.  In order to remove through-going cosmic-ray muons and backgrounds from dijet events that generate falsely-identified muon candidates (fake muons), only events with a dimuon
opening-angle less than 120 degrees in the plane transverse to the beam line were 
passed to L3.  The full event reconstruction employed in L3 
performed a full track fit and required CMU muon candidates to have $p_T > 1.5$~GeV/$c$, 
CMX muon candidates to have $p_T > 2.0$~GeV/$c$, the scalar sum $p_T$ for the
two muon candidates to exceed $5$~GeV/$c$, the dimuon mass to be less than 
$6$~GeV/$c^2$, and the difference in the $z$ coordinates of the muon tracks at the 
point of closest approach to the beam line to satisfy $|\Delta z_0| < 5$~cm.  At the
highest instantaneous luminosities, the accept rate of this dimuon trigger path was too high and events were randomly discarded with a frequency that depended on the instantaneous luminosity.
This reduction in rate discarded 
approximately 10\% of the total dimuon candidate events.  In the full Run~II data set
collected by CDF, $822\,740$ ($498\,443$) events satisfied the CMU-CMU (CMU-CMX) trigger path with $\Mmm > 4.669$~GeV/$c^2$
and formed the initial data sample for the CC (CF) channel.

\section{Monte Carlo simulation}
\label{sec:MC}
We employ Monte Carlo (MC) simulations of \bsmm\ and \bdmm\ decays together with a 
full CDF detector simulation to estimate signal efficiencies not measurable with data control samples. 
In addition to the \bsdmm\ MC sample we also produce a sample of \bjk\ simulated 
events for modeling cross-checks. A MC sample of \bsmm\ is generated using 
{\sc{Pythia}}~\cite{ref:pythia} and EvtGen~\cite{ref:EvtGen} with the underlying event 
modeling tuned to reproduce minimum-bias events in \pp\ collisions at $\sqrt{s}=1.96$~TeV~\cite{minBiasUnder}. 
We generate simulations of $b\overline{b}$ pair production and their subsequent hadronization. One of the resulting $B$ hadrons is required to be a \bs\ or \bd\ meson that decays to two muons.  There are no requirements on the second $B$ hadron, which is allowed to decay inclusively. The MC events are also run through a detailed CDF II detector simulation~\cite{cdfSim} that accounts for resolution and occupancy effects in all the subdetector systems. 
The MC events are required to meet all the baseline requirements discussed in 
Sec.~\ref{sec:baseline} with the exception of the \dedx\ and muon likelihood requirements, which are omitted because data-driven estimates of their efficiency are used instead. 
We weight the \bs -meson $p_T$ and isolation (cf. Sec.~\ref{sec:baseline}) distributions to match 
the measured spectra obtained from \bjk\ data and \bsjp\ data, respectively. 
The \bu\ $p_T$ spectrum is expected to be similar to that of the \bs\
for $p_T>4.0$ GeV/$c$ and provides a significantly larger sample size. The isolation spectra, however, may differ between \bu\ and \bs\ mesons due to the participation 
of $\overline{u}$ and $\overline{s}$ quarks in the hadronization processes, respectively.  For the \bdmm\ search the \bd -meson $p_T$ and isolation distributions are weighted using \bjk\ data.

We use simulated \bsmm\ decays to estimate a mass resolution 
of $24\:\MeV/c^2$ for events passing the baseline and vertex requirements described in Sec.~\ref{sec:baseline}.
Comparisons between data and MC using $\jp\to\mm$ and \bjkmm\ samples reveal a 10\% discrepancy in mass resolution, which is propagated as a systematic uncertainty and negligibly affects the efficiencies discussed in Sec.~\ref{sec:signalEff}.

\section{Event selection}
\label{sec:evtSel}
This analysis searches for \bsmm\ and \bdmm\ decays using the full 10~\fb\ CDF II 
data set. The same analysis methods are used for both decays.  The signal-search 
dimuon-mass range is adapted to the different \bd\ and \bs\ pole masses and corresponds
to $\pm 2.5$ times the two-track mass resolution. The branching fractions are measured
relative to a \bjkmm\ normalization mode. This mode, together with directly produced 
$\jp\to\mm$ decays, are used to estimate signal efficiencies and perform consistency checks. Initially we apply baseline requirements (described in 
Sec.~\ref{sec:baseline}) on all data and MC samples. An artificial neural
network (NN) classifier is then applied to enhance the expected signal-to-background-ratio. 
In order to avoid inadvertent biases, the data in an extended mass-signal region,
$5.169 < \Mmm < 5.469~\mathrm{GeV}/c^2$, are kept hidden until all selection criteria
are finalized.  An unbiased optimization of the analysis is performed using mass-sideband data as a model of the combinatorial background and MC events as a model of the peaking backgrounds and signal.

Major background processes include Drell-Yan dimuon production ($q\bar{q} \rightarrow \mu^+\mu^-$) processes 
through virtual $\gamma$ and $Z$-boson states, double semileptonic decay 
($b\bar{b}, c\bar{c} \rightarrow \mu^+\mu^-X$), and sequential semileptonic 
decay ($b \rightarrow c\mu^-X\rightarrow s\mu^-\mu^+X'$) of $b$ and $c$ quarks. 
A combination of a semileptonic decay and a fake muon or two fake muons from two-body 
hadronic decays of $B$ hadrons (\bhh\ where $h$ and $h^{\prime}$ are charged pions and kaons), can also be a source of background. Fake muons are tracks 
from pions and kaons that have a matching muon stub and are falsely identified as muons. The backgrounds can be divided into a combinatorial dimuon background and a peaking \bhh\ background, which are estimated separately.
Backgrounds are studied in detail in statistically-independent control samples with 
various baseline requirements inverted or relaxed to enhance the background contribution.

The search for a \bsdmm\ signal is performed in bins of NN output.  The NN binning is determined by an {\it{a priori}} optimization, discussed in detail 
in Sec.~\ref{sec:nnOpt}, that uses the expected \brbsmm\ limit in absence of signal 
as a figure of merit, resulting in eight NN bins.  Additionally, the signal region 
is divided into five mass bins centered on the world average \bs\ and \bd\ masses. This
yields a total of 80 single-bin counting experiments corresponding to the CC and CF 
topologies, each with eight NN bins, and five mass bins.

Once the signal efficiencies and background estimates are well understood, a thorough statistical analysis of the result is performed. 
The sections below discuss the analysis methodology and results in more detail.

\subsection{Baseline event selection}
\label{sec:baseline}
Except where specifically discussed, all samples used
in this analysis are required to pass a set of baseline requirements that
consist of kinematic-, particle-identification-, and vertex-related requirements discussed in this section.

Muon-candidate tracks are required to be matched with a muon identified by the trigger and have $p_T>2.0$ GeV/$c$ and
$p_T>2.2$ GeV/$c$ for CMU and CMX muons, respectively. 
Tracks are required to be fiducial to the COT, by demanding that the absolute value of the $z$ coordinate be less than
155 cm at the COT exit radius ($r=136$ cm). This ensures that
tracks traverse the full radial extent of the COT. Tracks are also required 
to have $r\phi$ hits in at least three L00+SVXII layers.

A likelihood method~\cite{gavril} together with a \dedx\ based selection~\cite{bhhprl} is used to
further suppress contributions from hadrons misidentified as muons. 
The muon likelihood is based on matching muon stubs to COT tracks in the $\phi$ and $z$ coordinates, 
energy deposits in the electromagnetic and hadronic calorimeters, information concerning the subsystems in which the muon is identified, 
and kinematic information.
We require a muon likelihood with a value greater than 0.1, which is approximately 99\% efficient for signal, while
rejecting 50\% of the combinatorial background, which contains a significant fraction of hadrons 
misidentified as muons. The \bhh\ decays are efficiently rejected as discussed in detail in Sec.~\ref{sec:peakBack}.

A calibration for the \dedx\ measurement is applied to ensure stability over the tracker volume and operational conditions. This calibration
corrects for effects dependent on instantaneous luminosity, local density of tracks, and kinematic information. We make the requirement 
$\ln\left(\frac{\dedx_{\mathrm{o}}}{\dedx_{\mathrm{e}}}\right) > -0.83$,
where $\ln$ is the natural logarithm, $\dedx_{\mathrm{o}}$ is the observed \dedx\ after the calibration has been applied, and $\dedx_{\mathrm{e}}$
is the expected \dedx\, estimated using the observed particle's momentum and the muon mass hypothesis.
This requirement is nearly $100\%$ efficient for muons, while rejecting about 50\% of kaons. 
Estimations of the signal efficiency and background rejection for 
the choice of likelihood and \dedx\ selection criteria are made
using $J/\psi \to \mm$ events compared to combinatorial background from the
dimuon-mass sidebands and kaons from \bjk\ decays.

Muon pairs are required to have an invariant mass in the range $4.669 < \Mmm < 5.969$ GeV/$c^2$.
Reconstructed $B$-meson candidates must also have $|\eta|<1$ and $p_T>4.0$ GeV/$c$. We reconstruct the \pp\ interaction point for each event by refitting track-helix parameters of all charged particles, after excluding the muon candidates, that have $z_0$ within 1 cm of the dimuon average $z_0$ and have $p_T>0.5$ GeV/$c$ to a common space point (vertex). 
If the fit fails, the primary vertex is determined by the beam line position, estimated using COT+SVXII information, at the average $z_0$ of the dimuon pair.

\begin{figure}
  \centering
  \includegraphics[width=3.2in]{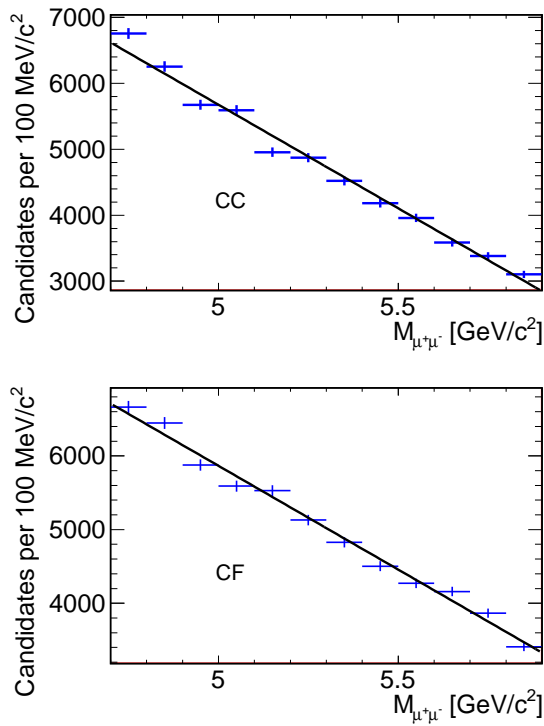}
  \caption{Dimuon invariant mass distribution for events
    satisfying the baseline requirements for the \bsmm\ and \bdmm\ search 
    sample with linear fit overlaid for the CC (top) and CF (bottom)
    channels.}
  \label{fig:bsmass_ux}
\end{figure}

In addition to the above requirements, we fit the two muon tracks to a common secondary vertex. 
Several demands on secondary-vertex-related variables are made. We define
a three-dimensional displacement length 
$L_{3D} = \vec{p}(B)\cdot \vec{x}_{B} / |\vec{p}(B)|$, where $\vec{p}(B)$ is the 
$B$-candidate-momentum vector estimated as the vector sum of the muon momenta 
and $\vec{x}_{B}$ is the secondary-vertex position vector determined relative to the primary-vertex position.
We estimate a proper decay time $t=L_{3D}\Mmm/|\vec{p}(B)|$, where \Mmm\ is the dimuon invariant mass, which in turn is used
to define the proper decay-length $\lambda=ct$. 
The baseline requirements demand that the measured proper decay-length of the {\it B} candidate, with its uncertainty, $\sigma_\lambda$, satisfy
$0 < \lambda < 0.3$ cm and $\lambda/\sigma_\lambda > 2$; the secondary vertex fit $\chi^2/N_{\mathrm{dof}}$, where $N_{\mathrm{dof}}$ is the number of degrees of freedom, must be less than 15;
the three-dimensional displacement length and its uncertainty $\sigma_{L_{3D}}$ satisfy
$L_{3D}<1.0$ cm and $\sigma_{L_{3D}}<0.015$ cm;
the three-dimensional opening angle between $\vec{x}_{B}$ and $\vec{p}(B)$, $\pting$, satisfy
$\pting < 0.7~\rm{rad}$; and the {\it B}-candidate track-isolation, \iso, satisfy $\iso > 0.50$. The isolation
is defined as $I=p_T(B)/( p_T(B)+\displaystyle\sum\limits_{i} p_T(i) )$, where the
sum goes over all charged particles with $p_T>500$ MeV/$c$ and within an $\eta\phi$ cone centered around the $B$-meson momentum with radius $R=\sqrt{(\Delta\eta)^2+(\Delta\phi)^2}<1.0$.
The trigger and baseline requirements result in a total of $60\,842$ CC 
and $64\,495$ CF muon pairs, shown in Fig.~\ref{fig:bsmass_ux}.  

For the final selection, we define search regions around the known \bs\
and \bd\ masses~\cite{PDG2010}. These regions correspond to approximately 
$\pm 2.5\sigma_{m}$, where $\sigma_{m}\approx24$~$\mathrm{MeV}/c^{2}$ 
is the two-track  mass resolution estimated from \bsmm\ MC events satisfying the 
trigger and baseline requirements.
The sideband regions $5.009 < \Mmm < 5.169\:\GeV/c^2$ and  $5.469 < \Mmm
< 5.969\:\GeV/c^2$ are used to estimate combinatorial backgrounds.
Dimuon candidates with mass smaller than $\Mmm=5.009$ GeV/$c^2$ are not used for background estimations due to
$b\to\mu^{+}\mu^{-}X$~\cite{btosmm} contributions but are used for the NN training. It was verified that inclusion of these candidates does not 
significantly affect the discriminating capabilities of the NN.
Backgrounds from \bhh\ decays, 
which peak in the signal mass region, are estimated separately.

\subsection{Normalization decay mode}
\label{sec:normBaseline}
\begin{figure}
  \includegraphics[width=3.2in]{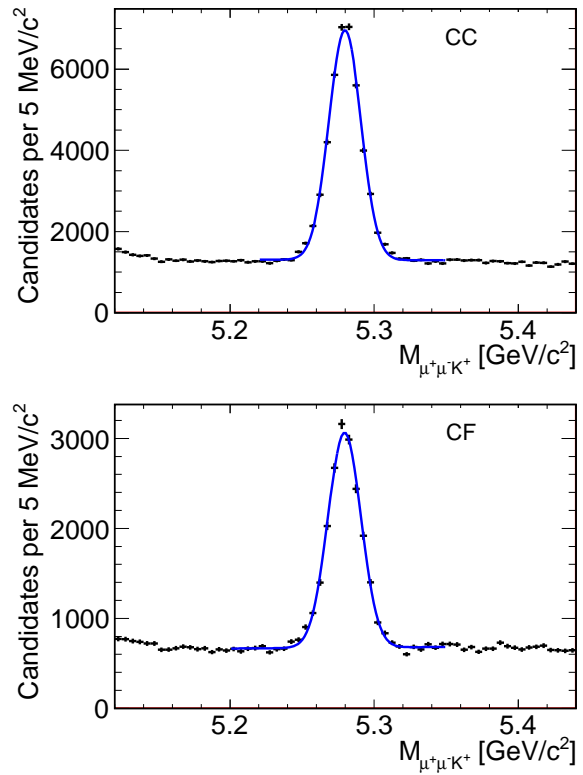}

  \caption{The \mmk\ invariant mass distribution for events
    satisfying the requirements for the \bjk\ sample with fit overlaid
    for the CC (top) and CF (bottom) channels.}
  \label{fig:BUmass}
\end{figure}

A sample of $\bjk$ events serves as a normalization decay mode. The $\bjk$ sample is collected 
using the same dimuon triggers and selection requirements as used for the signal sample so that common systematic 
uncertainties are suppressed.  The kaon candidate must satisfy the same COT and 
L00+SVXII requirements as the muon candidates and must have $p_T>1~\GeV/c$, 
a regime for which the COT tracking efficiency is well understood. For each kaon candidate the $\mu^+$, $\mu^-$, and $K^+$ tracks are constrained to originate from a common vertex in three dimensions. 
The $\chi^2$ probability of the vertex fit is required to be greater than $10^{-5}$.
Additionally, the dimuon invariant mass is required to be consistent with the world average \jp\ mass ($3.017 < \Mmm < 3.177\:~\GeV/c^2$).  The \mmk\ mass distribution of candidates satisfying these criteria is shown in Fig.~\ref{fig:BUmass}.
A signal mass region, $|M_{\mmk}-M_{\bu}| < 35\:\: \MeV/c^2$, is used together with mass sidebands to estimate a \bjk\ yield of $28\,081 \pm 219$ and $12\,144 \pm 153$ in the CC and CF channels, respectively, using simple 
sideband subtraction.  These yields include a 0.14\% correction for 
$B^+\to J/\psi\,\pi^+$ contributions. The correction factor is determined 
by comparing the relative geometric acceptance, reconstruction efficiency, mass window efficiency, and branching fraction of \bjk\ and $B^+\to J/\psi\, \pi^+$ decays.  The uncertainties on the \bjk\ yields are due to the limited size of the \mmk\ sample. The shape of the \mmk\ mass distribution is parametrized
using a Gaussian summed with a first-order polynomial and a fit to the data yields a Gaussian mean of $5280\pm11\:\MeV/c^2$ and $5274\pm11\:\MeV/c^2$ for CC and CF, respectively, consistent with the world average \bd\ mass. 

\subsection{The neural network discriminator}
\label{sec:NN}
We search for \bsmm\ (\bdmm ) decays in a narrow 120 MeV/$c^2$ mass window centered around the world average \bs\ (\bd ) mass.  After application of 
the baseline selection criteria the data sample is dominated by combinatorial background and has a signal-to-background ratio of approximately $10^{-4}$.  We significantly improve the discrimination power between signal and combinatoric background by combining kinematic, isolation-related, and lifetime-related variables using a neural-network classifier. A key feature of the NN is that it is designed to be independent of the dimuon mass.  This allows an estimation of the combinatorial background in the signal mass region by interpolation from mass sidebands after the NN has been applied to the data.

Fourteen variables describing the measured kinematics-, isolation-, and lifetime-related properties of the signal are selected to construct a NN
discriminant \nn.
These variables are selected based on a study of the physics characteristics
of the signal and background using training samples described below.
Variables that are poorly modeled by the MC or correlated
to dimuon mass are excluded from the NN.
The signal and background samples are separated into CC and CF
data sets that are exclusively independent.
Trainings are performed separately using the same input variables for the CC and CF data sets.

We chose the {\sc neurobayes} NN package~\cite{ref:neurobayes1,ref:neurobayes2} for the construction of the NN.
Using this multivariate analysis technique, we achieve a background
rejection of 99.9\%, while maintaining a signal efficiency of 40\%.

\subsubsection{Background training sample}
\label{sec:bkgSample}
 We define two dimuon mass sidebands, 
$4.669 < \Mmm < 5.169$ GeV/$c^2$ (lower SB) and 
$5.469 < \Mmm < 5.969$~GeV/$c^2$ (upper SB), which are used to 
construct a combinatorial background sample used as one of
the NN training samples. Although this analysis is based on 10~\fb, the NN optimization was done, {\it{a priori}}, during the previously published analysis based on 7~\fb\ \cite{CDFPRL2011}.  
A total of 36\,329 CC and 39\,657 CF mass-sideband events survive the baseline requirements in 7~fb$^{-1}$ of integrated luminosity and are used as the background sample for the NN training and testing.
%
%

\subsubsection{Signal training sample}

The simulated \bsmm\ signal sample used for the NN training is described in Sect.~\ref{sec:MC}.
The dimuon mass range used for this signal sample corresponds to the
search region defined in Sec.~\ref{sec:baseline}.
A sub-sample of MC signal events equal in size to the sum of the background samples is randomly chosen for NN training while the full MC sample is used to estimate the efficiency of the neural network.

\subsubsection{Input parameters}
\label{sec:nnInput}
Candidates from combinatorial backgrounds, when compared to signal events,
tend to have smaller dimuon mass, shorter proper-decay-lengths, a softer $p_T$ spectrum, and a higher density of tracks near the \bs\ candidate. We investigated a variety of kinematic, isolation, and lifetime variables as inputs to the NN.  After removing variables that were poorly modeled or were found to cause a correlation between 
\nn\ and the dimuon mass, an initial set of 20 discriminating variables remained.  Multiple neural networks were trained varying the number and combination of input parameters employed.  The performances of the trained networks were compared in the plane of combinatorial-background acceptance versus signal acceptance.  A NN using 14 input variables was found to offer excellent background discrimination and is employed as the final discriminant in this analysis.  The 14 variables employed are listed here in order of descending discrimination power between signal and background:
\begin{description}
  \label{desc:nnIn}
\item[$\boldsymbol{\Delta\Omega}$:] Three-dimensional angle
  between the \bs\ momentum and the vector pointing from the primary to 
  secondary vertex.

\item[$\boldsymbol{I}$:] Isolation of the \bs\ candidate as defined  in 
  Sec.~\ref{sec:baseline}. 

\item[Larger $\boldsymbol{d_0(\mu)}$:] For the muon pair, the impact 
  parameter of the muon with the larger value. 

\item[$\boldsymbol{d_0(\bs)}$:] Impact parameter of \bs\ candidate with 
  respect to the primary event vertex. 

\item[$\boldsymbol{L_{T}/\sigma_{L_{T}}}$:] Significance of the transverse 
  decay length $L_{T}$, where
  $L_{T}  =  \vec{p_T}(B)\cdot \vec{x}_T / |\vec{p_T}(B)|$ and 
  $\vec{x}_T$ is the secondary-vertex-position vector relative to the 
  primary-vertex position in the plane transverse to the beam line. 

\item[$\boldsymbol{\chi^2}$:] $\chi^2$ per degree of freedom of the 
  secondary-vertex fit. 

\item[$\boldsymbol{L_{3D}}$:] Three-dimensional vertex displacement
  as defined in Sec.~\ref{sec:baseline}. 

\item[Lower $\boldsymbol{p_T(\mu)}$:] For the muon pair, the transverse 
  momentum of the muon with the lower value. 

\item[Significance of smaller $\boldsymbol{d_0(\mu)}$:] 
  $d_0(\mu)/\sigma_{d_0(\mu)}$ of the muon with 
  smaller impact parameter, where $\sigma_{d_0(\mu)}$ is the estimated 
  uncertainty of $d_0(\mu)$.

\item[$\boldsymbol{\lambda/\sigma_{\lambda}}$:]
  Significance of $\lambda$. 

\item[Smaller $\boldsymbol{|d_0(\mu)|}$:] For the muon pair, the impact
  parameter of the muon with the smaller value. 

\item[$\boldsymbol{\lambda}$:] Three-dimensional proper decay length
  defined in Sec.~\ref{sec:baseline}. 

\item[$\boldsymbol{\Delta\Omega_{T}}$:]
  Angle between the \bs\ momentum and the vector pointing from the primary 
  to secondary vertex in the plane transverse to the beam line. 

\item[Significance of larger $\boldsymbol{d_0(\mu)}$:]
  $d_0(\mu)/\sigma_{d_0(\mu)}$ of the muon with larger impact parameter. 
\end{description}

\begin{figure}
  \begin{center}
    \includegraphics[width=3.5in]{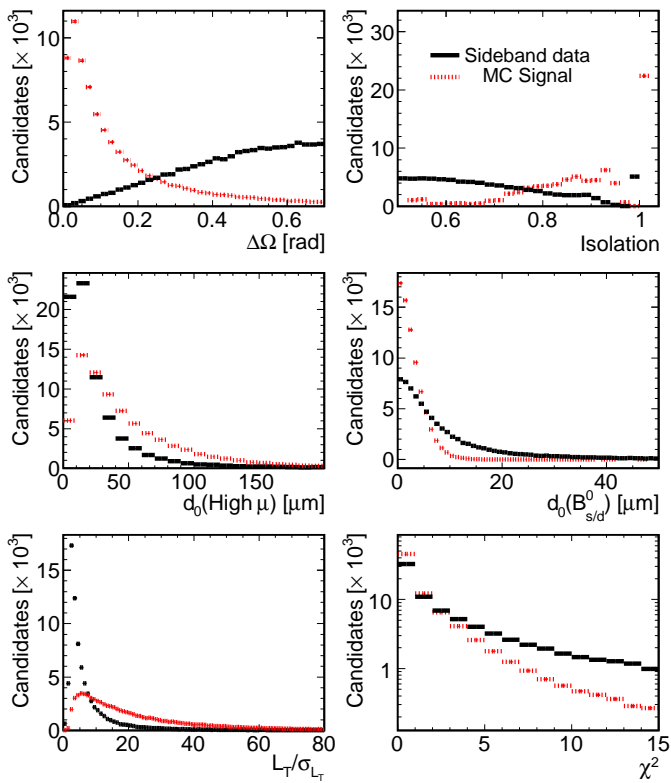}
  \end{center}
  \caption{Comparison of the combinatorial background distribution (solid), taken from dimuon mass sideband events in the first 7~\fb\ of data, 
  to the signal distribution (dashed), taken from MC events, for the six
  most discriminating of the NN input variables.}
  \label{fig:nnInMain}
\end{figure}

When available, the silicon $rz$ tracking information strongly discriminates against combinatorial and partially reconstructed backgrounds. Tracks from combinatorial background are less likely to originate from a common vertex in three dimensions and are suppressed by the vertex-fit quality information used in the multivariate discriminate. Similarly, precise three-dimensional reconstruction
of the primary and secondary vertices allows the comparison of the $B$-candidate flight direction and the secondary-vertex vector, which rejects both combinatorial and partially-reconstructed background.
Comparisons of the distributions of background and signal samples
for the input variables with the greatest discriminating power 
are shown in Fig.~\ref{fig:nnInMain}.

\subsubsection{Discriminant output}
\label{sec:nnOutput}
\begin{figure}
  \begin{center}
    \includegraphics[width=3in]{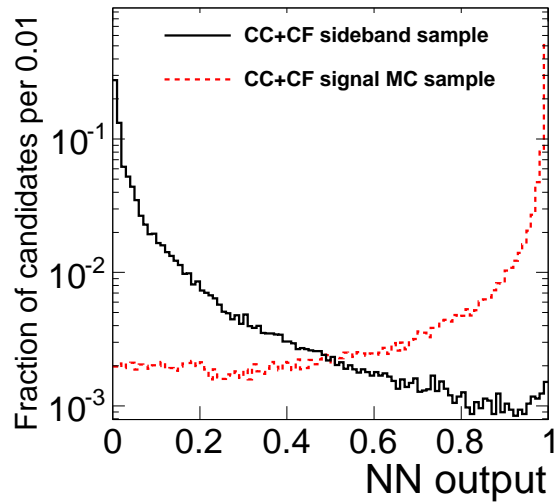}
  \end{center}
  \caption{
  Distributions of \nn\ for signal and background samples. The background 
  sample consists of dimuon mass sideband events from the first 7~\fb\ of data.
  }
  \label{fig:NNeff_NB}
\end{figure}

When training the NN, 80\% of each training sample is used for the actual training, while the remaining 20\% is used for validation and over-training tests.
The trained NN takes the input parameters for every event and returns
an output value \nn\  in the range [$0$ (background-like), 1 (signal-like)].
The combined CC and CF \nn\ distribution is shown in Fig.~\ref{fig:NNeff_NB} for
background and signal separately.   For various \nn\ requirements, the resulting signal and background efficiencies are given in 
Table~\ref{tab:eff_NB_CCCF}.

\begin{table*}
  \begin{center}
  \caption{ \label{tab:eff_NB_CCCF} Efficiency for various \nn\ 
   requirements, prior to MC weighting, signal (Sig.) and background 
   (Bgd.) events for the CC and CF trainings separately.  Uncertainties
   include only the statistical component.}
    \begin{tabular}{rrr}
      \hline\hline
      & \multicolumn{2}{c}{CC training} \\ 
      $< \nn$ & Sig. (\%) & Bgd (\%) \\ \hline
      0.999\:\: & $17.72 \pm 0.25$\:\: & $0.01 \pm 0.01$ \\
      0.998\:\: & $29.13 \pm 0.33$\:\: & $0.03 \pm 0.01$ \\
      0.997\:\: & $34.34 \pm 0.37$\:\: & $0.05 \pm 0.01$ \\
      0.996\:\: & $38.69 \pm 0.39$\:\: & $0.07 \pm 0.01$ \\
      0.995\:\: & $42.18 \pm 0.42$\:\: & $0.08 \pm 0.02$ \\
      0.994\:\: & $44.79 \pm 0.43$\:\: & $0.10 \pm 0.02$ \\
      0.993\:\: & $46.77 \pm 0.45$\:\: & $0.12 \pm 0.02$ \\
      0.992\:\: & $48.38 \pm 0.46$\:\: & $0.13 \pm 0.02$ \\
      0.991\:\: & $49.99 \pm 0.47$\:\: & $0.14 \pm 0.02$ \\
      0.990\:\: & $51.50 \pm 0.48$\:\: & $0.16 \pm 0.02$ \\
      0.980\:\: & $60.87 \pm 0.53$\:\: & $0.32 \pm 0.03$ \\
      0.970\:\: & $64.29 \pm 0.55$\:\: & $0.44 \pm 0.03$ \\
      0.960\:\: & $66.64 \pm 0.57$\:\: & $0.52 \pm 0.04$ \\
      0.950\:\: & $68.91 \pm 0.58$\:\: & $0.63 \pm 0.04$ \\
      \hline\hline
    \end{tabular}
    \begin{tabular}{cc}
      \hline\hline
      \multicolumn{2}{c}{CF training} \\ 
      Sig. (\%) & Bgd. (\%) \\ \hline
      $11.18 \pm 0.21$\:\: & $0.01 \pm 0.01$ \\
      $24.71 \pm 0.33$\:\: & $0.02 \pm 0.01$ \\
      $30.97 \pm 0.38$\:\: & $0.03 \pm 0.01$ \\
      $37.37 \pm 0.43$\:\: & $0.05 \pm 0.01$ \\
      $42.34 \pm 0.47$\:\: & $0.07 \pm 0.01$ \\
      $44.54 \pm 0.48$\:\: & $0.08 \pm 0.01$ \\
      $45.74 \pm 0.49$\:\: & $0.10 \pm 0.02$ \\
      $46.60 \pm 0.50$\:\: & $0.11 \pm 0.02$ \\
      $47.57 \pm 0.50$\:\: & $0.11 \pm 0.02$ \\
      $48.47 \pm 0.51$\:\: & $0.12 \pm 0.02$ \\
      $57.59 \pm 0.57$\:\: & $0.25 \pm 0.03$ \\
      $62.14 \pm 0.60$\:\: & $0.37 \pm 0.03$ \\
      $64.90 \pm 0.62$\:\: & $0.48 \pm 0.03$ \\
      $67.17 \pm 0.64$\:\: & $0.55 \pm 0.04$ \\
      \hline\hline
    \end{tabular}
  \end{center}
\end{table*}

\subsection{Neural network consistency checks}
We perform several consistency checks of the NN, including tests for \Mmm -\nn\ correlations, tests of over-training, and studies of MC mis-modeling of the 
\nn\ signal distribution.

\subsubsection{Check for correlations between \nn\ and dimuon mass}
\label{sec:XCheckCorrelation}
\begin{figure*}
  \begin{center}
    \includegraphics[width=0.35\textwidth]{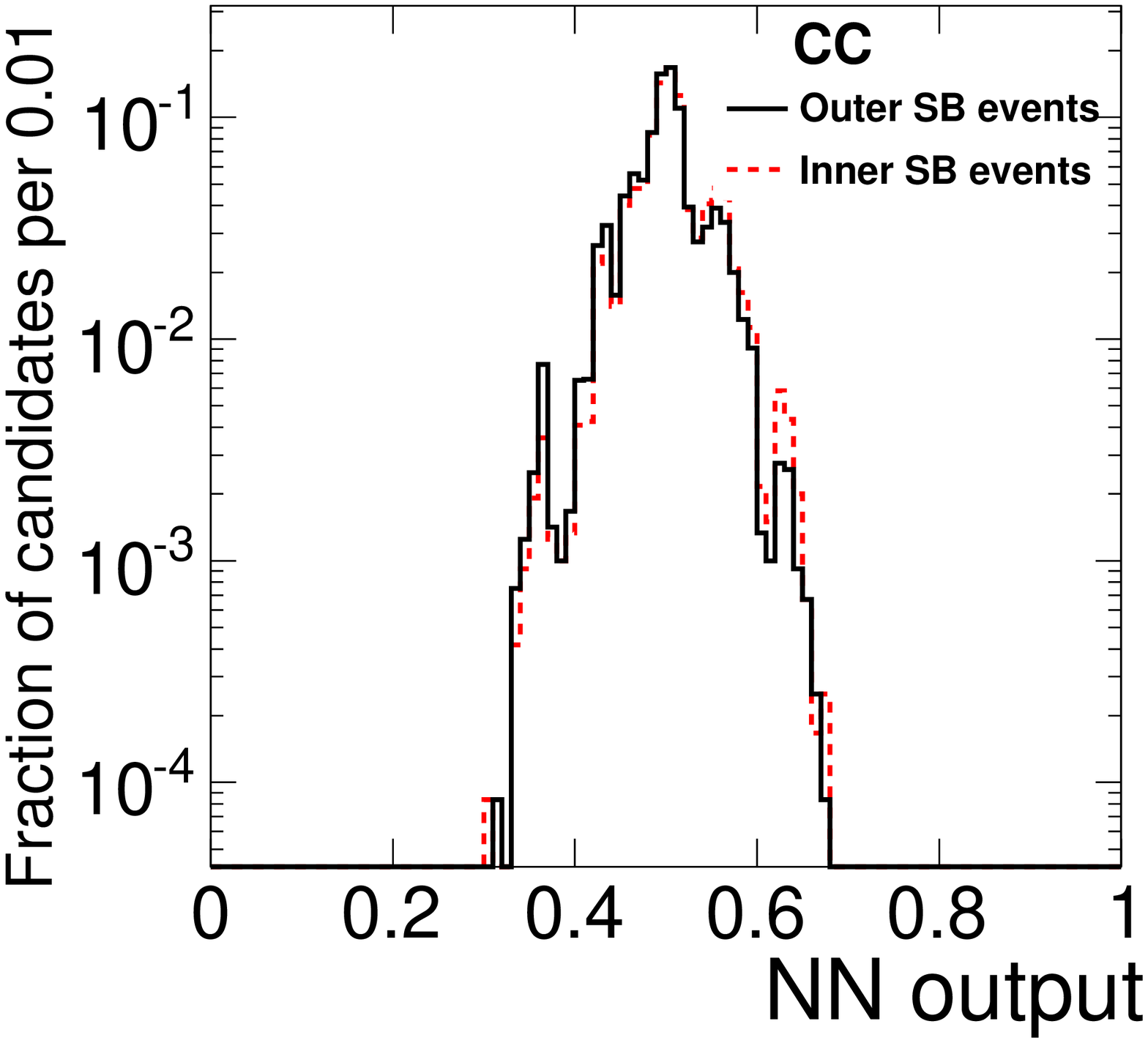}
    \includegraphics[width=0.35\textwidth]{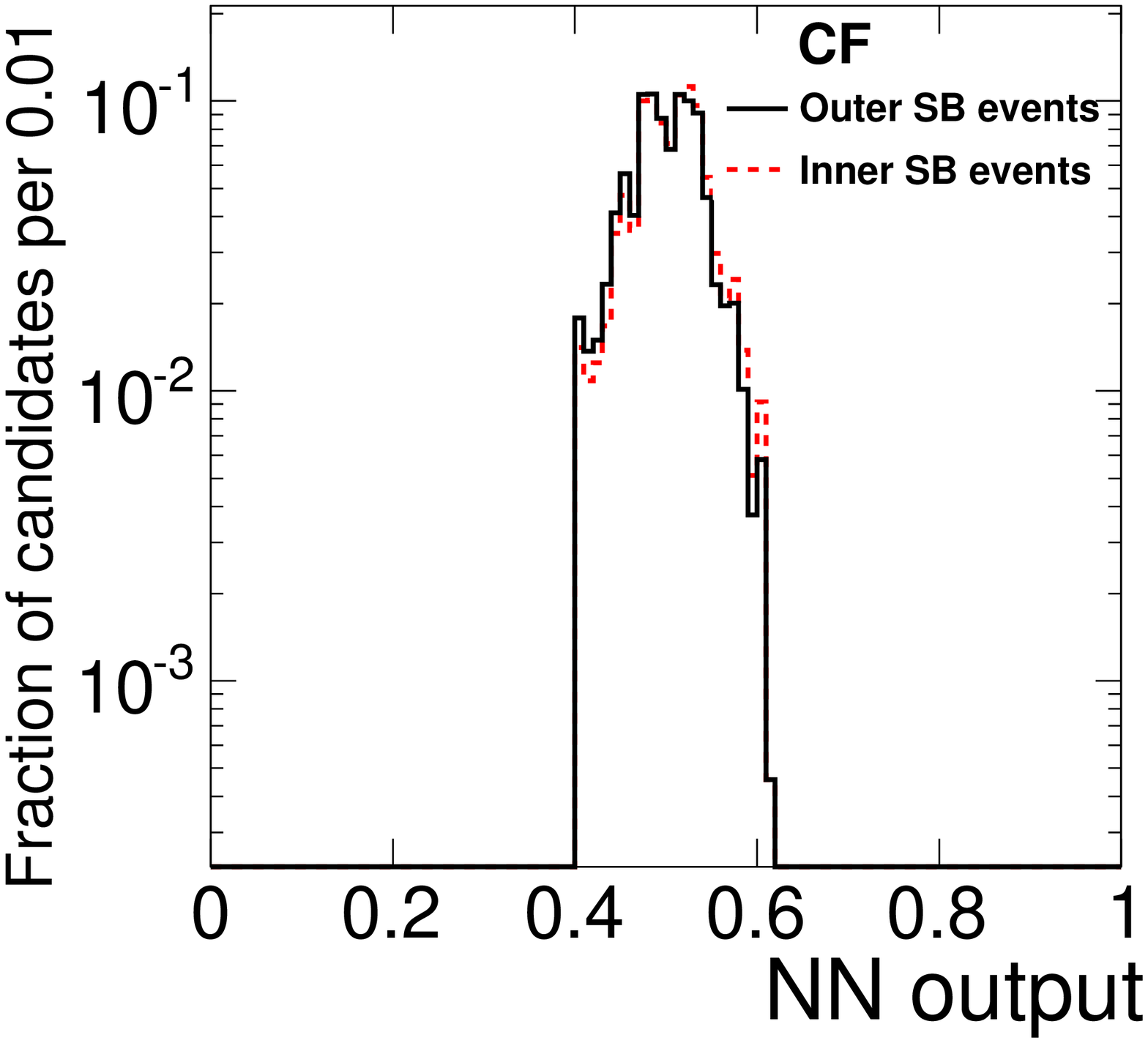}
  \end{center}
  \caption{
  Comparison of the \nn\ distribution of inner sideband events to outer
  sideband events using the full 10~\fb\ of data and a custom-trained NN 
  as described in Sec.~\ref{sec:XCheckCorrelation}.  The CC and CF channels 
  are shown separately.
  }
  \label{fig:NB_biascheck_CCCF}
\end{figure*}

We estimate the dominant combinatorial background in the dimuon mass signal region by linear interpolation from the dimuon mass sideband region.  
An unbiased estimate of the resulting combinatorial background requires that 
\nn\ is independent of the dimuon mass. We perform several studies of the 
\Mmm -\nn\ dependence. For example, we divide the sideband sample into an 
``inner'' region and an ``outer'' region according to dimuon mass.
The inner region is defined as $5.002 < \Mmm < 5.169$ GeV/$c^2$ and
$5.496 < \Mmm < 5.636$~GeV/$c^2$, and is used as a ``signal''
sample.  The outer sample is formed by the remaining events in the sideband, and is treated as a ``background'' sample.
Using these ``inner signal'' and ``outer background'' samples, the NN training is repeated with the same set of 14 input variables listed in Sec.~\ref{sec:nnInput} and the resulting \nn\ distributions are compared.  This check is based on the observation that event properties are nearly identical for events in the inner and outer regions and thus only differ by dimuon mass.  The 
resulting \nn\ distributions for the inner and outer samples are compared in
Fig.~\ref{fig:NB_biascheck_CCCF} for the CC and CF channels separately.  No significant difference in the \nn\ distribution is found, indicating that \nn\ is independent of dimuon mass.  This is 
strong evidence that the NN cannot use the 14 input variables to infer 
anything about the dimuon mass and that \Mmm\ and \nn\ are uncorrelated.

\begin{figure}
  \begin{center}
    \includegraphics[width=3.2in]{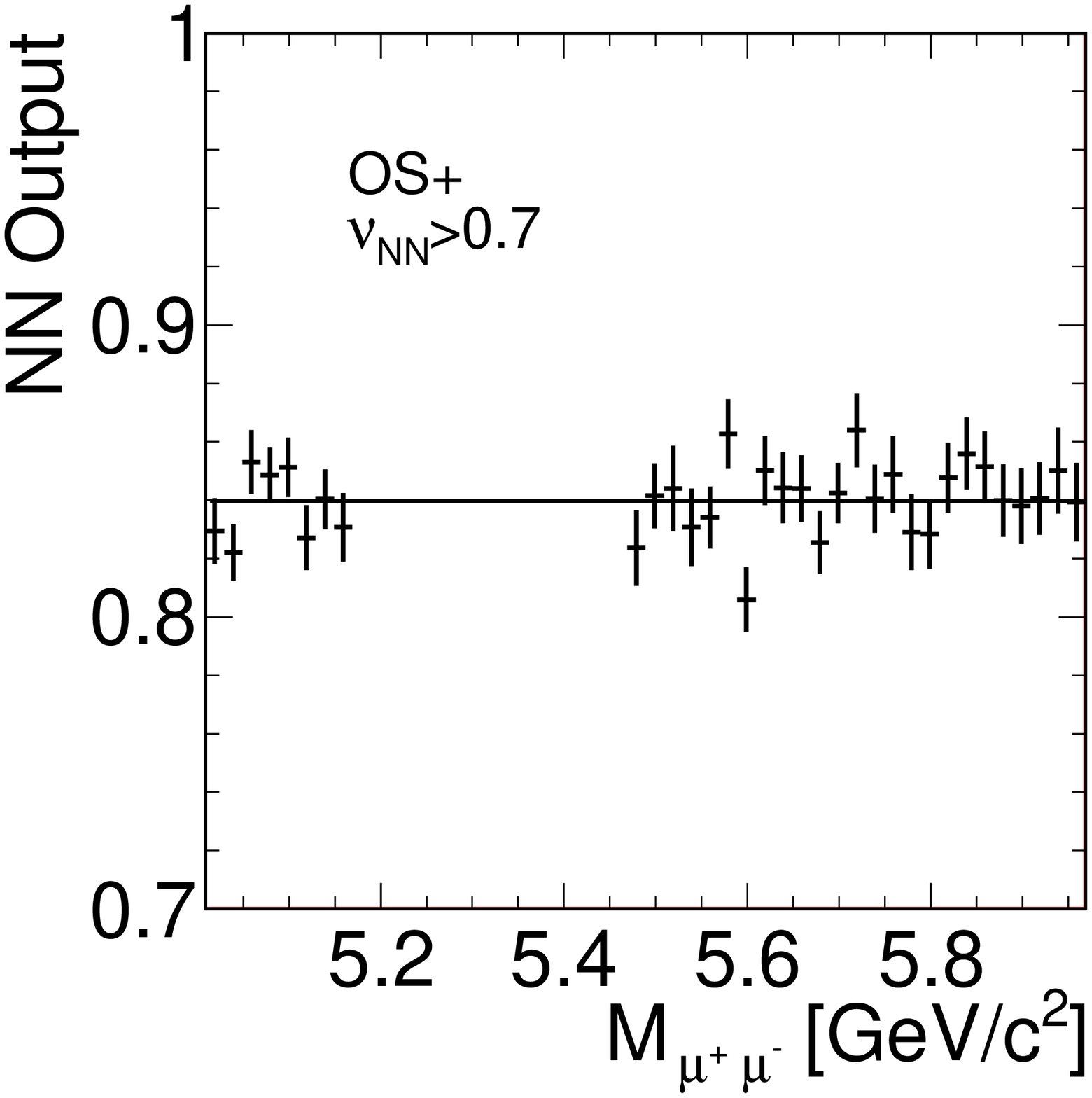}\\
    \includegraphics[width=3.2in]{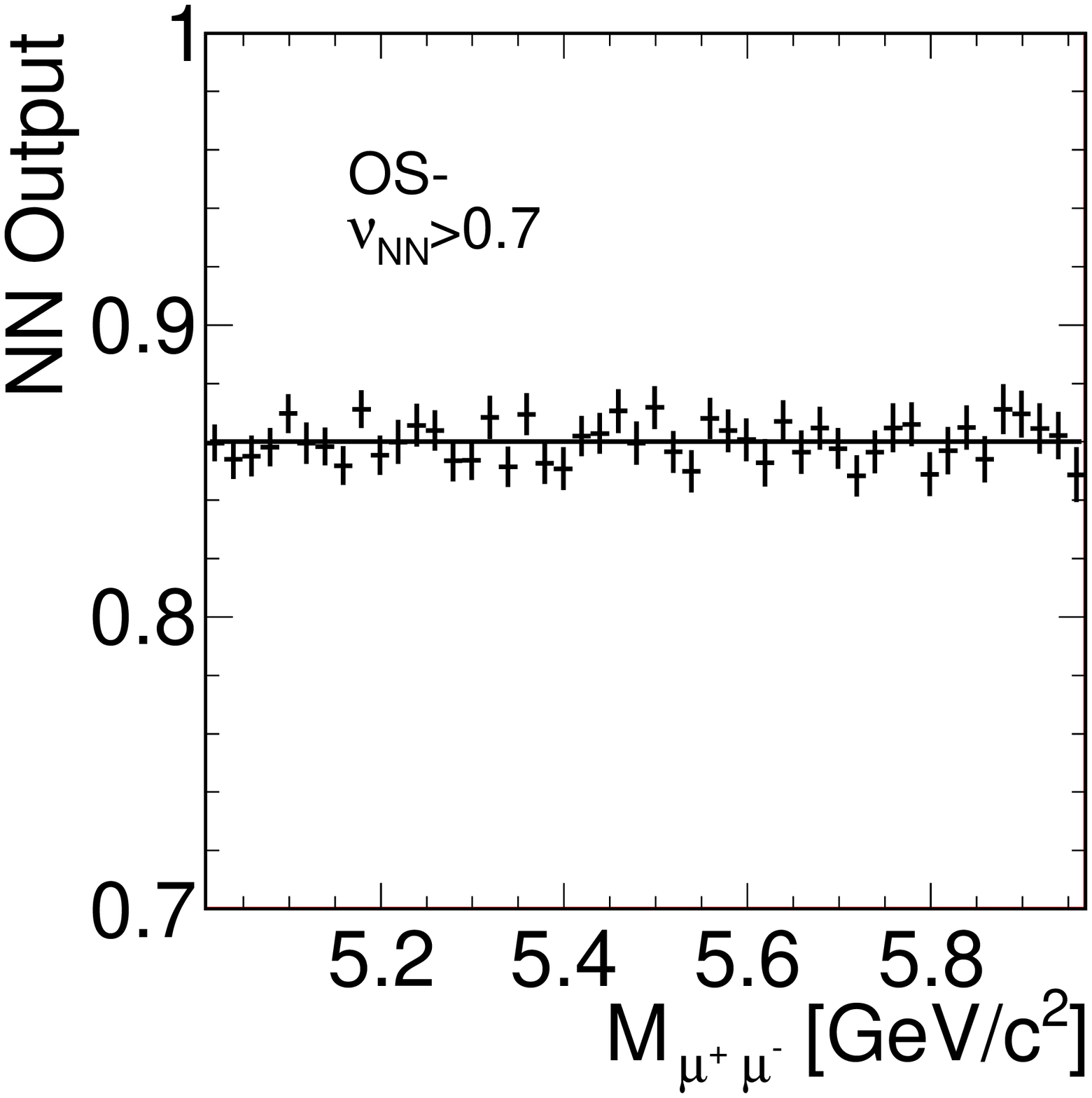}
  \end{center}
  \caption{
    Result of the \Mmm -\nn\ correlation check using the first 7~\fb\ of data.
    Top: correlation between average \nn\ and dimuon mass in the OS+ sample for which we keep the \bsmm\ $+$ \bdmm\ signal region blinded.
    Bottom: correlation between average \nn\ and dimuon mass in background-dominated OS$-$ sample.
  }
  \label{fig:NB_biascheck}
\end{figure}

As a further investigation of \Mmm -\nn\ dependencies, we study the average 
value of \nn\ as a function of the dimuon mass.  We use both the signal-sideband data, with two oppositely-charged muons and $\lambda>0$ (OS+), and a control sample consisting of events with two oppositely-charged muons but with proper-decay length $\lambda<0$ (OS$-$). The OS$-$ control sample is discussed further in Sec.~\ref{sec:controlSamples} and is dominated by prompt combinatorial background.   The resulting distributions, 
shown in Fig.~\ref{fig:NB_biascheck}, are consistent with a flat line, again indicating that \Mmm\ and \nn\ are independent.

\begin{figure}
  \begin{center}
    \includegraphics[width=0.40\textwidth]{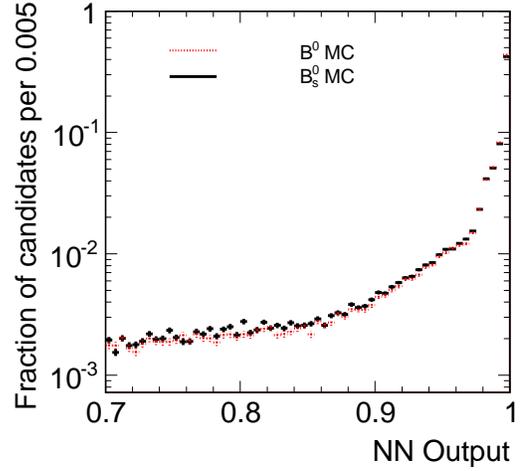}
  \end{center}
  \caption{Comparison of the \nn\ distribution for simulated samples of \bs\ and \bd\ signal events. The distributions have been 
  normalized to the same area over the entire NN output range, $0<\nn<1$.
  }
  \label{fig:bsBdNNOutput}
\end{figure}

To check for correlations of \nn\ with small variations of mass -- of the order of the mass difference between \bd\ and \bs -- a separate \bd\ MC sample is produced. 
The resulting \nn\ distribution is compared with that obtained from the \bs\ MC sample in Fig.~\ref{fig:bsBdNNOutput}. No significant difference between the \bd\ and 
\bs\ \nn\ distributions is found, again indicating the \Mmm\ and \nn\ are
independent.

We conclude that the choice of discriminating variables yields a NN classifier with excellent signal-to-background discrimination, while remaining independent of \Mmm\ and leaving the shape of the dimuon mass distribution unchanged.

\subsubsection{Check for NN over-training}
\label{sec:nnChecks}
\begin{figure}
  \begin{center}
    \includegraphics[width=0.40\textwidth]{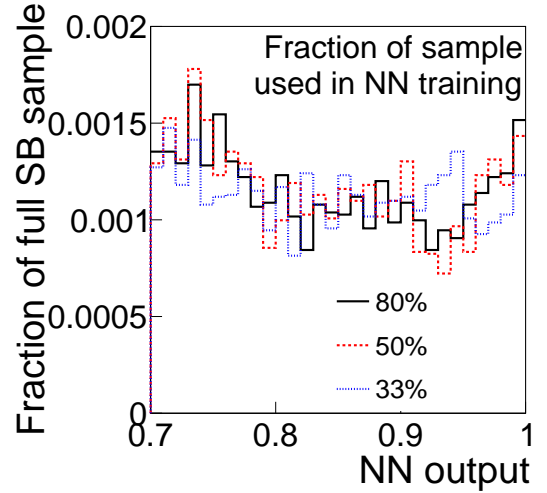}
  \end{center}
  \caption{The \nn\ distributions for networkks trained with 33\%, 50\%, 
    and 80\% of the input background sample taken from the dimuon mass 
    sideband events reconstructed in 7~\fb\ of data.}
  \label{fig:nnFrac}
\end{figure}

A portion of the background and signal training samples is set aside for internal validation tests.  These tests include metrics provided by the {\sc neurobayes} package that are sensitive to over-training.  All the metrics show that no over-training occurred for any of the neural nets used in this analysis.  As a further test of possible over-training, we repeat the NN optimization using 33\% and 50\% of the input background sample
in the training.  The resulting \nn\ distribution for sideband events is compared among these two trainings and the default training in 
Fig.~\ref{fig:nnFrac} and no significant differences are observed.  We conclude there is no evidence of NN over-training.

\subsubsection{Validation of the discriminant distributions}
\label{sec:nnValidation}
\begin{figure*}
  \begin{center}
    \includegraphics[width=0.35\textwidth]{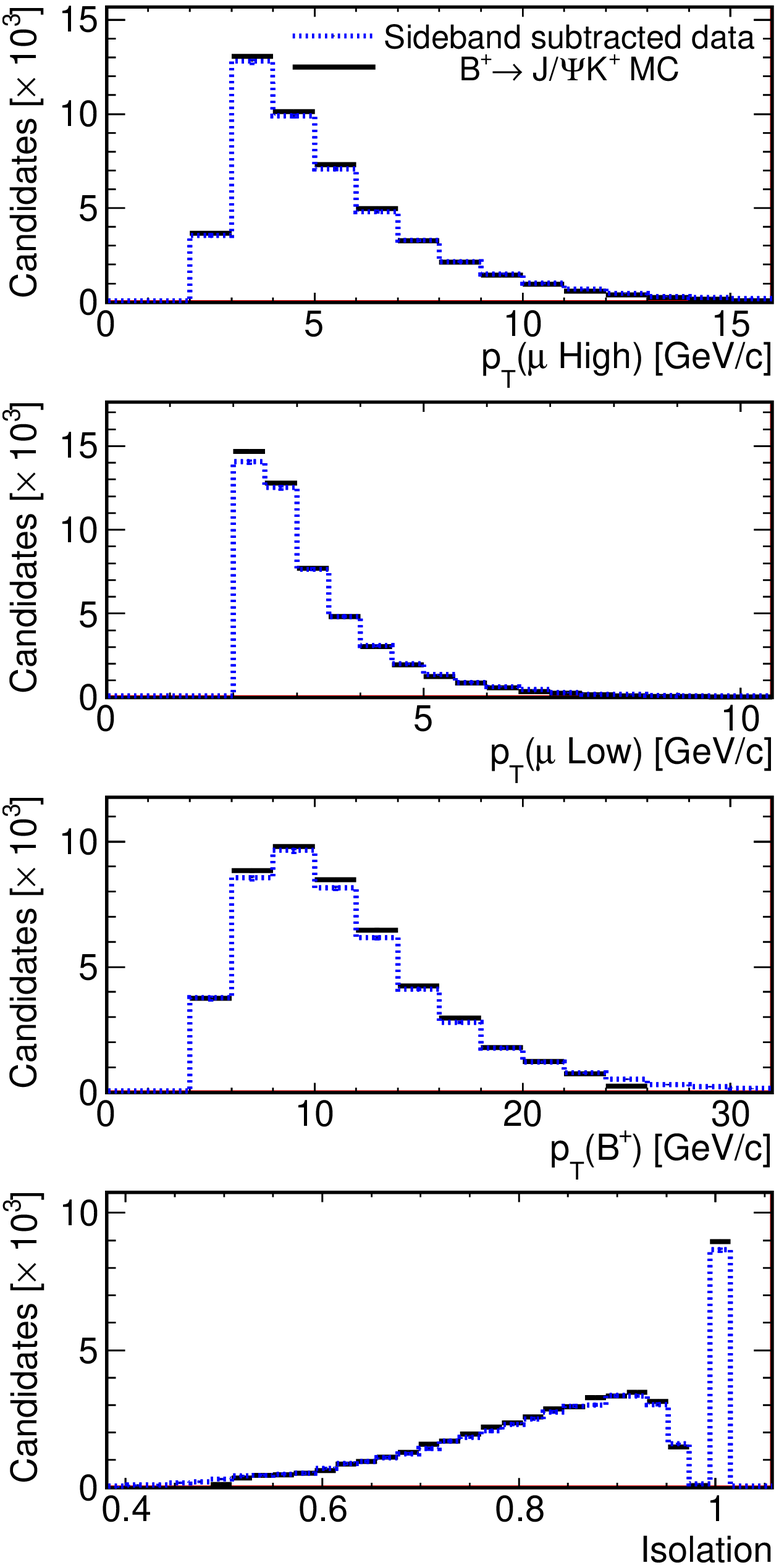}
    \includegraphics[width=0.35\textwidth]{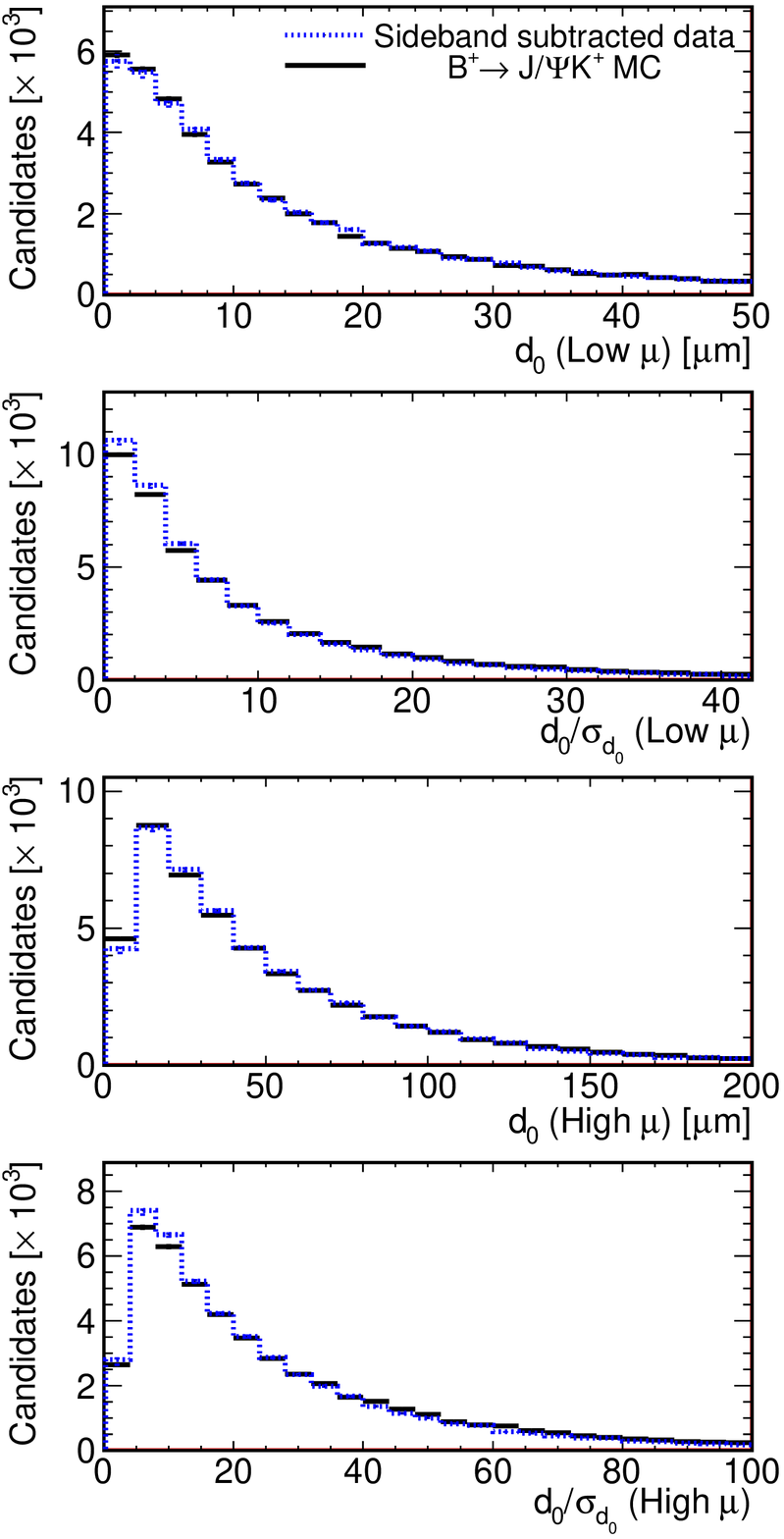}
  \end{center}
  \caption{ A comparison of \bjk\ sideband-subtracted data to simulated 
  \bjk\ events for
  a variety of kinematic and lifetime-related distributions, including some
  used as input to the neural net.}
  \label{fig:bPNNInputMCDt1}
\end{figure*}

\begin{figure*}
  \begin{center}
    \includegraphics[width=0.35\textwidth]{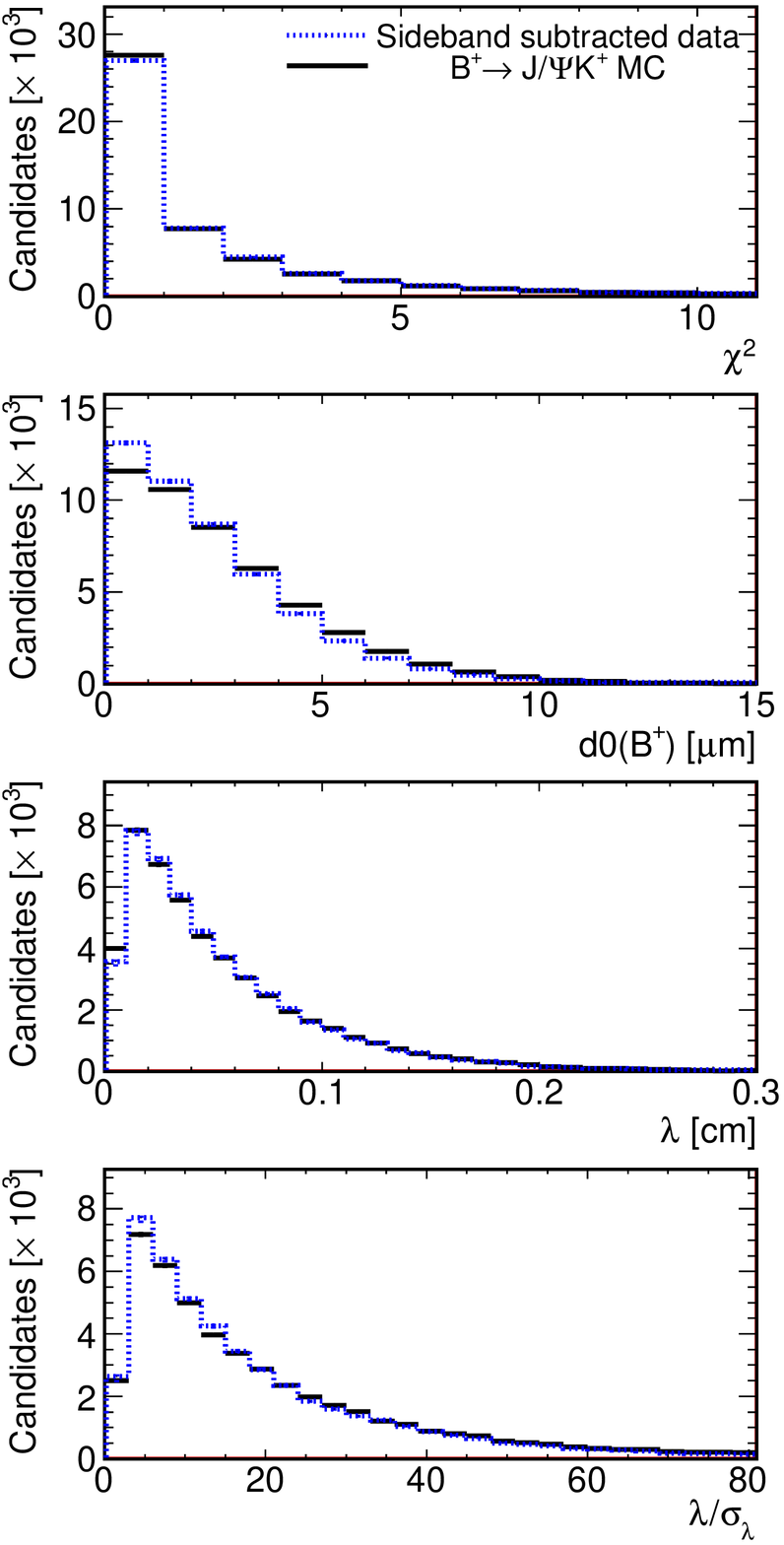}
    \includegraphics[width=0.35\textwidth]{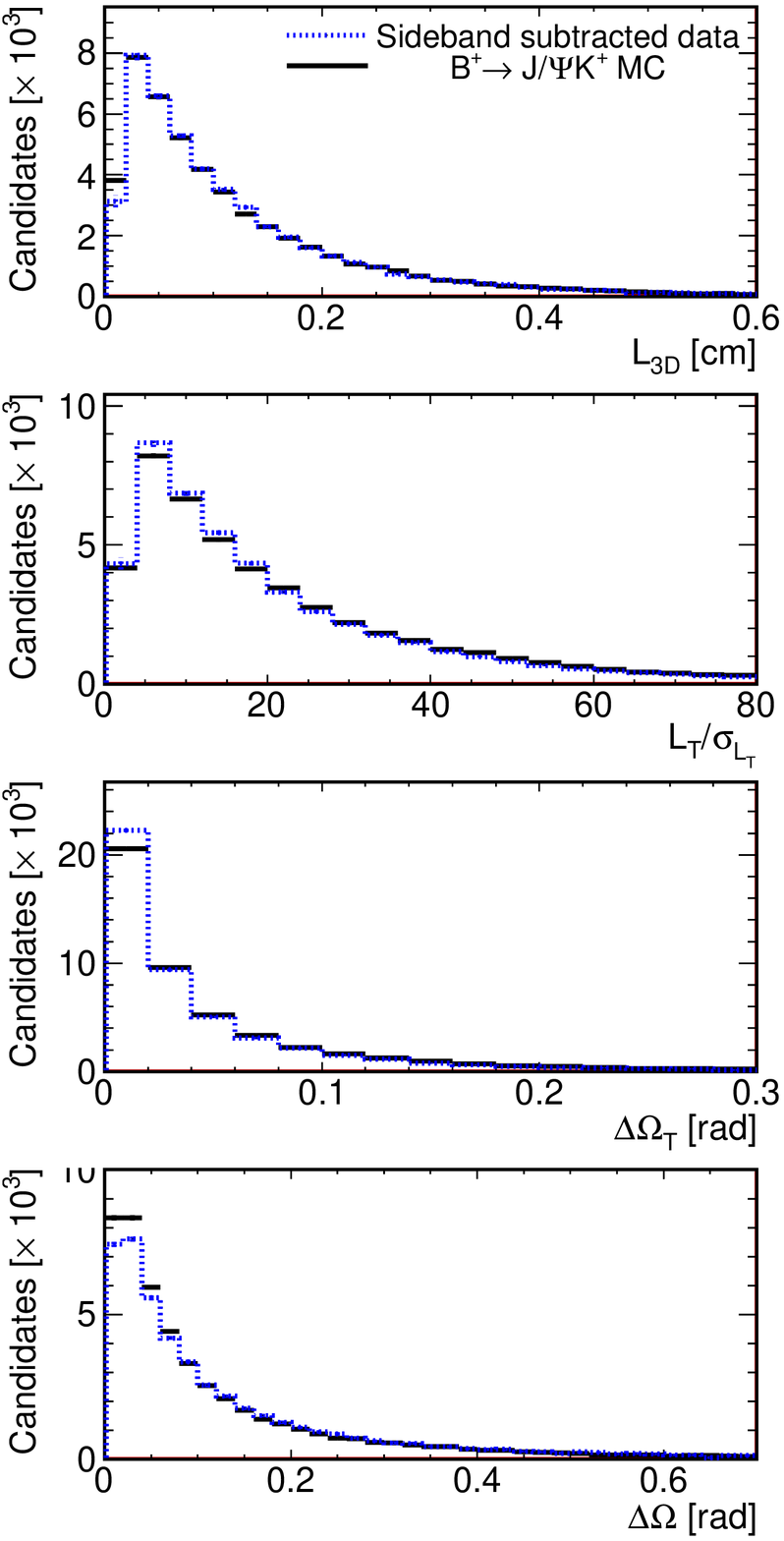}
  \end{center}
 \caption{ A comparison of \bjk\ sideband-subtracted data to simulated 
  \bjk\ event for several variables used as input to the neural net.}
  \label{fig:bPNNInputMCDt2}
\end{figure*}

\begin{figure}
  \begin{center}
    \includegraphics[width=0.40\textwidth]{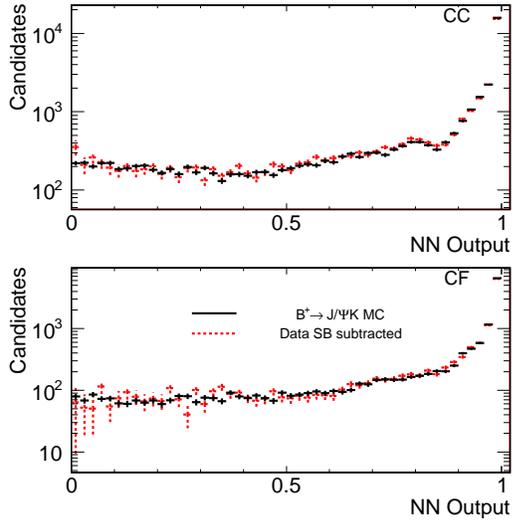}
  \end{center}
  \caption{Comparison of the \nn\ distribution for \bjk\ sideband-subtracted data to that obtained using simulated \bjk\ events for the CC and CF channels separately.  The MC is normalized to the observed number data events.}
  \label{fig:bpNNOutCC}
\end{figure}

\begin{figure}
  \begin{center}
    \includegraphics[width=0.40\textwidth]{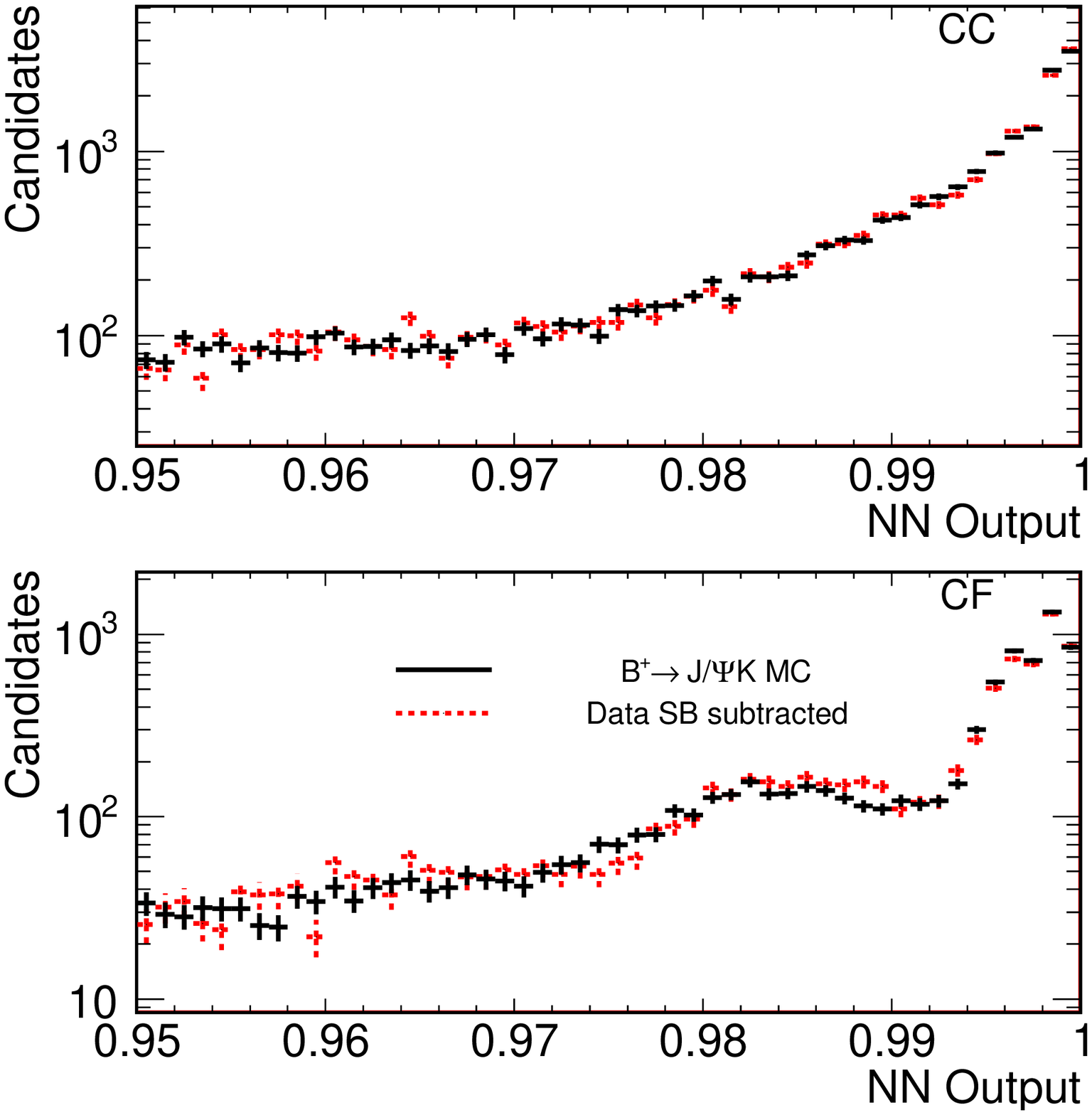}
  \end{center}
  \caption{Comparison of the $\nn > 0.95$ distribution for \bjk\ sideband-subtracted data to that obtained using simulated \bjk\ events for the CC and CF channels separately. The MC is normalized to the observed number of data events over the range $0 < \nn < 1$.}
  \label{fig:bpNNOutCF}
\end{figure}

We check the MC modeling of the 14 input variables and additional kinematic and lifetime variables using the sample of $B^+ \rightarrow J/\psi K^+$ events. Distributions from \bjk\ data and MC are compared in Figs.~\ref{fig:bPNNInputMCDt1}~and~\ref{fig:bPNNInputMCDt2}; in order to better mimic the
resolutions relevant for $B^0 \rightarrow \mu^+ \mu^- $ decays, the vertex variables use only the two muons from the $J/\psi$ of the \bjk\ decay while the $B$-hadron momentum variables, $p(B)$ and $p_T(B)$, and isolation variables use the three-track information.  The \nn\ distribution obtained from the \bjk\ MC is compared to that obtained using the sideband-subtracted \bjk\ data in Figures~\ref{fig:bpNNOutCC} and~\ref{fig:bpNNOutCF} for the CC and CF channels, respectively.  The small discrepancies observed are used to assign systematic uncertainties as discussed in Sec.~\ref{sec:nnEff}.


\section{Determining the \bsmm\ Branching Fraction}
\label{sec:signalEff}

The branching fraction \brbsdmm\ is determined using Eq.~(\ref{eq:brbsmm}), where 
$\alpha_{\bsd}$ is the geometrical and kinematic acceptance of the triggers
employed to collect the dimuon data set; $\epsilon_{\bsd}^{\mathrm{trig}} $ is the trigger efficiency for \bsd\ hadrons decaying to \mm\ within the acceptance; $\epsilon_{\bsd}^{\mathrm{reco}}$ is the efficiency of the reconstruction, baseline, and mass requirements for \mm\ pairs satisfying the 
trigger requirements; and $\epsilon_{\bsd}^{\mathrm{NN}} $ is the efficiency of the NN selection for events satisfying the trigger and baseline requirements and a given set of \nn\ requirements.  The equivalent efficiencies and acceptance for the normalization mode are indicated with the \bu\ subscript. No $\epsilon_{\bu}^{\mathrm{NN}}$ term appears since the NN is not applied to the normalization mode. The \bsdmm\ and \bjk\ acceptance and efficiencies are estimated separately to account for kinematic differences arising from the differing $B$-hadron decays.  The $b$-quark fragmentation-fraction-ratio 
$f_{u} / f_{s}$, the relevant product branching fractions for the normalization
mode $\mathcal{B}(\bjk\ra\mmk)$, and their uncertainties are taken from 
Ref.~\cite{PDG2010}. 
For the \bd\ search, $f_s$ is replaced by $f_d$ and the fragmentation ratio is 
set to unity.  Due to their differing sensitivities, the CC and CF channels are treated separately and then combined to yield the final result.
Normalizing the observed signal rate to the rate of an abundant, well-known, and kinematically-similar decay results in a significant 
reduction to the total uncertainty since systematic effects largely cancel in the acceptance and efficiency ratios of Eq.~(\ref{eq:brbsmm}).

 \begin{widetext}
  \begin{equation}
    \label{eq:brbsmm}
    \brbsdmm = \frac{N_{\bsd}}{N_{\bu}}\cdot\frac{\alpha_{\bu}}{\alpha_{\bsd}}
    \cdot\frac{\epsilon_{\bu}^{\mathrm{trig}} }{\epsilon_{\bsd}^{\mathrm{trig}} }
    \cdot\frac{\epsilon_{\bu}^{\mathrm{reco}} }{\epsilon_{\bsd}^{\mathrm{reco}} }
    \cdot\frac{1}{\epsilon_{\bsd}^{\mathrm{NN}} }
    \cdot\frac{f_{u}}{f_{s}}\cdot \mathcal{B}(\bjk\ra\mm K^{+})
  \end{equation}
 \end{widetext}

We define the single-event-sensitivity (SES) as the branching fraction determined
from Eq.~(\ref{eq:brbsmm}) when setting $N_{\bsd}=1$. The SES approximates the 
smallest signal branching-fraction to which the analysis is sensitive.  The methods 
used to estimate the inputs to Eq.~(\ref{eq:brbsmm}) are described below and the 
results are summarized in Table~\ref{tab:sigSummary}.  Combining the CC and CF 
channels yields a SES for the \bsmm\ search of $1.4\times 10^{-9}$ with an 18\% total uncertainty.  The combined SES for the \bdmm\ search is a factor of $3.5$ smaller with a reduced uncertainty of about 12\% since the 
fragmentation ratio $f_d/f_u$ is not relevant.

\subsection{Acceptance}
\label{sec:acc}
The acceptances are determined using \bsmm\ and \bjk\ MC simulations for \bs\ and \bu\ mesons that satisfy $|y|<1.0$ and $p_T(\bs)>4$ GeV/$c$,
where $y=\frac{1}{2} \ln\left(\frac{E+p_z}{E-p_z}\right)$ is the rapidity. Both muons must satisfy the fiducial and kinematic requirements of the trigger discussed in Sec.~\ref{sec:muonTrig}. Muons are required to have 
$p_T>2.0 $ GeV/$c$ if detected in the CMU, and $p_T>2.2 $ GeV/$c$ if detected in the CMX and their 
trajectories must extrapolate to the active fiducial-volumes of the muon, COT, and L00+SVXII 
systems. The kaon in the normalization channel must have $p_T>1.0$ GeV/$c$ and its trajectory must extrapolate to the active fiducial-volumes of the COT and L00+SVXII systems.
Effects from COT-track reconstruction, multiple scattering, and stub-track matching are included in the reconstruction efficiencies discussed below. 
Systematic uncertainties on the acceptances are assessed by varying the $b$-quark mass, fragmentation modeling, and renormalization and factorization scales in the MC generation by one standard-deviation-uncertainty and quantifying the resulting change in the acceptance ratio. Additionally, we assign variations of the acceptance due to changes in the size of the \pp\ luminous region as a systematic uncertainty. The observed differences are added in quadrature. The final acceptance ratio including statistical and systematic uncertainties is given in Table~\ref{tab:sigSummary} for the CC and CF channels separately.

\begin{table*}[ht]
  \begin{center}
\caption{\label{tab:sigSummary} A summary of the inputs used in 
  Eq.~(\ref{eq:brbsmm}) to determine \brbsmm\ in the CC and CF channels separately.
  The relative uncertainties are given parenthetically. The uncertainties for 
  the trigger efficiency ratios are significantly smaller than 1\% and are 
  denoted by $0.00$. The single-event-sensitivities (SES) for $\nn > 0.70$ are 
  given in the last row. Combining CC and CF gives a SES for the \bsmm\ search
  of $1.4\times 10^{-9}\;( \pm 18\% )$.  The combined SES for the \bdmm\ search is
  $3.9\times 10^{-10}\;( \pm 12\% )$.
}
    \begin{tabular}{ccrcr} \hline\hline
      & \multicolumn{2}{c}{CC}  
      & \multicolumn{2}{c}{CF}  \\ \hline
      $\alpha_{\bu} / \alpha_{\bs}$
      & $0.31\pm0.02$ &($\pm6\%$)\:\:\: 
      & $0.20\pm0.01$ &($\pm7\%$) \\
      $\epsilon_{\bu}^{\mathrm{trig}}  / \epsilon_{\bs}^{\mathrm{trig}} $
      & $1.00\pm 0.00$ &( - )\:\:\: 
      & $0.98\pm 0.00$ &( - ) \\
      $\epsilon_{\bu}^{\mathrm{reco}}  / \epsilon_{\bs}^{\mathrm{reco}} $
      & $0.85\pm0.06$ &($\pm 8\%$)\:\:\: 
      & $0.84\pm0.06$ &($\pm 9\%$) \\
      $\epsilon_{\bs}^{\mathrm{NN}} (\nn>0.70)$
      & $0.92\pm0.04$ &($\pm 4\%$)\:\:\: 
      & $0.86\pm0.04$ &($\pm 4\%$) \\
      $\epsilon_{\bs}^{\mathrm{NN}} (\nn>0.995)$
      & $0.46\pm0.02$ &($\pm 5\%$)\:\:\: 
      & $0.47\pm0.02$ &($\pm 8\%$) \\
      $N_{\bu}$
      & $28081\pm 219$     &($\pm1\%$)\:\:\: 
      & $12144\pm 153$     &($\pm1\%$) \\
      $f_{u} / f_{s}$
      & $3.55\pm0.47$   &($\pm13\%$)\:\:\: 
      & $3.55\pm0.47$   &($\pm13\%$) \\
      $\mathcal{B}(\bjkmm)$
      & \multicolumn{4}{c}{$(6.01\pm0.21)\times10^{-5}$\:\:\: ($\pm4\%$)} \\
      & & & & \\
      SES ($\nn > 0.70$)
      & $2.2\times 10^{-9}$   &($\pm 18\%$)\:\:\: 
      & $3.3\times 10^{-9}$   & ($\pm 18\%$)\\ 
      \hline\hline
    \end{tabular}
    
  \end{center}
\end{table*}

\subsection{Trigger efficiencies}
\label{sec:trigEff}
The trigger efficiency is measured separately for the L1, L2, and L3 dimuon triggers  using data control samples collected using unbiased 
trigger selections identifying a high-purity sample of dimuon events that satisfy 
the criteria of the trigger under study.  The efficiency is determined from the fraction of events in this sample for which also the trigger under study fires.  Dimuon data collected requiring only trigger L1 are used to measure the L2 and L3 efficiencies, which exceed 99\%.  The L1 dimuon trigger efficiency is determined as the product of the L1 efficiency for each muon separately.  The single-muon L1 efficiencies are measured using a tag-and-probe method on samples of high-purity $\jp\ra\mm$ events collected with a trigger L1 that requires only one muon.  The muon firing the L1 trigger used to collect the \jp\ control sample is identified as the ``tag'', while the second muon from the \jp\ decay is unbiased with respect
to the trigger whose efficiency we are studying and can be used a ``probe''.  Dimuon mass-sideband subtraction is used to remove effects of the small background present in the sample. The single-muon 
L1 efficiencies are parametrized as a function of the date the data were recorded and the track $p_T$, $| \eta |$, and $\phi$ for CMU and CMX muons separately.  The parametrization by date accounts for significant changes in detector operating conditions arising from variations in trigger configuration, COT performance, or Tevatron beam parameters.  The parametrization in $p_T$ describes the rapidly changing L1 efficiency near the trigger $p_T$ threshold.
The parametrization in $| \eta |$ describes changes in the L1 efficiency due to the increased ionization path-lengths of tracks traversing the COT at large 
$\left|\eta\right|$, which may increase the probability for the corresponding hits to exceed the noise threshold and fire the trigger.  The parametrization in $\phi$ is primarily important for 
a small amount ($0.2$~fb$^{-1}$) of early data  for which the COT gain was temporarily degraded in the bottom portion of the chamber~\cite{COTaging}.  The resulting single-muon 
L1 efficiency is about 96\% for the muons relevant for this analysis and plateaus at 
about 99\% for muons with $p_T > 5$~GeV/$c$.  The dimuon L1 efficiency is estimated by
convolution of the single-muon efficiencies with the 
$( p_T^{\mu^+},\; | \eta^{\mu^+} |,\; \phi^{\mu^+},\; p_T^{\mu^-},\; | \eta^{\mu^-} |,\; \phi^{\mu^-} )$ distribution obtained from \bsmm\ and \bjk\ MC for events in the
geometrical and kinematic acceptance of the trigger.  This convolution yields a dimuon L1 efficiency of about 93\% for both the \bsmm\ and \bjk\ samples in each of the CC and CF
channels.  The total systematic uncertainty for the trigger L1 efficiency is in the 1--2\% range and is dominated by
variations of the single-muon efficiency as a function of the isolation of the muon and by differences between muons detected in the $\eta >0$ and $\eta<0$ volumes of the detector.  Double-muon correlations are studied and found to be negligible.  When propagating the single-muon uncertainties to the dimuon efficiency, the uncertainties are taken to be 100\% correlated between the two muons.

The total trigger efficiency is determined as the product of the L1, L2, and L3 trigger efficiencies for the \bsmm\ and \bjk\ modes separately. The total uncertainty of the efficiency ratio is much less than $0.01$ and is estimated by treating the \bsmm\ and \bjk\ total uncertainties as 100\% correlated. The final trigger efficiency ratios for the CC and CF channels and their associated uncertainties are shown in Table~\ref{tab:sigSummary}.

\subsection{Reconstruction efficiencies}
\label{sec:recoEff}
The total reconstruction efficiency is factorized into the efficiency for 
COT-track reconstruction, muon-stub reconstruction, association of L00+SVXII hits to the COT track, the dimuon-vertex reconstruction, and the
dimuon-mass reconstruction efficiencies. 
For the normalization mode, the kaon COT, the kaon silicon, and the \bu\ vertex reconstruction efficiencies are also evaluated.

\subsubsection{COT track reconstruction efficiency}
\label{sec:cotEff}
The probability for identifying a single track in the COT is evaluated by embedding 
MC tracks into real data events and measuring the fraction of those tracks that are successfully reconstructed and satisfy the CDF standard-track-quality requirements.  
Hits from MC charged-particle tracks are inserted into events from beam data. ÊIn readout channels where MC
hits overlap with hits in the event record, the hits are merged. ÊThe MC
is tuned so that quantities such as deposited charge per hit and
single-hit position resolution match those observed in the data.
The dependencies of these quantities on various parameters, such as the track $\eta$, the electric charge of the track, and the track isolation, are also accurately reproduced in the MC.  For charged particles fully fiducial to the COT and 
with $p_T>1.5$ GeV/$c$, the COT reconstruction is consistent with being fully efficient.  Systematic uncertainties include variations of the MC parameters affecting the COT hit distributions and account for small ($<1\%$) observed variations as a function of track isolation, $p_T$, and $\eta$.  The dimuon efficiency is taken to be the square of the single-track efficiency and the resulting dimuon efficiency ratio is $1.00\pm0.01$ for both the CC and CF channels.

\subsubsection{Muon stub reconstruction efficiency}
\label{sec:stubEff}
The muon-stub reconstruction efficiencies are determined from \jpmm\ events using a tag-and-probe method similar to that described in Sec.~\ref{sec:trigEff}.  The ``tag'' muon candidate is required to have a COT track satisfying the relevant trigger and COT baseline requirements of Sec.~\ref{sec:baseline} matched to a muon stub, while the other muon is only required to pass the COT baseline requirements.  The muon-stub reconstruction 
efficiency is determined using the fraction of events for which the second muon is also 
matched to a muon stub. The efficiency is estimated both with data and MC samples and the 
efficiency measured with data is divided by the efficiency measured with MC to account for geometric losses already included in the acceptance estimate.
As a consistency check of the efficiency estimates, the same procedure is used
to determine the efficiency of high-$p_T$ muons using a sample of high-purity $Z^0\to\mm$ decays. The difference in efficiencies is taken as the systematic
uncertainty. The ratio of dimuon efficiencies for both the CC and CF is $1.00\pm0.03$.

\subsubsection{Muon ID efficiency}
\label{sec:muidEff}
The combined efficiency of the \dedx\ and muon-likelihood requirements is determined using a \jpmm\ sample by comparing the signal yield with and without the application of
these muon identification criteria.  The single-muon efficiency is determined as a function of $p_T$.  The total dimuon efficiency is evaluated by convolution of the 
$( p_T^{\mu^+},\; p_T^{\mu^-} )$ distribution from \bsmm\ and \bjk\ MC for events within the acceptance and satisfying the trigger, COT, and muon-stub reconstruction requirements.   This efficiency is cross-checked {\it in situ} using \bjk\ data events. 
The difference between the efficiency in the \bjk\ MC events and in the \bjk\ 
sideband-subtracted data is assigned as a systematic uncertainty.  The final dimuon 
efficiency ratio for the muon identification requirements is
$1.01\pm0.03$ for both the CC channel and CF channels.

\subsubsection{L00+SVXII association efficiency}
\label{sec:svxEff}
The efficiency of associating L00 and SVXII hits to muon tracks reconstructed in the COT is estimated using \jpmm\ events in a manner similar to that described in Sec.~\ref{sec:stubEff}.  The \jp\ sample includes only muons reconstructed as a 
track in the COT matched to a muon stub in the CMU or CMX and surviving the muon 
identification requirements.  The efficiency for associating at least three $r\phi$ hits from the L00+SVXII silicon layers is evaluated as the fraction of this sample for 
which the muon tracks satisfy that criteria.  The variation of the efficiency on 
track $p_T$ , isolation, and azimuthal dimuon opening angle is used to assign systematic uncertainties.  The resulting dimuon efficiency ratio is $1.00\pm0.03$ for both the CC and CF channels.

\subsubsection{Dimuon vertex efficiency}
\label{sec:dimuvtxEff}
The efficiency of the dimuon-vertex requirements specified in Sec.~\ref{sec:baseline} is estimated using simulated \bjk\ and \bsmm\ samples.   The resulting efficiency is found to be consistent with the efficiency determined using sideband-subtracted \bjk\ data events, thus verifying the accuracy of the MC modeling. The resulting ratio of efficiencies for the dimuon-vertex requirements is $0.99\pm0.01$ for the CC and CF channels.

\subsubsection{Dimuon mass efficiency}
\label{sec:massEff}
The efficiency of the dimuon-mass requirements is estimated using simulated \bsmm\ and \bjk\ events surviving the baseline, vertex, and trigger requirements.  Comparisons of the mean and width of the invariant mass distribution using data and MC samples of $\jp\to\mm$ and \bjkmm\ events reveal discrepancies at the 10\% level for the width, which are used to assign systematic uncertainties.  Since the signal-search mass-windows correspond to $\pm2.5\sigma_{m}$ and have high efficiency, the systematic uncertainties negligibly affect the efficiency ratio, which is $1.00$.

\subsubsection{\bjk\ reconstruction efficiency}
\label{sec:bjkEff}
In addition to the dimuon efficiencies discussed above, the total reconstruction efficiency for the \bjk\ events also includes the efficiency for reconstructing the kaon as a COT track, for associating L00+SVXII 
hits to the COT track, and for the \mmk\ vertex requirements.
The kaon COT efficiency is estimated using the method of Sec.~\ref{sec:cotEff} and
is $0.964\pm0.016$ for both the CC and CF channels. This value is lower than that for
muons due to inefficiencies resulting from kaon interaction with matter and
due to the lower transverse-momentum threshold ($p_T^K > 1.0$~GeV/$c$) employed for the
kaons.  The uncertainty includes variations of the relevant kaon-matter interaction 
cross sections; of the material modeling in the MC; and efficiencies as a function of kaon isolation, $p_T$, and $\eta$.
The efficiency for associating at least three $r\phi$ hits from L00+SVXII to a good kaon
COT track is evaluated using \bjk\ data events.  The \bjk\ signal 
yield is compared before and after applying the L00+SVXII requirements on the kaon track and
the ratio is used as a measure of the efficiency.  The \mm\ pair in the events must satisfy all the relevant dimuon baseline requirements of Sec.~\ref{sec:baseline}, including the silicon requirements discussed in Sec.~\ref{sec:svxEff} and the dimuon vertex requirements 
of Sec.~\ref{sec:dimuvtxEff}.  The resulting efficiencies are $0.942\pm0.002$ and $0.948\pm0.003$ for the CC and CF channels, respectively. 
The efficiency of the baseline requirements relevant for the \mmk\ vertex is 
also directly determined using sideband-subtracted \bjk\ data and is
$0.938\pm0.006$ and $0.919\pm0.010$ for the CC and CF channels, respectively.

\subsubsection{Total reconstruction efficiency}
\label{sec:totalEff}
The total dimuon-reconstruction efficiency-ratio is the product over all the dimuon efficiency ratios described above.  The final efficiency ratio relevant for 
Eq.~(\ref{eq:brbsmm}) is obtained by including the product of the kaon reconstruction
efficiencies and the \bjk\ vertex efficiency.  The resulting ratios for the CC and CF 
channel and their associated total uncertainties are shown in Table~\ref{tab:sigSummary}.

\subsection{Efficiency of the NN selection}
\label{sec:nnEff}
The final NN selection criteria divide the surviving
events into eight bins in \nn .  The bin boundaries are determined in an optimization 
described in Sec.~\ref{sec:nnOpt}. The efficiency associated with each \nn\ bin is estimated from simulated
\bsmm\ events meeting all other selection criteria.  In Table~\ref{tab:sigSummary} 
the efficiency summed over all eight bins ($\nn > 0.70$) and for the highest bin alone 
($\nn > 0.995$) are shown for the CC and CF channels.  For the likelihood fits described 
in Sec.~\ref{sec:nnOpt}, the efficiency given in Table~\ref{tab:nnEff} is used in each bin separately. The NN efficiency determined from the \bdmm\ MC sample 
is consistent with the results of Table~\ref{tab:nnEff}. Recall that the NN is not applied in selecting 
the \bjk\ sample relevant for the normalization in Eq.~(\ref{eq:brbsmm}). 

The MC modeling of the NN efficiency is checked by comparing the signal efficiencies for \bjk\ decays reconstructed in data and MC using the \bjk\ MC sample described in
Sec.~\ref{sec:nnValidation}.  The comparisons are shown in Figs.~\ref{fig:bpNNOutCC} 
and~\ref{fig:bpNNOutCF} and are quantified in Table~\ref{tab:nnEffXCheck}. Some difference in performance is observed and likely arises
from radiation-damage-induced silicon-sensor degradation that is not completely simulated.
For most NN bins the difference between MC and sideband-subtracted data does not exceed $2.5$ times the associated statistical uncertainty.  The most significant
deviation occurs in the $\nn > 0.995$ bin and is 3.4\% and 7.0\% for the CC and CF, respectively.  These differences are assigned as systematic uncertainties to this bin. 

The distributions of simulated \bjk\ and \bsmm\ events are weighted to match the $p_T$ and isolation distributions
from \bjk\ and \bsjp\ data, respectively. A systematic uncertainty of 4\% is assigned 
as determined by varying the observed $B$-meson $p_T$ and isolation distributions 
within their statistical uncertainties, repeating the weighting, and quantifying
the resulting changes in the NN efficiencies. 

\begin{table*}[htb]
  \begin{center}
    \caption{\label{tab:nnEff} The NN efficiency for each \nn\ bin for the
      CC and CF channels after $p_T$ and isolation weighting.  The 
      uncertainties include only the statistical component.}
    \begin{tabular}{crr} \hline\hline
      NN bin & \multicolumn{1}{c}{CC} & \multicolumn{1}{c}{CF} \\ \hline
      $0.700<\nn<0.760$\:\: & 2.2$\pm$0.1 & 2.3$\pm$0.1\\
      $0.760<\nn<0.850$\:\: & 4.1$\pm$0.1 & 4.4$\pm$0.1\\
      $0.850<\nn<0.900$\:\: & 2.9$\pm$0.1 & 3.4$\pm$0.1\\
      $0.900<\nn<0.940$\:\: & 4.5$\pm$0.1 & 4.7$\pm$0.1\\
      $0.940<\nn<0.970$\:\: & 8.3$\pm$0.1 & 6.2$\pm$0.1\\
      $0.970<\nn<0.987$\:\: & 10.9$\pm$0.1 & 10.2$\pm$0.1\\
      $0.987<\nn<0.995$\:\: & 12.5$\pm$0.1 & 8.5$\pm$0.1\\
      $0.995<\nn<1.000$\:\: & 46.1$\pm$0.3 & 46.8$\pm$0.3\\\hline\hline
    \end{tabular}
  \end{center}
\end{table*}

\begin{table*}[htb]
  \caption{\label{tab:nnEffXCheck} Relative difference in NN bin efficiency 
   between \bjk\ data and MC. A positive (negative) difference indicates that 
   the MC efficiency is higher (lower) than the data efficiency. The differences 
   normalized to the associated statistical uncertainty 
   are given in parenthesis. 
  }
  \begin{tabular}{crrrr}
    \hline\hline
    NN bin &\multicolumn{2}{c}{CC} & \multicolumn{2}{c}{CF} \\ 
    \hline    
    $0.700<\nn<0.760$\:\: &--8.3\% &(--1.6$\sigma$)\:\: & --5.3\% &(--0.7$\sigma$) \\  
    $0.760<\nn<0.850$\:\: &--8.5\% &(--2.3$\sigma$)\:\: & --7.9\% &(--1.4$\sigma$) \\  
    $0.850<\nn<0.900$\:\: &4.0\% &(+0.9$\sigma$)\:\: & --8.2\% &(--1.3$\sigma$) \\  
    $0.900<\nn<0.940$\:\: &--0.5\% &(--0.1$\sigma$)\:\: & 2.4\% &(+0.5$\sigma$) \\  
    $0.940<\nn<0.970$\:\: &0.1\% &(+0.1$\sigma$)\:\: & --6.1\% &(--1.4$\sigma$) \\  
    $0.970<\nn<0.987$\:\: &2.9\% &(+1.1$\sigma$)\:\: & 0.3\% &(+0.1$\sigma$) \\  
    $0.987<\nn<0.995$\:\: &4.4\% &(+2.1$\sigma$)\:\: & --4.1\% &(--1.0$\sigma$) \\  
    $0.995<\nn<1.000$\:\: &3.4\% &(+2.6$\sigma$)\:\: & 7.0\% &(+3.7$\sigma$) \\  
    \hline \hline
  \end{tabular}
  
\end{table*}

\subsection{Standard model signal expectations}
The expected SM \bsmm\ signal yield for each NN bin is given in 
Table~\ref{tab:smExp} and is estimated using Eq.~(\ref{eq:brbsmm}), the SM value of \brbsmm~\cite{smbr}, the quantities from 
Table~\ref{tab:sigSummary}, and the NN efficiencies of 
Table~\ref{tab:nnEff} to solve for $N_{\bs}$.  Combining all NN bins, 
approximately 1.4 and 1.0 SM \bsmm\ events are expected in the CC and CF 
channels, respectively. The expected SM \bdmm\ yield is a factor of 
$( f_s / f_u )\times( \brbsmm / \brbdmm ) \approx 9$ smaller.

\begin{table*}[htb]
  \begin{center} 
    \caption{The SM expected \bsmm\ signal contribution in each NN bin for 
      the CC and CF channels separately.}
    \label{tab:smExp}
    \begin{tabular}{crr}
      \hline\hline
      NN bin & \multicolumn{1}{c}{CC} & \multicolumn{1}{c}{CF} \\ \hline
      $0.700<\nn<0.760$\:\:\: & $0.04\pm0.01$\:\:\: & $0.03\pm0.01$\\ 
      $0.760<\nn<0.850$\:\:\: & $0.07\pm0.01$\:\:\: & $0.05\pm0.01$\\
      $0.850<\nn<0.900$\:\:\: & $0.05\pm0.01$\:\:\: & $0.04\pm0.01$ \\
      $0.900<\nn<0.940$\:\:\: & $0.07\pm0.01$\:\:\: & $0.05\pm0.01$ \\
      $0.940<\nn<0.970$\:\:\: & $0.10\pm0.02$\:\:\: & $0.07\pm0.01$  \\
      $0.970<\nn<0.987$\:\:\: & $0.13\pm0.02$\:\:\: & $0.11\pm0.02$\\
      $0.987<\nn<0.995$\:\:\: & $0.20\pm0.04$\:\:\: & $0.09\pm0.02$ \\
      $0.995<\nn<1.000$\:\:\: & $0.75\pm0.13$\:\:\: & $0.52\pm0.10$\\
      \hline\hline
 \end{tabular}
   
  \end{center}
\end{table*}

\section{Background estimation}
\label{sec:background}
The background falls into two classes. The dominant source of background in the \bsmm\ search comes from accidental combinations of muon candidates that meet the selection requirements (combinatorial background). In addition, a peaking background from \bhh\ decays, where $h$ and $h^{\prime}$ are either a pion or kaon, contributes. These
two-body charmless $B$ decays are a more significant background for the \bdmm\ search 
due to the downward shift in \Mmm\ caused by assuming the muon mass for both charged particles. 
The two classes of backgrounds are estimated for each $( \nn , \Mmm )$ bin for the CC and CF channels separately.

\subsection{Combinatorial backgrounds}
\label{sec:combBack}

\begin{figure*}
  \centering
  \includegraphics[width=0.5\textwidth]{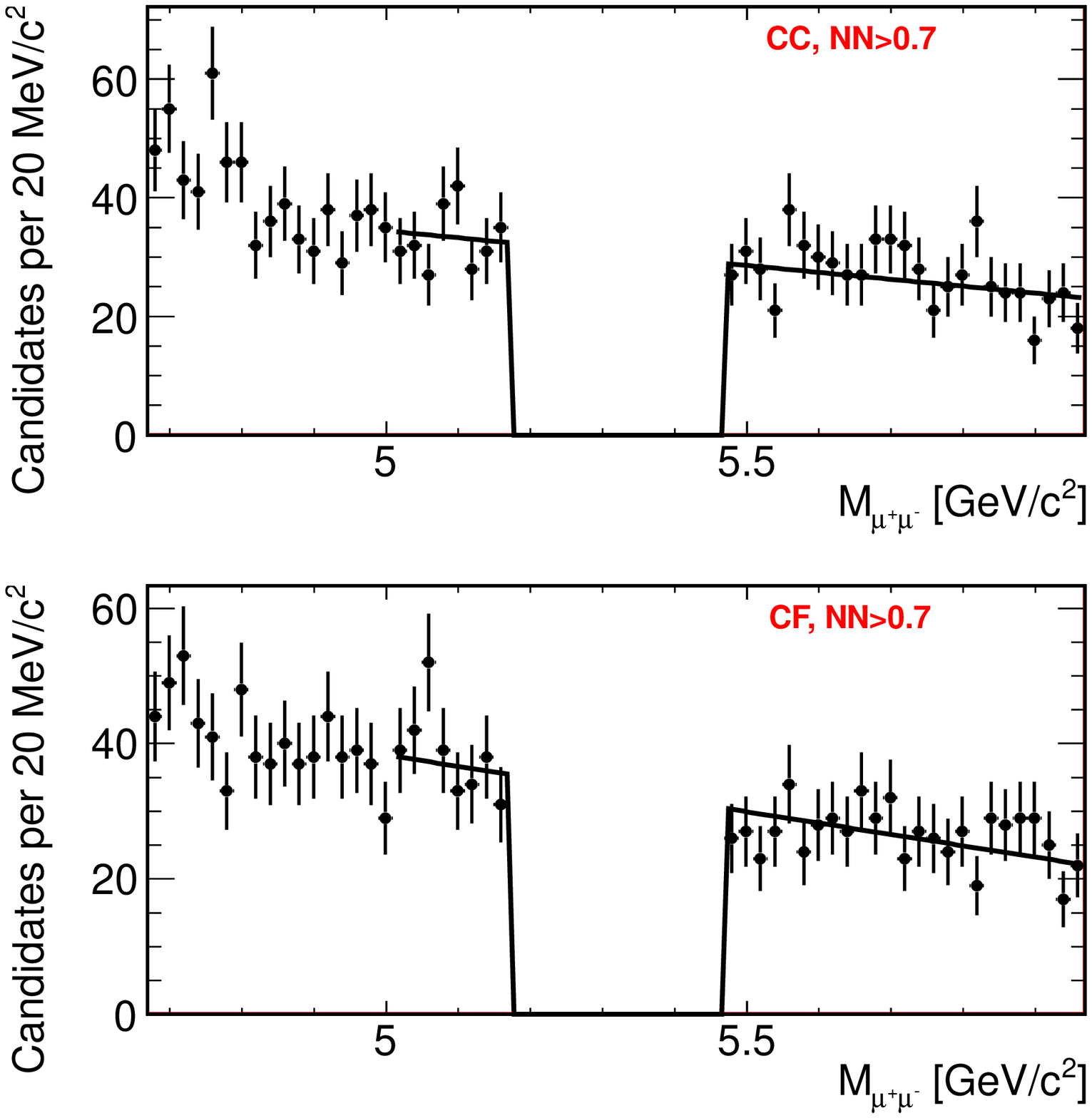}
  \caption{Dimuon-mass distributions for $\nn>0.7$ for the CC and CF channels with 
    extended signal-region blinded.  The slopes from these fits are fixed and used 
    to estimate the combinatorial background in each NN bin.}
  \label{fig:allNNBinFits}
\end{figure*}

\begin{figure*}
  \centering
  \includegraphics[width=0.40\textwidth]{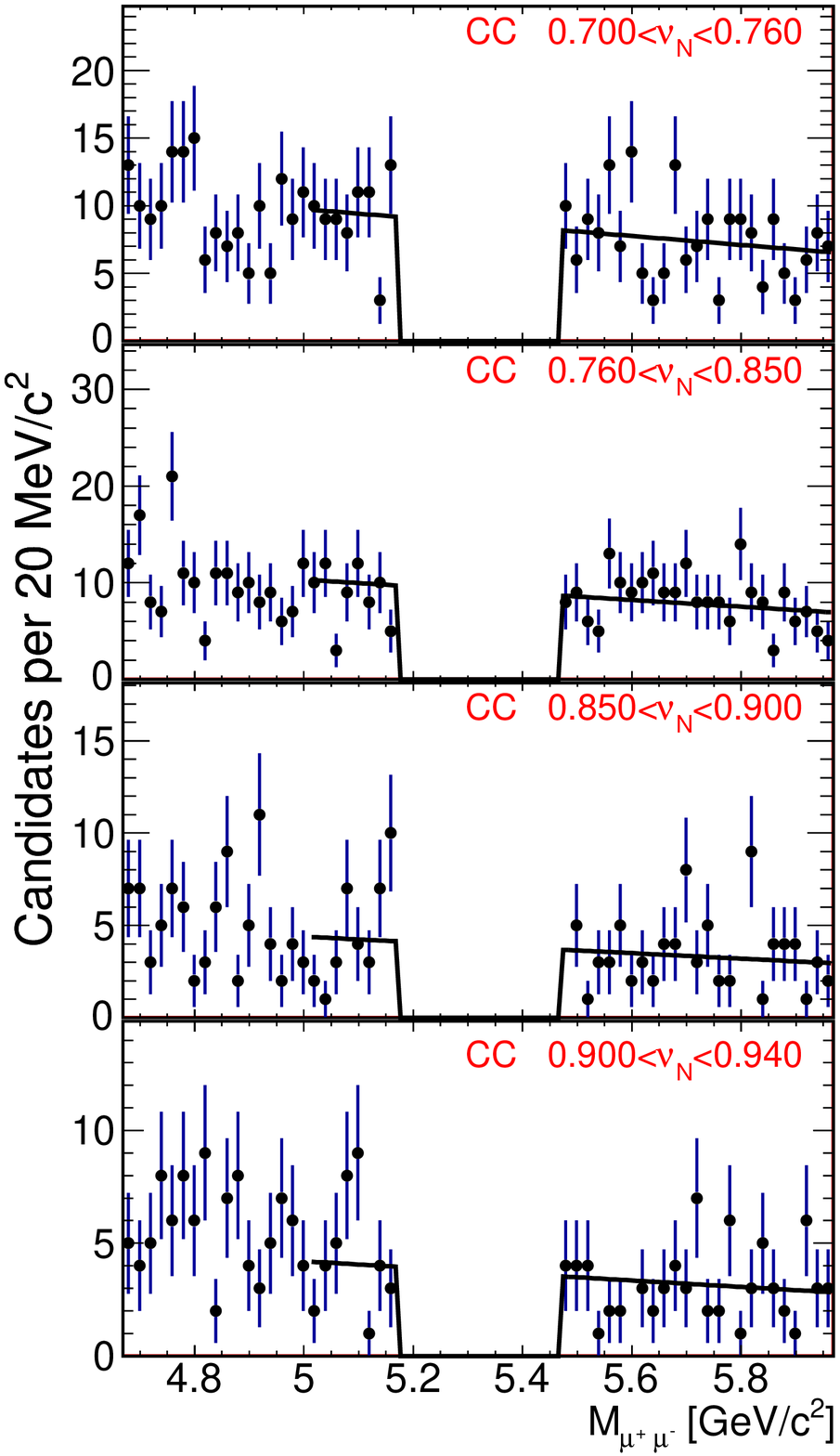} 
  \includegraphics[width=0.40\textwidth]{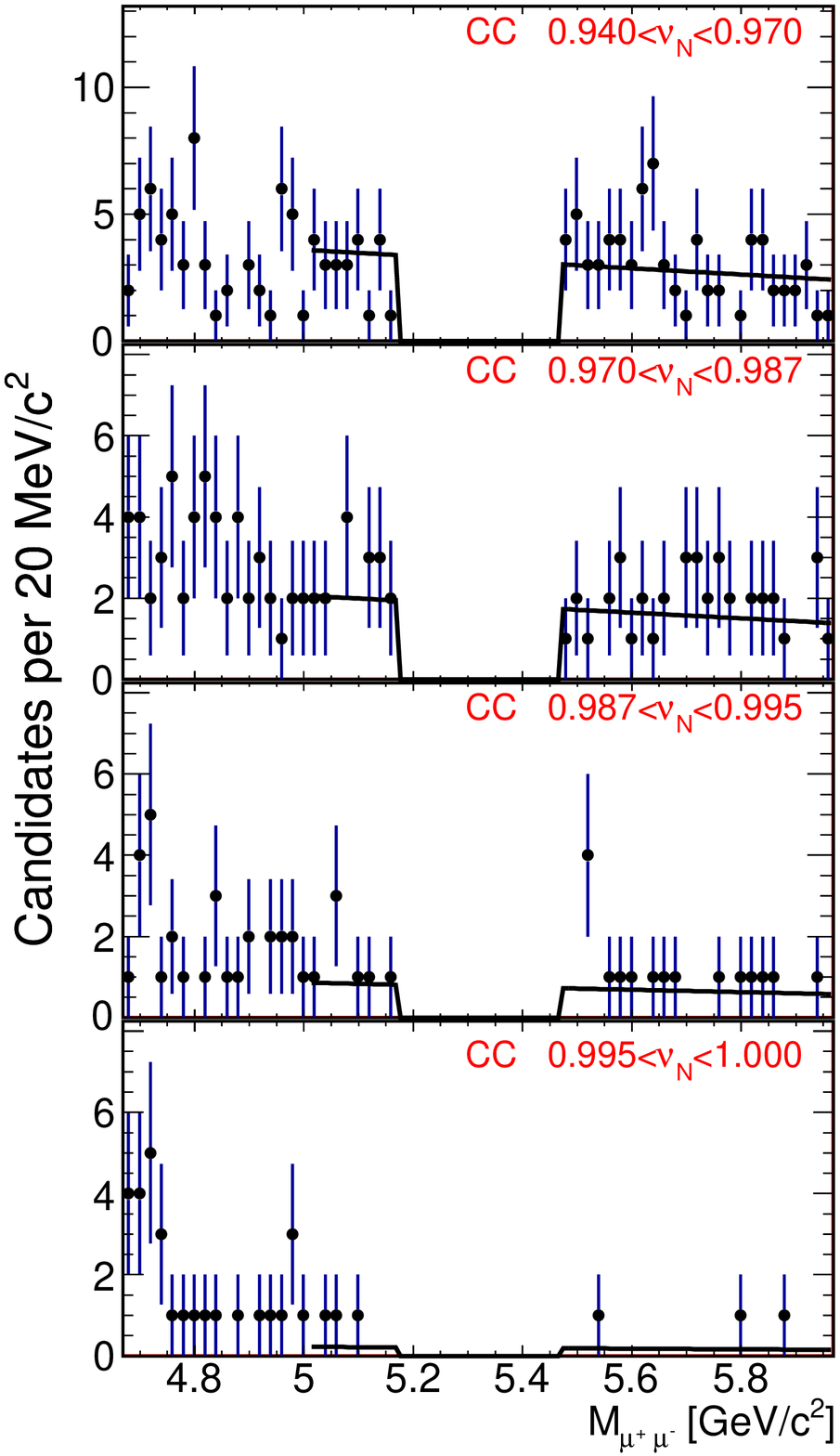}
  \caption{Dimuon-mass distributions with fit overlaid for each of the 
    eight NN bins for the    
    CC channel with the extended signal-region blinded.  The slope of each 
    curve is taken from the fit in Fig.~\ref{fig:allNNBinFits} while the 
    normalization is determined in each bin separately.}
  \label{fig:ccMmm}
\end{figure*}

\begin{figure*}
  \centering
  \includegraphics[width=0.40\textwidth]{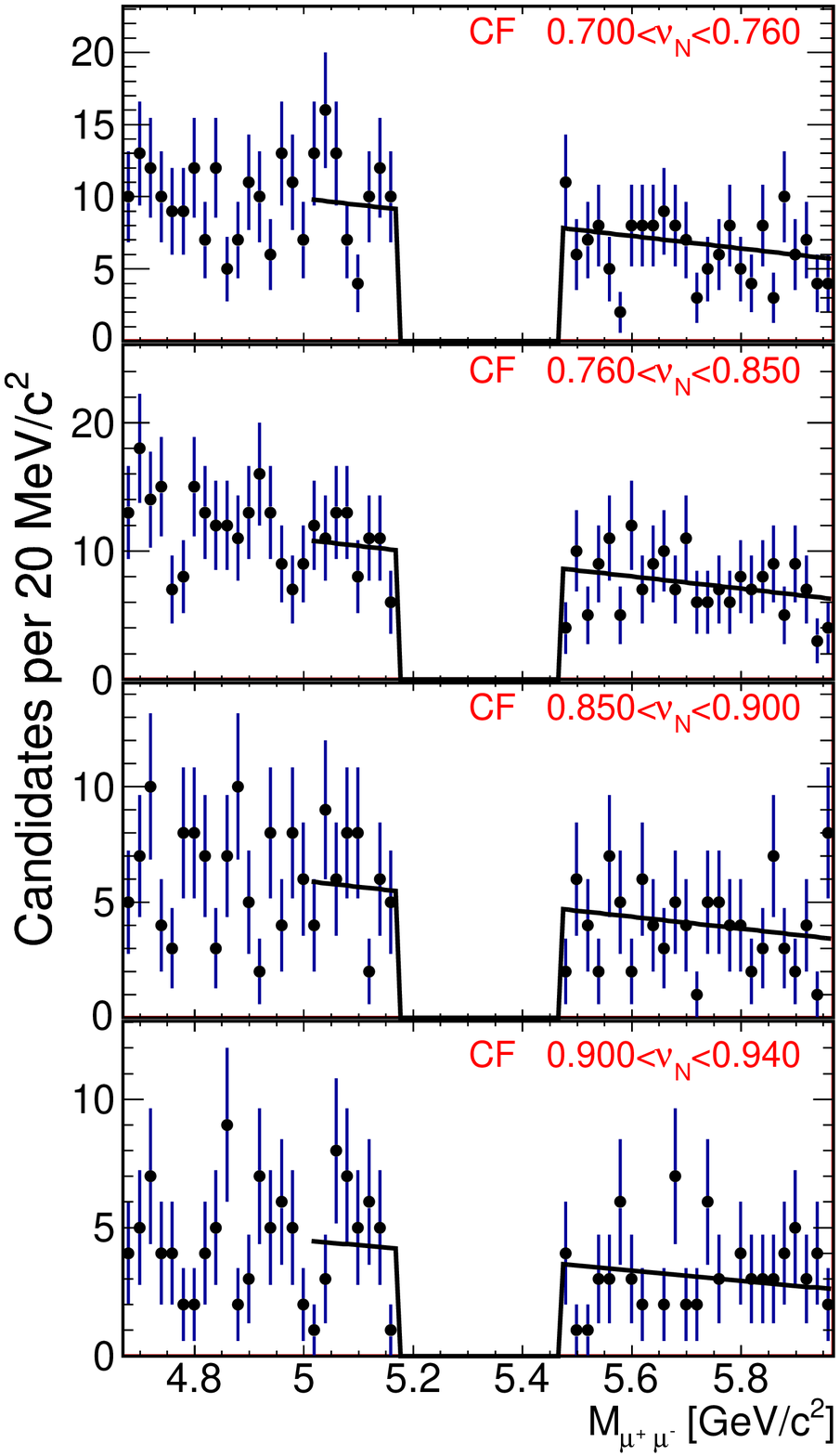} 
  \includegraphics[width=0.40\textwidth]{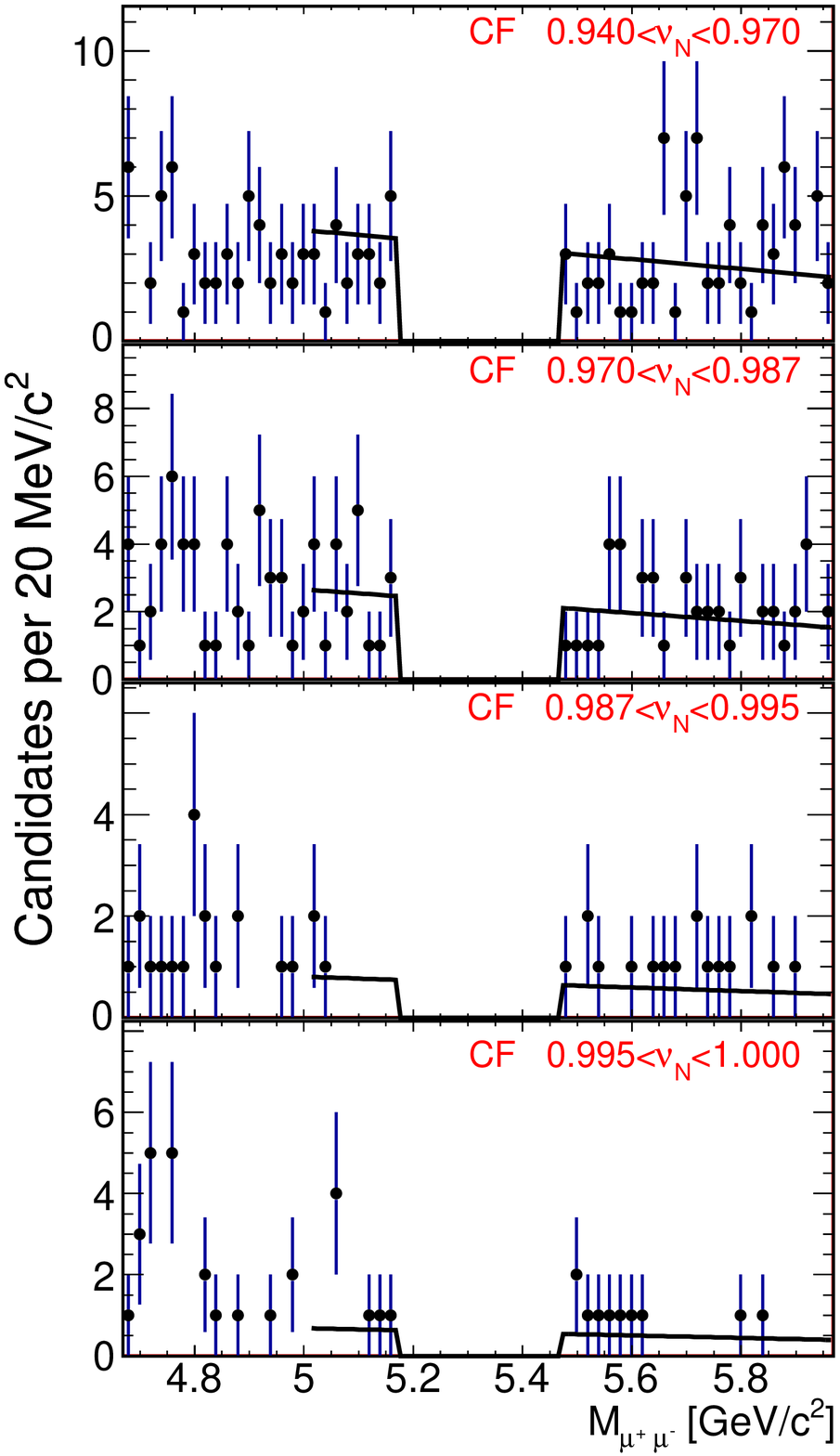}
  \caption{Dimuon-mass distributions with fit overlaid for each of the 
    eight NN bins for the    
    CF channel with the extended signal-region blinded.  The slope of each 
    curve is taken from the fit in Fig.~\ref{fig:allNNBinFits} while the 
    normalization is determined in each bin separately.}
  \label{fig:cfMmm}
\end{figure*}

The combinatorial background is estimated using data events in the 
$5.009<\Mmm<5.169$ GeV/$c^2$ and $5.469<\Mmm<5.969$ GeV/$c^2$ sidebands. A fit to a straight line for the sum of sideband events over all 
eight \nn\ bins, $\nn >0.70$, is shown in Fig.~\ref{fig:allNNBinFits} and is used to determined a fixed slope.  For each \nn\ bin the mass sidebands are fit to a straight line using the fixed slope but with
a free floating normalization.  The resulting function is integrated over the relevant mass signal region to estimate the combinatorial background in that \nn\ and \Mmm\ bin. The fits in each NN bin are shown in Figures~\ref{fig:ccMmm} and~\ref{fig:cfMmm} for the CC and CF channels, respectively.  This fixed-slope methodology significantly 
reduces the uncertainty on the combinatorial background for the highest \nn\ bins and 
is possible because \Mmm\ and \nn\ are independent (cf. Sec.~\ref{sec:XCheckCorrelation}).
We assign a 6\% systematic uncertainty associated with the statistical uncertainty
on the fixed slope. The statistical uncertainty from the normalization is also 
propagated into the background-estimate uncertainties and is a dominant contribution 
to the total uncertainty in the most sensitive NN bins. The final combinatorial background
estimates are given in Tables~\ref{tab:combBs} and~\ref{tab:combBd}.


\begin{figure}
  \centering
  \includegraphics[width=3.5in]{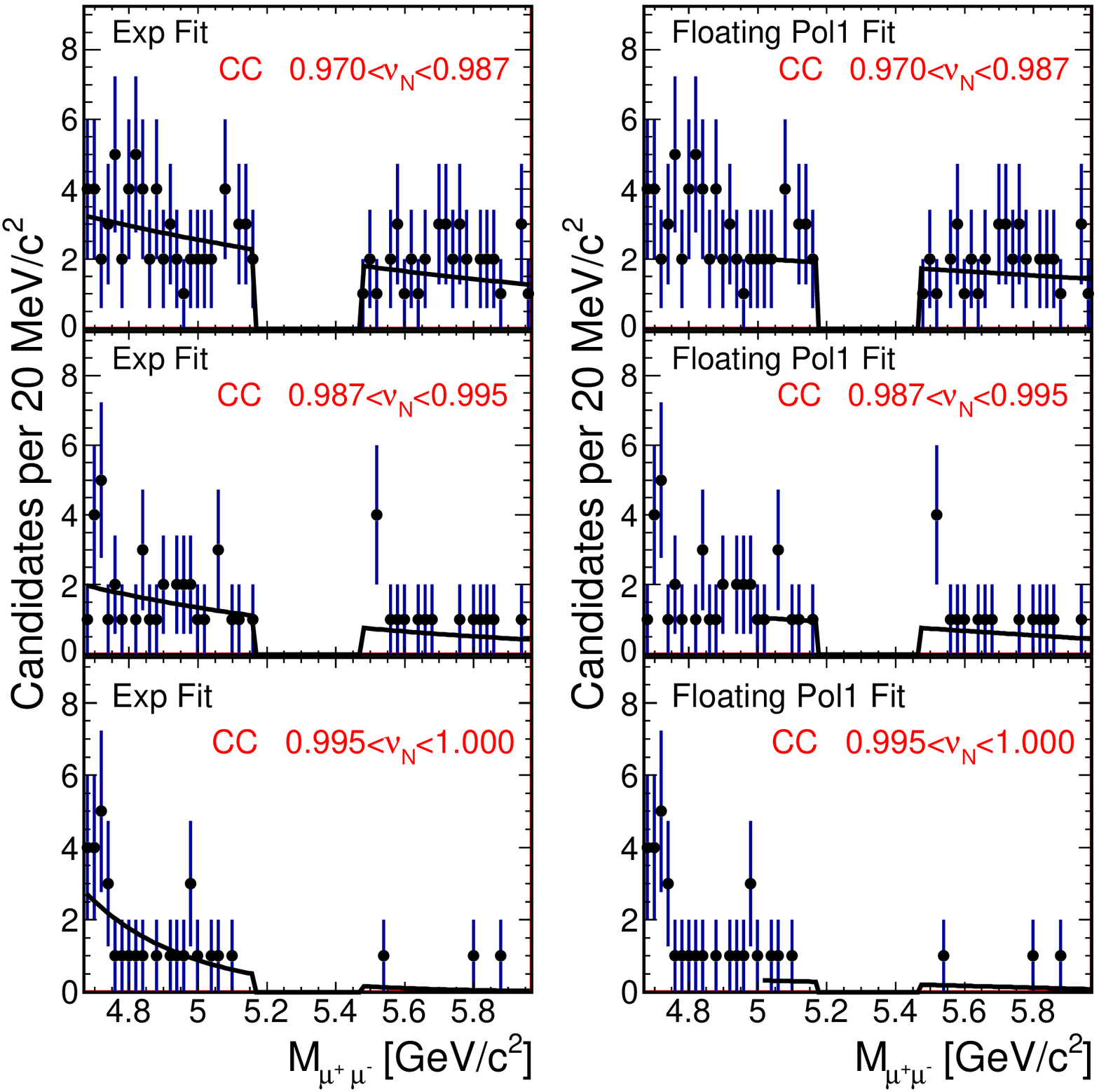}
  \caption{Dimuon mass-sideband distributions with alternative fits 
    overlaid for the three highest NN bins in the CC channel with the 
    extended signal region blinded.}
  \label{fig:altFitsCC}
\end{figure}

\begin{figure}
  \centering\
  \includegraphics[width=3.5in]{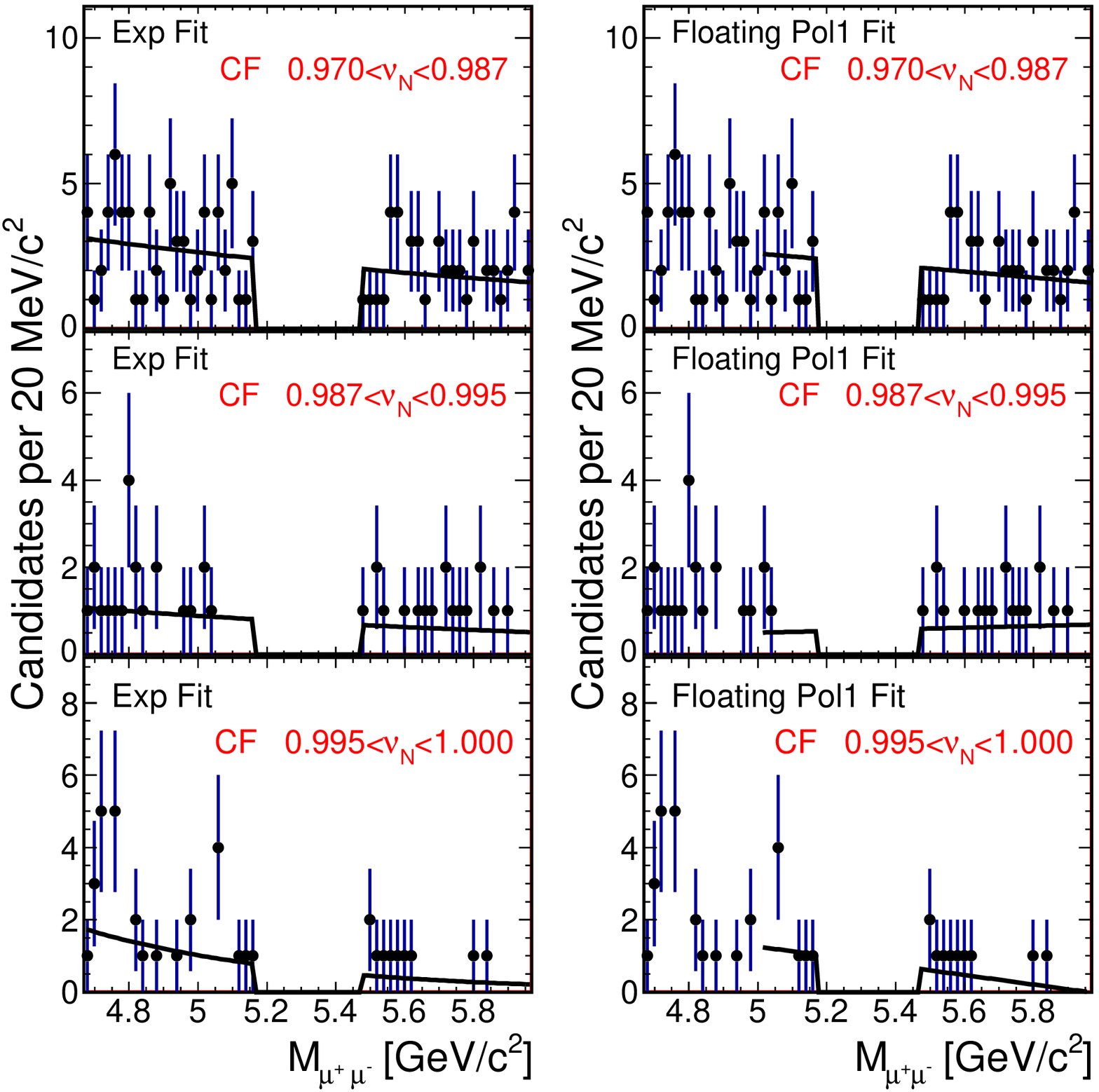}
  \caption{Dimuon mass-sideband distributions with alternative fits 
    overlaid for the three highest NN bins in the CF channel with the 
    extended signal region blinded.}
  \label{fig:altFitsCF}
\end{figure}

Fully reconstructed \bhh\ and partially reconstructed $b\to\mm X$ 
decays can have kinematic properties similar to \bsdmm\ decays and thus
can obtain large \nn\ values.  If these background processes contribute 
significantly to the sideband regions, they would invalidate the 
fixed-sloped methodology.  The estimated \bhh\ contribution to the 
sideband regions is less than $0.1$ event in the CC and CF channels each.
The partially reconstructed decays are constrained to have 
$\Mmm < M_{\bs}$, and the lower edge of the lower sideband ($5.009$~GeV/$c^2$) is chosen to
largely eliminate these events.  To account for the possibility that 
these background processes are affecting the combinatorial background estimates in the highest \nn\ bins, an additional systematic uncertainty is assessed for the three highest bins using alternative fits to the \Mmm\ sideband distribution.  An alternative straight-line fit is performed as described above except that all parameters of the fit are left floating.    A second alternative fit is performed over the extended dimuon-mass 
sideband region, $4.669<\Mmm<5.169$ GeV/$c^2$ and 
$5.469<\Mmm<5.969$ GeV/$c^2$.  By decreasing the lower edge of the lower sideband, a non-negligible number of partially reconstructed decays are 
admitted to the sideband sample in the highest \nn\ bins and an exponential
fit function is used.   The alternative fits are shown in 
Figs.~\ref{fig:altFitsCC} and~\ref{fig:altFitsCF} for the CC and CF channels, respectively. For each alternative, the resulting combinatorial-background estimate is compared to the default estimates in Tables~\ref{tab:combBs} and ~\ref{tab:combBd} and the largest observed difference is assigned as an additional systematic uncertainty for each bin. In the highest \nn\ bins, these systematic uncertainties are of the same magnitude as the statistical uncertainty on the normalization.
The final systematic uncertainties range from $\pm19\%$ 
($\pm3\%$) to $\pm43\%$ ($\pm42\%$) for the CC (CF) channel.

\begin{table*}
  \begin{center}
    \caption{\label{tab:combBs} Estimated number of combinatorial 
      background events for the \bs\ dimuon-mass signal-region for 
      each $( \Mmm , \nn )$ bin and the associated
      statistical uncertainty. }

    {\scriptsize

      \begin{tabular}{cccccc}
        \hline\hline
        \backslashbox{NN bin}{\:\:\:\: Mass bins (GeV/$c^2$)} & 5.310$-$5.334& 5.334$-$5.358& 5.358$-$5.382& 5.382$-$5.406& 5.406$-$5.430\\ 
        \hline 
        \textbf{CC}&&&&& \\ 
        
        0.700$<\nn<$0.760 & 10.42$\pm$0.72 & 10.33$\pm$0.71 & 10.23$\pm$0.70 & 10.14$\pm$0.70 & 10.04$\pm$0.69\\ 
        0.760$<\nn<$0.850 & 11.02$\pm$0.74 & 10.92$\pm$0.74 & 10.82$\pm$0.73 & 10.72$\pm$0.72 & 10.62$\pm$0.71\\ 
        0.850$<\nn<$0.900 & 4.69$\pm$0.46 & 4.65$\pm$0.45 & 4.61$\pm$0.45 & 4.56$\pm$0.44 & 4.52$\pm$0.44\\ 
        0.900$<\nn<$0.940 & 4.49$\pm$0.45 & 4.45$\pm$0.44 & 4.41$\pm$0.44 & 4.37$\pm$0.43 & 4.33$\pm$0.43\\ 
        0.940$<\nn<$0.970 & 3.85$\pm$0.41 & 3.81$\pm$0.41 & 3.78$\pm$0.40 & 3.74$\pm$0.40 & 3.71$\pm$0.39\\ 
        0.970$<\nn<$0.987 & 2.21$\pm$0.30 & 2.19$\pm$0.30 & 2.17$\pm$0.30 & 2.14$\pm$0.30 & 2.12$\pm$0.29\\ 
        0.987$<\nn<$0.995 & 0.92$\pm$0.19 & 0.91$\pm$0.19 & 0.91$\pm$0.19 & 0.90$\pm$0.19 & 0.89$\pm$0.19\\ 
        0.995$<\nn<$1.000 & 0.24$\pm$0.10 & 0.24$\pm$0.10 & 0.24$\pm$0.10 & 0.23$\pm$0.10 & 0.23$\pm$0.10\\ 
        
        \textbf{CF}&&&&& \\ 
        
        0.700$<\nn<$0.760 & 10.18$\pm$0.72 & 10.05$\pm$0.71 & 9.93$\pm$0.70 & 9.80$\pm$0.69 & 9.68$\pm$0.68\\ 
        0.760$<\nn<$0.850 & 11.21$\pm$0.76 & 11.08$\pm$0.75 & 10.94$\pm$0.74 & 10.80$\pm$0.73 & 10.66$\pm$0.72\\ 
        0.850$<\nn<$0.900 & 6.11$\pm$0.54 & 6.03$\pm$0.53 & 5.96$\pm$0.52 & 5.88$\pm$0.52 & 5.81$\pm$0.51\\ 
        0.900$<\nn<$0.940 & 4.65$\pm$0.46 & 4.59$\pm$0.46 & 4.54$\pm$0.45 & 4.48$\pm$0.44 & 4.42$\pm$0.44\\ 
        0.940$<\nn<$0.970 & 3.94$\pm$0.42 & 3.90$\pm$0.42 & 3.85$\pm$0.41 & 3.80$\pm$0.41 & 3.75$\pm$0.40\\ 
        0.970$<\nn<$0.987 & 2.74$\pm$0.35 & 2.71$\pm$0.34 & 2.67$\pm$0.34 & 2.64$\pm$0.34 & 2.61$\pm$0.33\\ 
        0.987$<\nn<$0.995 & 0.83$\pm$0.19 & 0.82$\pm$0.18 & 0.81$\pm$0.18 & 0.80$\pm$0.18 & 0.79$\pm$0.18\\ 
        0.995$<\nn<$1.000 & 0.71$\pm$0.17 & 0.70$\pm$0.17 & 0.69$\pm$0.17 & 0.68$\pm$0.17 & 0.67$\pm$0.16\\ 
        \hline \hline
      \end{tabular}
      
    }
  \end{center}
\end{table*}

\begin{table*}[htb]
  \begin{center}
    \caption{\label{tab:combBd} Estimated number of combinatorial 
      background events for the \bd\ dimuon-mass signal-region for 
      each $( \Mmm , \nn )$ bin and the associated
      statistical uncertainty. }

    {\scriptsize

      \begin{tabular}{cccccc}
        \hline\hline
        \backslashbox{NN bin}{\:\:\:\: Mass bin (GeV/$c^2$)} & 5.219$-$5.243& 5.243$-$5.267& 5.267$-$5.291& 5.291$-$5.315& 5.315$-$5.339\\ 
        \hline 
        \textbf{CC}&&&&& \\ 
        0.700$<\nn<$0.760 & 10.78$\pm$0.74 & 10.69$\pm$0.74 & 10.59$\pm$0.73 & 10.50$\pm$0.72 & 10.40$\pm$0.72\\ 
        0.760$<\nn<$0.850 & 11.41$\pm$0.77 & 11.30$\pm$0.76 & 11.21$\pm$0.76 & 11.10$\pm$0.75 & 11.00$\pm$0.74\\ 
        0.850$<\nn<$0.900 & 4.85$\pm$0.47 & 4.81$\pm$0.47 & 4.77$\pm$0.46 & 4.72$\pm$0.46 & 4.68$\pm$0.46\\ 
        0.900$<\nn<$0.940 & 4.64$\pm$0.46 & 4.60$\pm$0.46 & 4.56$\pm$0.45 & 4.52$\pm$0.45 & 4.48$\pm$0.44\\ 
        0.940$<\nn<$0.970 & 3.98$\pm$0.42 & 3.95$\pm$0.42 & 3.91$\pm$0.42 & 3.88$\pm$0.41 & 3.84$\pm$0.41\\ 
        0.970$<\nn<$0.987 & 2.28$\pm$0.32 & 2.26$\pm$0.31 & 2.24$\pm$0.31 & 2.22$\pm$0.31 & 2.20$\pm$0.30\\ 
        0.987$<\nn<$0.995 & 0.95$\pm$0.20 & 0.95$\pm$0.20 & 0.94$\pm$0.20 & 0.93$\pm$0.20 & 0.92$\pm$0.19\\ 
        0.995$<\nn<$1.000 & 0.25$\pm$0.10 & 0.25$\pm$0.10 & 0.24$\pm$0.10 & 0.24$\pm$0.10 & 0.24$\pm$0.10\\ 
      
        \textbf{CF}&&&&& \\ 
        0.700$<\nn<$0.760 & 10.65$\pm$0.75 & 10.52$\pm$0.74 & 10.40$\pm$0.73 & 10.27$\pm$0.72 & 10.15$\pm$0.72\\ 
        0.760$<\nn<$0.850 & 11.73$\pm$0.80 & 11.60$\pm$0.79 & 11.46$\pm$0.78 & 11.32$\pm$0.77 & 11.18$\pm$0.76\\ 
        0.850$<\nn<$0.900 & 6.39$\pm$0.56 & 6.31$\pm$0.55 & 6.24$\pm$0.55 & 6.16$\pm$0.54 & 6.09$\pm$0.53\\ 
        0.900$<\nn<$0.940 & 4.87$\pm$0.48 & 4.81$\pm$0.48 & 4.75$\pm$0.47 & 4.70$\pm$0.47 & 4.64$\pm$0.46\\ 
        0.940$<\nn<$0.970 & 4.13$\pm$0.44 & 4.08$\pm$0.44 & 4.03$\pm$0.43 & 3.98$\pm$0.43 & 3.94$\pm$0.42\\ 
        0.970$<\nn<$0.987 & 2.87$\pm$0.36 & 2.83$\pm$0.36 & 2.80$\pm$0.35 & 2.77$\pm$0.35 & 2.73$\pm$0.35\\ 
        0.987$<\nn<$0.995 & 0.87$\pm$0.20 & 0.86$\pm$0.19 & 0.85$\pm$0.19 & 0.84$\pm$0.19 & 0.83$\pm$0.19\\ 
        0.995$<\nn<$1.000 & 0.74$\pm$0.18 & 0.73$\pm$0.18 & 0.72$\pm$0.18 & 0.71$\pm$0.17 & 0.70$\pm$0.17\\ 
        \hline\hline
      \end{tabular}
    }
    
  \end{center}
\end{table*}

\subsection{Peaking backgrounds}
\label{sec:peakBack}
Background from \bhh\ decays must be estimated 
separately since they produce a peak in the \Mmm\ distribution and are not included in the combinatorial-background estimates described in Sec.~\ref{sec:combBack}.  Decays involving $B$ baryons, such as 
$\Lambda_b\to p \pi^{-}$, are more heavily suppressed than the $B$-meson background due to the significantly lower rate at which protons reach the muon detectors and satisfy the muon-identification requirements and due to smaller production cross-sections~\cite{bhhprl}. 

The \bhh\ contribution to the dimuon-mass signal region is estimated using data to determine the pion- and kaon-misidentification rates and the methods of Sec.~\ref{sec:acc} to determine the remaining acceptances and efficiencies.  The resulting \bhh\ background is about a factor of ten smaller than the combinatorial background in the \bs\ search, while comprising about half the total background in the \bd\ search. 

The probability for pions and kaons to meet the muon identification requirements is extracted with a pure sample of kaons and pions from $D^*$-tagged $D^0\to K^-\pi^+$ decays. These decays yield two same-sign pions, one from the $D^{*+}\to D^0 \pi^+$ decay and one from the subsequent $D^0$ decay, and an oppositely signed kaon. These charge 
correlations are used to identify the pions and kaons unambiguously without needing to
rely on any particle identification criteria. 
The $D^*$-tagged event sample is collected using the first 7~\fb\ of data with a trigger that requires two charged particles displaced from the beam line, each with $p_T > 2$~GeV/$c$, that reconstruct to a secondary vertex~\cite{DppDkk}.  We further require that the trigger particles have opposite charge and satisfy the baseline tracking criteria described in Sec.~\ref{sec:baseline}.  A kinematic fit is performed, constraining the two tracks to a common vertex, that must satisfy $\chi^2/N_{\mathrm{dof}} < 15$.  The resulting $D^0\to K^-\pi^+$ candidate must have $|\eta|<1$, an impact parameter less than $100\:\mu$m, and 
$1.77<M_{K\pi}<1.97$ GeV/$c^2$.  For surviving events we associate a third track, the soft pion from the $D^{*+}\to D^0 \pi$ decay, with $p_T>0.4$ GeV/$c$, $|z_0|<1.5$ cm, and 
$|d_0|<600$ $\mu$m and form a three-track vertex that must satisfy
$144<M_{K\pi\pi}-M_{K\pi}<147$ MeV/$c^2$.  The resulting sample of kaons and pions from the $D^0$ decay has a purity of $>99\%$ and is used to estimate the efficiency of the muon-identification requirements for kaons and pions, or fake rates.

\begin{figure}
  \centering
  \includegraphics[width=3in]{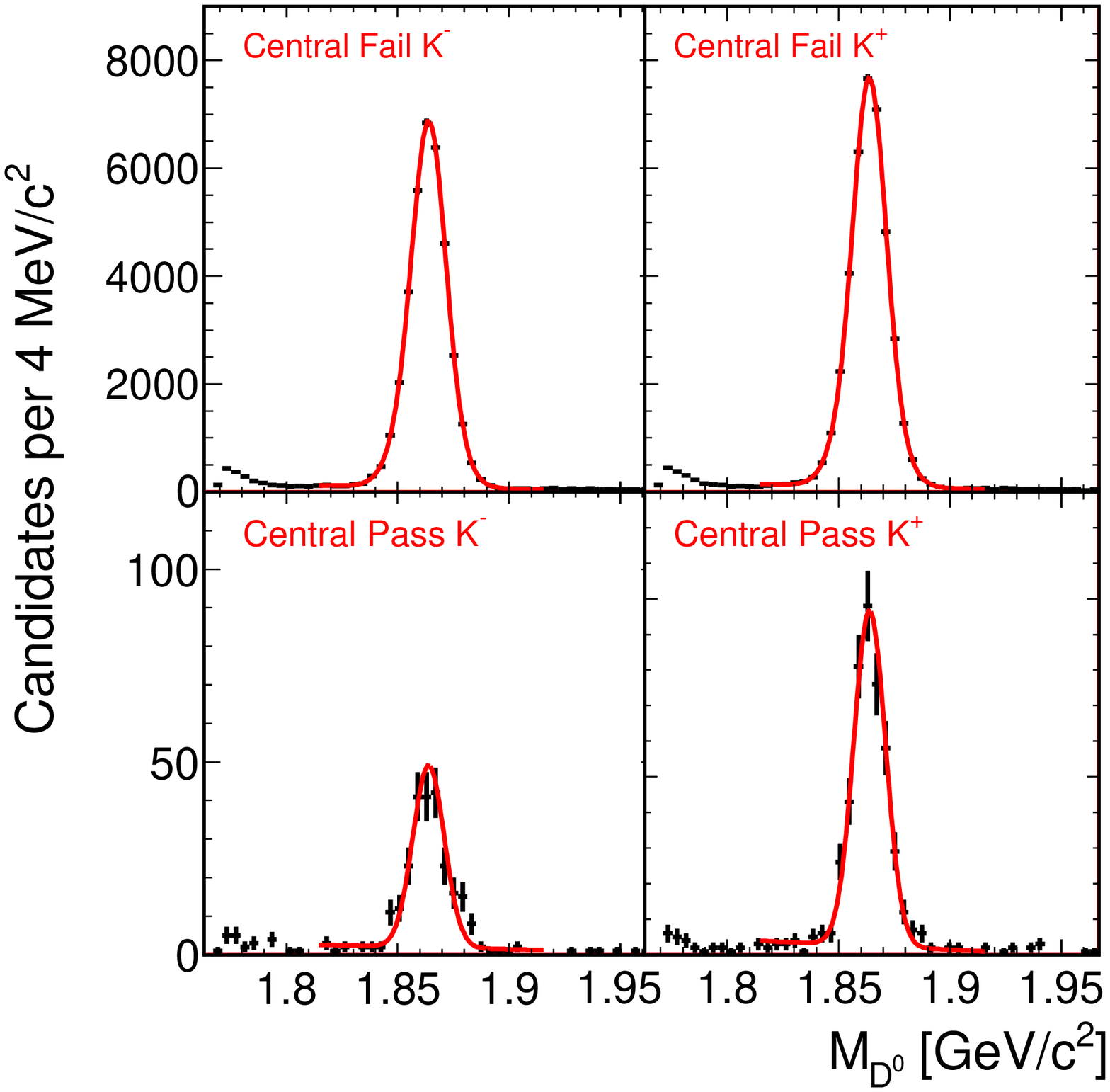}
  \caption{The $M_{K\pi}$ distributions with fits overlaid for 
    central kaons with $2.0<p_T<2.8$ GeV/$c$. The top panels contain the 
    distributions for $D^0$ candidates with kaons that fail the muon 
    requirements, while the bottom panels contain the distributions for
    $D^0$ candidates with kaons that meet the muon ID requirements.
    Distributions for $K^-$ ($K^+$) are on the left (right).}
  \label{fig:d0MassLowPt}
\end{figure}

\begin{figure}
  \centering
  \includegraphics[width=3in]{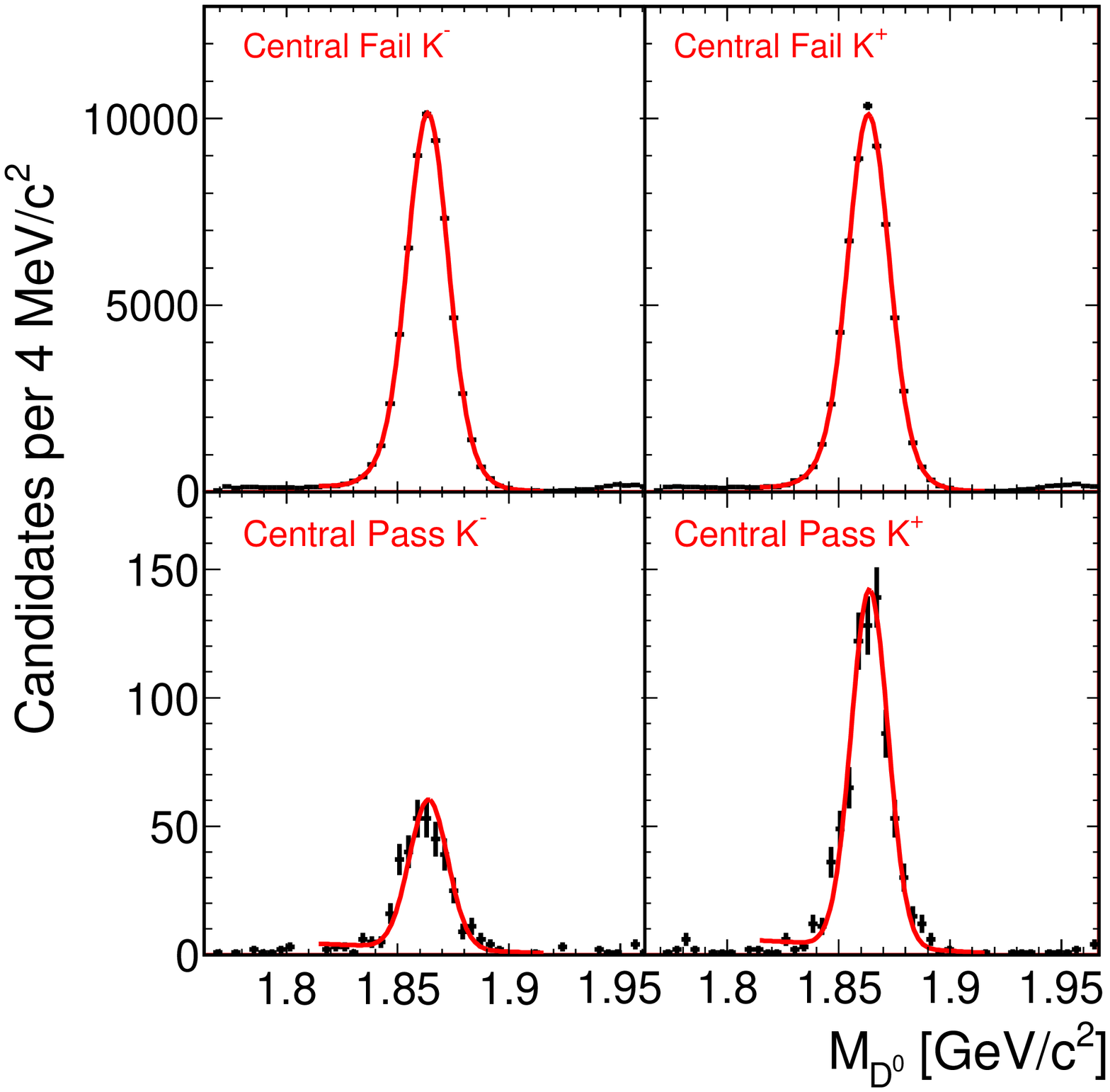}
  \caption{The $M_{K\pi}$ distributions with fits overlaid for 
    central kaons with $6.0<p_T<8.0$ GeV/$c$. The top panels contain the 
    distributions for $D^0$ candidates with kaons that fail the muon 
    requirements, while the bottom panels contain the distributions for
    $D^0$ candidates with kaons that meet the muon ID requirements.
    Distributions for $K^-$ ($K^+$) are on the left (right).}
  \label{fig:d0MassHighPt}
\end{figure}

The pion (kaon) fake rate is determined as the ratio of the number of $D^0$ candidates with a pion (kaon) track that satisfies the muon identification criteria of 
Sec.~\ref{sec:baseline} to the total number of $D^0$ candidates. 
The fake rates for $\pi^+$, $\pi^-$, $K^+$, and $K^-$ are measured as a function of instantaneous luminosity and $p_T$ for central and forward muons separately.  The $\pi^+$ and $\pi^-$ rates are found to be consistent and are combined to yield an average $\pi^\pm$ fake rate.  The number of events in which one track meets the muon selection criteria is estimated by fitting the $M_{K\pi}$ mass distribution to a function that is the sum of a Gaussian distribution and first-order polynomial. The number of events where both tracks fail the muon-selection criteria is determined by fitting the $M_{K\pi}$ distribution to the sum of two Gaussian distributions and a first-order polynomial. 
%
%
Figures~\ref{fig:d0MassLowPt} and~\ref{fig:d0MassHighPt} show the $M_{K\pi}$ distributions with fits overlaid for a lower and higher kaon $p_T$ bin for the central-muon-detector system combining all luminosity bins.

%


\begin{figure}
  \centering
  \includegraphics[width=3in]{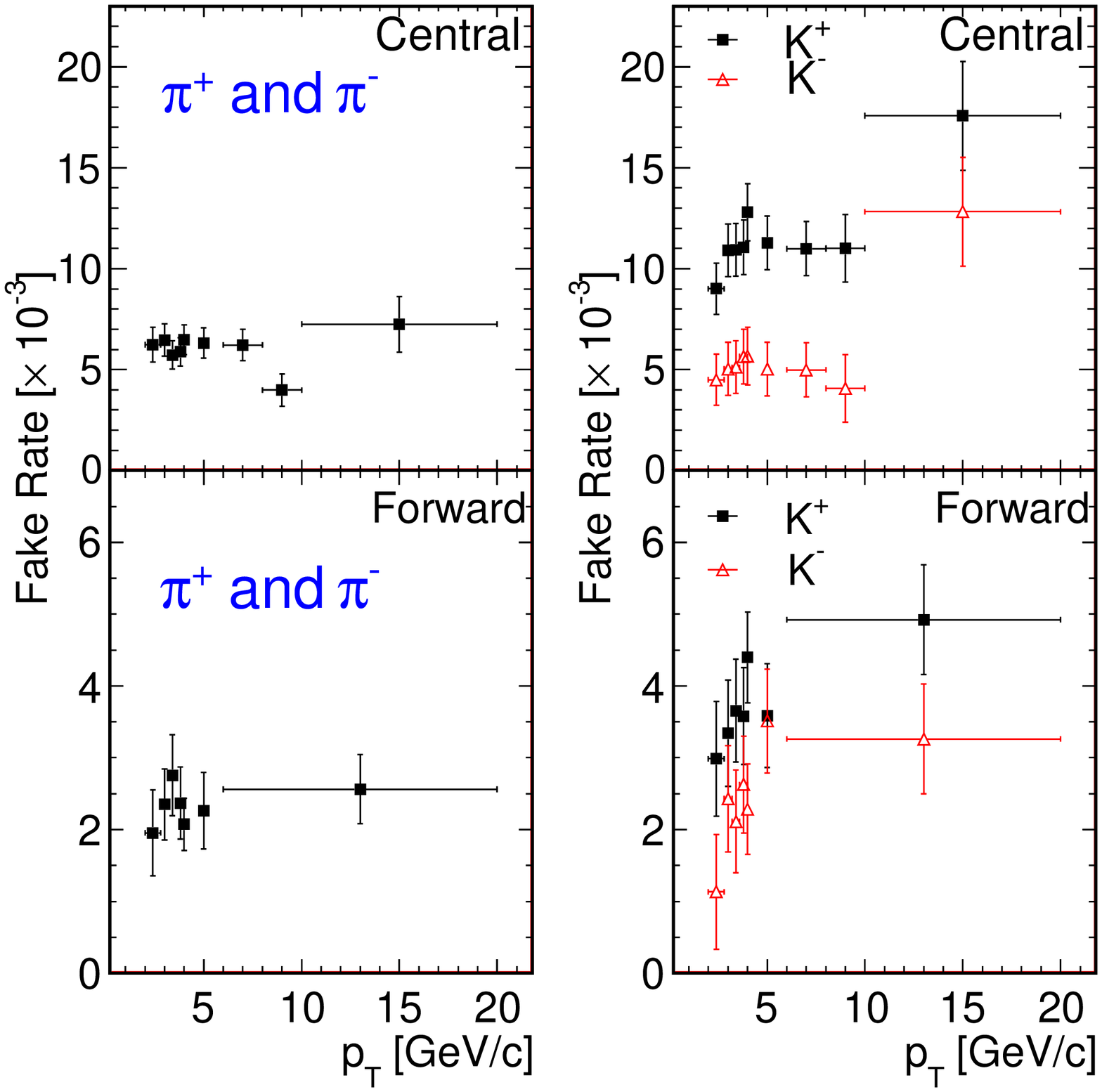}
  \caption{Fake rates as a function of $p_T$ and averaged over 
  instantaneous luminosity for central (top) and forward (bottom) muon 
  detectors. The left panels show the fake rates for pions while the right
  panels show the fake rate for kaons. }
  \label{fig:fakeR}
\end{figure}

Time-dependent changes in fake rates can occur due to changes in the accelerator performance affecting the instantaneous luminosity or due to differences in detector performance associated with aging or changes in the operational configuration.  The instantaneous luminosity was found to be the primary source of the fake-rate time dependence.  We perform a 
consistency check to investigate other sources.  The fake rates, binned in $p_T$ and 
instantaneous luminosity, are applied to the $D^*$-tagged sample as 
weights. The sum of weights is then compared to the actual number of fakes in bins of calendar date resulting in differences of up to 20\%. This 20\% difference is assigned as a systematic uncertainty and accounts for the largest contribution to the fake-rate uncertainty.  In the final determination of the total \bhh\ contribution a weighted average of fake rates is used, based on the instantaneous-luminosity profile of the dimuon-mass-sideband events. The luminosity-averaged fake rates are shown in Fig.~\ref{fig:fakeR}.

The expected number of peaking background events for a
specific two-body hadronic decay is given by
%
\begin{equation*}
  N_b = F_{sb}~\frac{\mathcal{B}_b}{S_s}~
        \frac{\epsilon_{b}^{\mathrm{reco}}}{\epsilon_{\bs}^{\mathrm{reco}}},
\end{equation*}
where $F_{sb}$ is the ratio of relevant fragmentation fractions depending
on which signal and background channels are being evaluated (i.e., $\bs\to \hh$ or $\bd\to\hh$) and on which signal-search mass-window is selected (i.e., \bsmm\ or \bdmm).
For \bd\ hadronic backgrounds in the \bd\ search and for \bs\ hadronic 
backgrounds in the \bs\ search this factor equals unity. For \bd\ hadronic backgrounds in the \bs\ search region $F_{sb}=f_d/f_s$, while for \bs\ 
hadronic backgrounds in the \bd\ search $F_{sb}=f_s/f_d$.  The value for $f_d/f_s$ is taken from Ref.~\cite{PDG2010} and has a 13\% uncertainty.
The value of the branching fraction $\mathcal{B}_b$ for a specific background mode and its associated uncertainty are also taken from 
Ref.~\cite{PDG2010}. For unobserved \bhh\ processes, the current branching-fraction upper-limits are used and a 100\% uncertainty is assigned. The $S_s$ term corresponds to the single-event-sensitivity for \bsmm\ as defined in Sec.~\ref{sec:signalEff} and is taken from Eq.~(\ref{eq:brbsmm}).  The last term
corrects $S_s$ for the differing reconstruction efficiencies between 
\bsmm\ ($\epsilon_{\bs}^{\mathrm{reco}}$) and the relevant \bhh\ 
decay ($\epsilon_{b}^{\mathrm{reco}}$).  In particular, the muon-stub reconstruction and identification efficiencies discussed in Secs.~\ref{sec:stubEff} and~\ref{sec:muidEff} are replaced by the relevant double-track pion and kaon fake-rate estimates.  The double-track fake rate is estimated from the convolution of the single-track fake rates with the 
$(p_T^{h^+}, p_T^{h^-})$ spectra obtained from a sample of \bhh\ MC events meeting the baseline criteria and using the instantaneous-luminosity distribution observed in the \Mmm\ sideband events.  A systematic uncertainty of 35\%  is assigned based on the 20\% single-track uncertainty. 
The efficiency for the reconstructed invariant mass falling into a given \Mmm\ bin is corrected for differences between \bsmm\ and \bhh\ decays using simulated \bhh\ events surviving the baseline criteria weighted by their double-track fake rates.  All other relevant efficiencies are found to be consistent between \bsmm\ and \bhh\ decays.
The final peaking background estimates are given in Tables~\ref{tab:bhhbs} and~\ref{tab:bhhbd}.

The \nn\ distribution is assumed to be the same for both \bhh\ and \bsmm\ since they feature similar kinematic properties and the \nn\ does not use any muon-identification criteria as input variables. We verify that the $p_T$
dependence of the fake rates negligibly affects the \nn\ distribution.
%

\begin{table*}[htb]
  \begin{center}
    \caption{\label{tab:bhhbs}The \bhh\ background estimates and their
    total uncertainty for the \bs\ signal window for each $( \Mmm , \nn )$ 
    bin. The contributions are negligibly small in all the lower \nn\ 
    bins. Uncertainties less than $0.001$ are given as $0.000$ in the table.}
    {\scriptsize
      \begin{tabular}{cccccc}
        \hline\hline
        \backslashbox{NN bin}{\:\:\:\: Mass bin (GeV/$c^2$)} & 5.310$-$5.334& 5.334$-$5.358& 5.358$-$5.382& 5.382$-$5.406& 5.406$-$5.430\\ 
        \hline
        \textbf{CC}&&&&& \\ 
     
        0.700$<\nn<$0.760 & 0.003$\pm0.000$ & 0.001$\pm0.000$ & - & - & -\\ 
        0.760$<\nn<$0.850 & 0.006$\pm$0.000 & 0.002$\pm0.000$ & 0.001$\pm0.000$ & - & -\\ 
        0.850$<\nn<$0.900 & 0.005$\pm$0.001 & 0.002$\pm0.000$ & 0.001$\pm0.000$ & - & -\\ 
        0.900$<\nn<$0.940 & 0.007$\pm$0.001 & 0.003$\pm0.000$ & 0.001$\pm0.000$ & - & -\\ 
        0.940$<\nn<$0.970 & 0.011$\pm$0.001 & 0.003$\pm0.000$ & 0.001$\pm0.000$ & - & -\\ 
        0.970$<\nn<$0.987 & 0.013$\pm$0.002 & 0.005$\pm$0.001 & 0.002$\pm0.000$ & 0.001$\pm0.000$ & -\\ 
        0.987$<\nn<$0.995 & 0.019$\pm$0.002 & 0.007$\pm$0.001 & 0.002$\pm0.000$ & 0.001$\pm0.000$ & -\\ 
        0.995$<\nn<$1.000 & 0.074$\pm$0.010 & 0.026$\pm$0.003 & 0.009$\pm$0.001 & 0.003$\pm0.000$ & 0.001$\pm0.000$\\ 
        \textbf{CF}&&&&& \\ 
        0.700$<\nn<$0.760 & 0.001$\pm0.000$ & - & - & - & -\\ 
        0.760$<\nn<$0.850 & 0.002$\pm0.000$ & 0.001$\pm0.000$ & - & - & -\\ 
        0.850$<\nn<$0.900 & 0.002$\pm0.000$ & 0.001$\pm0.000$ & - & - & -\\ 
        0.900$<\nn<$0.940 & 0.002$\pm0.000$ & 0.001$\pm0.000$ & - & - & -\\ 
        0.940$<\nn<$0.970 & 0.003$\pm0.000$ & 0.001$\pm0.000$ & - & - & -\\ 
        0.970$<\nn<$0.987 & 0.004$\pm$0.001 & 0.002$\pm0.000$ & 0.001$\pm0.000$ & - & -\\ 
        0.987$<\nn<$0.995 & 0.004$\pm$0.001 & 0.002$\pm0.000$ & 0.001$\pm0.000$ & - & -\\ 
        0.995$<\nn<$1.000 & 0.021$\pm$0.003 & 0.009$\pm$0.001 & 0.003$\pm0.000$ & 0.002$\pm0.000$ & -\\ 
        \hline\hline
      \end{tabular}
      
    }
  \end{center}
\end{table*}

\begin{table*}[htb]
  \begin{center}
      \caption{\label{tab:bhhbd}The \bhh\ background estimates and their 
      total uncertainty for the \bd\ signal window for each 
      $( \Mmm , \nn )$ bin. Uncertainties less than $0.001$ are given as $0.000$ in the table.}
    {\scriptsize
      \begin{tabular}{cccccc}
        \hline\hline
        \backslashbox{NN bin}{\:\:\:\: Mass bin (GeV/$c^2$)} & 5.219$-$5.243& 5.243$-$5.267& 5.267$-$5.291& 5.291$-$5.315& 5.315$-$5.339\\ 
        \hline 
        \textbf{CC}&&&&& \\
        0.700$<\nn<$0.760 & 0.015$\pm$0.002 & 0.013$\pm$0.001 & 0.011$\pm$0.001 & 0.006$\pm$0.001 & 0.002$\pm0.000$\\ 
        0.760$<\nn<$0.850 & 0.027$\pm$0.003 & 0.027$\pm$0.003 & 0.019$\pm$0.002 & 0.011$\pm$0.002 & 0.004$\pm$0.001\\ 
        0.850$<\nn<$0.900 & 0.022$\pm$0.002 & 0.019$\pm$0.002 & 0.014$\pm$0.002 & 0.008$\pm$0.001 & 0.003$\pm0.000$\\ 
        0.900$<\nn<$0.940 & 0.030$\pm$0.003 & 0.029$\pm$0.003 & 0.022$\pm$0.003 & 0.013$\pm$0.002 & 0.004$\pm$0.001\\ 
        0.940$<\nn<$0.970 & 0.047$\pm$0.005 & 0.039$\pm$0.004 & 0.031$\pm$0.004 & 0.016$\pm$0.002 & 0.005$\pm$0.001\\ 
        0.970$<\nn<$0.987 & 0.060$\pm$0.006 & 0.052$\pm$0.006 & 0.040$\pm$0.005 & 0.023$\pm$0.003 & 0.009$\pm$0.001\\ 
        0.987$<\nn<$0.995 & 0.084$\pm$0.008 & 0.083$\pm$0.009 & 0.061$\pm$0.008 & 0.033$\pm$0.004 & 0.011$\pm$0.001\\ 
        0.995$<\nn<$1.000 & 0.325$\pm$0.032 & 0.298$\pm$0.031 & 0.221$\pm$0.028 & 0.126$\pm$0.017 & 0.050$\pm$0.006\\ 
        \textbf{CF}&&&&& \\ 
        0.700$<\nn<$0.760 & 0.004$\pm0.000$ & 0.004$\pm0.000$ & 0.003$\pm0.000$ & 0.002$\pm0.000$ & 0.001$\pm0.000$\\ 
        0.760$<\nn<$0.850 & 0.007$\pm$0.001 & 0.008$\pm$0.001 & 0.006$\pm$0.001 & 0.004$\pm$0.001 & 0.001$\pm0.000$\\ 
        0.850$<\nn<$0.900 & 0.006$\pm$0.001 & 0.006$\pm$0.001 & 0.005$\pm$0.001 & 0.003$\pm0.000$ & 0.001$\pm0.000$\\ 
        0.900$<\nn<$0.940 & 0.008$\pm$0.001 & 0.008$\pm$0.001 & 0.007$\pm$0.001 & 0.004$\pm$0.001 & 0.001$\pm0.000$\\ 
        0.940$<\nn<$0.970 & 0.011$\pm$0.001 & 0.011$\pm$0.001 & 0.009$\pm$0.001 & 0.005$\pm$0.001 & 0.002$\pm0.000$\\ 
        0.970$<\nn<$0.987 & 0.017$\pm$0.002 & 0.018$\pm$0.002 & 0.015$\pm$0.002 & 0.008$\pm$0.001 & 0.003$\pm0.000$\\ 
        0.987$<\nn<$0.995 & 0.014$\pm$0.001 & 0.015$\pm$0.002 & 0.012$\pm$0.002 & 0.007$\pm$0.001 & 0.002$\pm0.000$\\ 
        0.995$<\nn<$1.000 & 0.078$\pm$0.008 & 0.084$\pm$0.009 & 0.065$\pm$0.008 & 0.038$\pm$0.005 & 0.014$\pm$0.002\\ 
        \hline\hline
      \end{tabular}
    }
  \end{center}
\end{table*}

\subsection{Background estimate checks with control samples}
\label{sec:controlSamples}
The methods used to predict the background rates are validated using statistically-independent background-dominated data samples.  The control samples are designed to reproduce the salient features of the combinatorial and \bhh\ backgrounds.  Given a signal sample that consists of two opposite-charge muons with $\lambda>0$, we form four independent control samples,
\begin{description}
\item[OS$-$:]  opposite-sign muon pairs, passing the baseline requirements with $\lambda < 0$;
\item[SS$+$:]  same-sign muon pairs with $\lambda > 0$ and relaxed 
trigger-matching to improve event sample size;
\item[SS$-$:]  same-sign muon pairs with $\lambda < 0$ and relaxed 
trigger-matching to improve event sample size;
\item[FM$+$:]  opposite-sign fake-muon-enhanced pairs, in which at least one track of which is required to {\it{fail}} the muon likelihood or \dedx\ requirement with $\lambda > 0$.
\end{description}

The OS$-$ sample is representative of combinatorial backgrounds with short lifetime, which have a symmetric lifetime distribution around zero.
The same-sign samples are dominated by events in which a muon from a semileptonic decay of a $B$ hadron is combined with a muon from the sequential semileptonic decay $b\to cX\to \mu X$ of the other $\overline{B}$ meson in the event and by events in which muons are combined from non-sequential processes. The FM$+$ sample is enriched in \bhh\ background due to the reversal of the muon-identification requirements.  
To mimic the \pting\ and $\lambda$ distributions of the $\lambda>0$ samples, we apply the transformations $\lambda\rightarrow -\lambda$ and $\Delta\Omega\rightarrow\pi - \Delta\Omega$ to the $\lambda<0$ samples.

The background contribution to each control sample is estimated using the same methods as described in Sec.~\ref{sec:combBack} and~\ref{sec:peakBack}.
For the OS$-$ and SS$\pm$ samples, only the combinatorial background is estimated due to the dominance of this component over the peaking background.  For the FM$+$ sample, we estimate both the combinatorial and the \bhh\ background. New fake-rates are evaluated using the relaxed muon-identification criteria and the method described in Sec.~\ref{sec:peakBack}. In all cases the backgrounds are evaluated for the extended \Mmm\ signal region for each \nn\ bin and then compared to the observed number of events. 
The Poisson probability for making an observation at least as large as found in the extended signal region, given the predicted background and its systematic uncertainty, is calculated.
These probabilities are expected to be uniformly distributed between 0 and 100\% for a set of independent samples. The resulting comparisons for all 
\nn\ bins and control samples are shown in Table~\ref{tab:control}.


\begin{table*}[htb]
  \begin{center}
    \caption{\label{tab:control} A comparison of the predicted (Pred.) and observed (Obs.)
      number of events in the extended \Mmm\ signal region as a function of 
      \nn\ bin for the various control samples. The uncertainties 
      correspond to the uncertainty on the mean of the background 
      prediction. The Poisson probability (Prob.) for making an observation at 
      least as large as the observed yield
      is also shown. In cases where no events 
      are observed, the probability is actually the Poisson probability to 
      observe zero events assuming a Poisson mean equal to the predicted mean.}
    \begin{tabular}{cccccccc} \hline\hline
      & & \multicolumn{3}{c}{CC}  
      & \multicolumn{3}{c}{CF}  \\
      Sample  &  NN bin  & Pred. & Obs.  & Prob.(\%) & Pred. & Obs. & Prob.(\%)\\ 
      
      \hline& $0.700<\nn<0.760$ & 268.8$\pm14.3$ & 249 & 82.3 & 261.8$\pm13.9$ & 241 & 84.1\\ 
      OS$-$ & $0.760<\nn<0.850$ & 320.8$\pm16.1$ & 282 & 95.1 & 399.0$\pm18.5$ & 397 & 53.2\\ 
      & $0.850<\nn<0.900$ & 150.3$\pm9.9$ & 156 & 36.5 & 193.5$\pm11.4$ & 185 & 68.3\\ 
      & $0.900<\nn<0.940$ & 146.2$\pm9.7$ & 158 & 23.0 & 177.4$\pm10.8$ & 183 & 37.7\\ 
      & $0.940<\nn<0.970$ & 146.2$\pm9.7$ & 137 & 72.9 & 156.8$\pm10.1$ & 143 & 81.2\\ 
      & $0.970<\nn<0.987$ & 100.4$\pm7.8$ & 98 & 58.3 & 112.6$\pm8.2$ & 110 & 58.3\\ 
      & $0.987<\nn<0.995$ & 78.8$\pm6.8$ & 59 & 97.0 & 53.3$\pm5.4$ & 68 & 6.5\\ 
      & $0.995<\nn<1.000$ & 41.2$\pm4.8$ & 42 & 47.2 & 48.2$\pm5.1$ & 44 & 70.0\\ 
      & & & & & & & \\
      & $0.700<\nn<0.760$ & 4.8$\pm1.2$ & 3 & 81.8 & 0.9$\pm0.5$ & 3 & 8.9\\ 
      SS$+$ & $0.760<\nn<0.850$ & 3.6$\pm1.0$ & 5 & 30.6 & 5.1$\pm1.2$ & 5 & 55.4\\ 
      & $0.850<\nn<0.900$ & 2.4$\pm0.8$ & 5 & 12.2 & 0.9$\pm0.5$ & 6 & 0.2\\ 
      & $0.900<\nn<0.940$ & 1.5$\pm0.7$ & 3 & 21.3 & 0.9$\pm0.5$ & 1 & 56.8\\ 
      & $0.940<\nn<0.970$ & 1.5$\pm0.7$ & 1 & 73.3 & 0.9$\pm0.5$ & 1 & 56.8\\ 
      & $0.970<\nn<0.987$ & 1.8$\pm0.7$ & 2 & 51.3 & 0.9$\pm0.5$ & 0 & 40.7\\ 
      & $0.987<\nn<0.995$ & 0.3$\pm0.3$ & 0 & 74.1 & 0.3$\pm0.3$ & 0 & 74.1\\ 
      & $0.995<\nn<1.000$ & 0.3$\pm0.3$ & 0 & 74.1 & 0.3$\pm0.3$ & 1 & 30.0\\ 
      & & & & & & & \\  
      & $0.700<\nn<0.760$ & 7.8$\pm1.5$ & 10 & 27.8 & 6.0$\pm1.3$ & 4 & 80.9\\ 
      SS$-$ & $0.760<\nn<0.850$ & 10.5$\pm1.8$ & 11 & 47.2 & 7.2$\pm1.5$ & 7 & 55.8\\ 
      & $0.850<\nn<0.900$ & 4.2$\pm1.1$ & 7 & 15.9 & 3.0$\pm0.9$ & 2 & 75.8\\ 
      & $0.900<\nn<0.940$ & 3.6$\pm1.0$ & 4 & 47.2 & 0.9$\pm0.5$ & 7 & 0.1\\ 
      & $0.940<\nn<0.970$ & 3.3$\pm1.0$ & 6 & 14.3 & 3.6$\pm1.0$ & 2 & 83.4\\ 
      & $0.970<\nn<0.987$ & 3.0$\pm0.9$ & 3 & 55.0 & 2.4$\pm0.8$ & 5 & 12.2\\ 
      & $0.987<\nn<0.995$ & 2.1$\pm0.8$ & 0 & 12.2 & 1.2$\pm0.6$ & 0 & 30.1\\ 
      & $0.995<\nn<1.000$ & 1.2$\pm0.6$ & 1 & 65.9 & 1.8$\pm0.7$ & 0 & 16.5\\ 
      & & & & & & & \\
      & $0.700<\nn<0.760$ & 152.2$\pm9.9$ & 161 & 29.6 & 66.5$\pm6.1$ & 88 & 2.5\\ 
      FM$+$ & $0.760<\nn<0.850$ & 140.9$\pm9.5$ & 157 & 15.3 & 81.7$\pm6.9$ & 76 & 70.0\\ 
      & $0.850<\nn<0.900$ & 65.2$\pm6.1$ & 50 & 94.4 & 44.7$\pm5.0$ & 34 & 91.6\\ 
      & $0.900<\nn<0.940$ & 48.7$\pm5.2$ & 40 & 85.8 & 24.4$\pm3.6$ & 38 & 2.3\\ 
      & $0.940<\nn<0.970$ & 27.7$\pm3.8$ & 24 & 73.1 & 12.7$\pm2.6$ & 20 & 7.1\\ 
      & $0.970<\nn<0.987$ & 10.9$\pm2.3$ & 12 & 41.4 & 7.7$\pm2.0$ & 13 & 8.8\\ 
      & $0.987<\nn<0.995$ & 11.0$\pm2.3$ & 4 & 98.3 & 2.7$\pm1.1$ & 3 & 48.3\\ 
      & $0.995<\nn<1.000$ & 28.3$\pm4.1$ & 32 & 30.6 & 4.4$\pm1.6$ & 8 & 13.0\\ 

      \hline\hline
    \end{tabular}
  \end{center}
\end{table*}

No large deviation between the predicted number of background events and 
the observed number of events is seen, providing confidence in the methods employed to estimate the background rates. 

\section{Systematic uncertainties}
\label{sec:systUncer}
The systematic uncertainties are either related to uncertainties in efficiencies, acceptances, normalization factors, peaking-background estimates, or combinatorial-background estimates. 
Table~\ref{tab:syst} summarizes all systematic uncertainties. 

The dominant systematic uncertainty among the efficiencies, acceptances, and normalization factors is the 13\% uncertainty on the ratio of $b$-quark fragmentation fractions, $f_u/f_s$~\cite{PDG2010}. The second largest systematic uncertainty is about a factor of two smaller and
is due to the acceptance ratio (Sec.~\ref{sec:acc}). 

An additional systematic uncertainty, the \bsd\ mass-shape uncertainty, 
is assigned based on the probability for a \bsd\ candidate to populate the dimuon-mass signal-search window.
This uncertainty is based on the world average \bsd -mass uncertainty and the COT momentum scale and resolution. The final uncertainty ranges between 1\% and 9\% depending on \Mmm\ bin.

The leading systematic uncertainty for the peaking background is the 35\% uncertainty assigned to the double-track fake rate as discussed in 
Sec.~\ref{sec:peakBack}.  In addition, an uncertainty associated with the branching fraction of each \bhh\ decay is taken from Ref.~\cite{PDG2010}. Branching fractions for which only upper limits are known are assigned a 100\% relative uncertainty. These decays, however, contribute a small fraction to the total \bhh\ background. The leading contribution to the \bhh\ background in the \bd\ signal-search mass-window
comes from $\bd\to K^+\pi^-$, $\bd\to\pi^+\pi^-$, and $\bs\to K^+ K^-$. The branching fractions of these decays
have relative uncertainties of 3\%, 4\%, and 16\%, respectively.
The \bhh\ background estimates that require $f_d/f_s$ are treated as correlated with the \bdmm\ normalization (cf. Eq.~(\ref{eq:brbsmm})).

The leading systematic uncertainty, up to 43\%, in the combinatorial background estimates is due to the dimuon-mass-sideband shape-uncertainty assigned to the three highest \nn\ bins. The dimuon-mass-sideband
shape-uncertainty is  discussed in~\ref{sec:combBack}. Another large source of systematic uncertainty, up to 42\% in the 
highest \nn\ bins, is due to finite sample-size in the \Mmm\ sidebands. A relatively small contribution ($\approx 6\%$) to the 
total-combinatorial-background systematic uncertainty arises from the 
uncertainty on the fixed slope used in the estimates.

In the statistical interpretation of results all the above systematic uncertainties are taken as nuisance parameters
with multidimensional Gaussian constraints that include correlations. Combinatorial background yields across all (\Mmm , \nn ) bins in the CC or CF channel are correlated because they are fit using the same slope.
Combinatorial background estimates in a given \nn\ bin are additionally correlated across the five mass bins because their normalization is determined from the same set of \Mmm\ sideband events. Peaking backgrounds are treated as correlated across all bins and across the CC and CF channels, 
due to the use of common fake-rates. All the acceptances and efficiencies of Eq.~(\ref{eq:brbsmm}) are treated as correlated across \nn\ and \Mmm\ bins and across the CC and CF channels due to the use of common MC and data control samples. 

\begin{table*}[htb]
  \begin{center}
    \caption{\label{tab:syst} Summary of systematic uncertainties.}
    \begin{tabular}{p{2.5cm}cccc} \hline\hline
      Category & Quantity  &  CC (\%) & CF (\%) & Source\\ 
      \hline
      \multirow{11}{*}{\parbox{2.5cm}{Efficiencies, acceptance, and normalization factors}} & $\alpha_{\bu}/\alpha_{\bs}$ & 6 & 7 & $b$ mass, renorm. scale, fragmentation modeling\\[.05in]
      & $\epsilon^{\mathrm{COT}}_{\bu}/\epsilon^{\mathrm{COT}}_{\bs}$ & 1 & 1 & Isolation, $p_T(B)$, detector effects\\[.05in]
      & $\epsilon^{\mu} _{\bu}/\epsilon^{\mu} _{\bs}$ & 3 & 3 & $Z^0\to\mm$ and $J/\psi\to\mm$ differences\\[.05in]
      & $\epsilon^{\mathrm{PID}}_{\bu}/\epsilon^{\mathrm{PID}} _{\bs}$ & 3 &3 & \bjk\ and $J/\psi\to\mm$ differences\\[.05in]
      & $\epsilon^{\mathrm{SVXII}}_{\bu}/\epsilon^{\mathrm{SVXII}}_{\bs}$ & 3 & 3 & $p_T(\mu)$, two-muon opening-angle, track isolation\\[.05in]
      & $\epsilon^{\mathrm{COT}}_{K}$ & 1.7 & 1.7 & Isolation, $p_T(B)$, detector effects\\[.05in]
      & $\epsilon^{\mathrm{NN}}$ & 4 & 4 & $B$ isolation, $p_T(B)$\\[.05in]
      & $\epsilon^{\mathrm{NN}}$ for $\nn>0.995$ & 3.4 & 7.0 & Data-MC differences\\[.05in]
      & $f_u/f_s$ & 13 & 13 & Ref.~\cite{PDG2010}\\[.05in]
      & $\mathcal{B}(\bjk\to \mm K^+)$ & 4 & 4 & Ref.~\cite{PDG2010}\\[.05in]
      & $B$-meson mass shape & 0.1--9 & 0.1--9 & Mass resolution, mass scale, Ref.~\cite{PDG2010}\\[.05in]
      & & & & \\[.1in]
      \multirow{2}{*}{\bhh} & Fake rates & 20 & 20 & Detector and luminosity effects\\[.05in]
      & $\mathcal{B}(\bhh)$ & 3--100 & 3--100 & Ref.~\cite{PDG2010}\\[.05in]     
      & & & & \\[.1in] 
      \multirow{3}{*}{Comb. Bkg} & Slope & 6 & 6 & Fit uncertainty\\
      & Normalization & 7--42 & 7--25& Sideband sample-size\\
      & Shape & 10--43 & 3--42 & Comparison of different fit functions\\
      \hline\hline
    \end{tabular}
  \end{center}
\end{table*}


\section{Analysis optimization}
\label{sec:nnOpt}
We optimize the \nn\ and \Mmm\ binning by minimizing the expected 95\% C.L. upper limit on \brbsmm\ assuming only background.  The expected limit is calculated using a modified frequentist methodology, the CLs method~\cite{cls1}, combining all the 
\nn\ and \Mmm\ bins in both the CC and CF channels, while taking correlations between bins into account.  

For events surviving all selection criteria, two likelihood fits are performed, one assuming a background-only hypothesis, with the likelihood $\mathcal{L}(b)$, and one assuming a signal-plus-background hypothesis, with likelihood $\mathcal{L}(s+b)$. A log-likelihood ratio $\rm{LLR}=-2\ln\left(\mathcal{L}(s+b)/\mathcal{L}(b)\right)$ is calculated.  The likelihoods are defined as the product of Poisson probabilities over all
$(\Mmm , \nn )$ bins in both the CC and CF channels.  For each bin, the mean number of expected events is estimated assuming only background for 
$\mathcal{L}(b)$ and assuming signal plus background for 
$\mathcal{L}(s+b)$, while the number of observed events is taken from the number of surviving events falling into that bin.  Systematic uncertainties are treated as nuisance parameters with Gaussian constraints.  At a fixed \brbsmm, the likelihoods are minimized by varying the nuisance parameters. 
We denote the minimum of the log-likelihood ratio as $\rm{LLR}_{\rm{min}}$.

Expected limits are calculated using an ensemble of background-only simulated data sets corresponding in size to the actual data set used in this analysis.  The number of contributing background events in each 
$( \Mmm , \nn )$ bin is drawn from a Poisson distribution whose mean corresponds to the values in Table~\ref{tab:combBs} for the combinatorial background and in Table~\ref{tab:bhhbs} for the peaking background.  The mean values shown in the tables are varied by their systematic uncertainties taking into account correlations between bins.  Once assembled, each simulated data set is treated just like the experimental data.  The median CLs as a function of assumed \brbsmm\ is used to determine the expected 95\% confidence level (C.L.) upper limit.  For alternative choices of the $(\Mmm , \nn )$ bins, the methods of Sec.~\ref{sec:combBack} and Sec.~\ref{sec:peakBack} are used to 
generate a mean background-expectation for each bin and to update the systematic uncertainties, while the methods of Sec.~\ref{sec:acc} are used to generate the corresponding signal acceptance.

The optimization process is iterative.  While fixing the \Mmm\ bins, it begins with many \nn\ bins and we then combine neighboring bins with similar expected signal-to-background ratios.  The bin boundaries of the resulting eight \nn\ bins are then varied to minimize the expected \brbsmm\ limit.  Finally the \Mmm\ bins are varied.  The resulting expected \brbsmm\ limit is not significantly dependent on the exact choice of bin boundaries and varies by less than 5\% over reasonable variations of the bin definitions.
The final configuration results in eight \nn\ bins and five \Mmm\ bins.  The \nn\ bins are $0.700<\nn<0.760$, $0.760<\nn<0.850$,
$0.850<\nn<0.900$, $0.900<\nn<0.940$, $0.940<\nn<0.970$, $0.970<\nn<0.987$, $0.987<\nn<0.995$, and $0.995<\nn<1.000$. The five mass bins for the \bs\ 
(\bd ) search are 
$5.310 < \Mmm < 5.334$, $5.334 < \Mmm < 5.358$, $5.358 < \Mmm < 5.382$, $5.382 < \Mmm < 5.406$, and $5.406 < \Mmm < 5.430$~GeV/$c^2$
($5.219 < \Mmm < 5.243$, $5.243 < \Mmm < 5.267$, $5.267 < \Mmm < 5.291$, $5.291 < \Mmm < 5.315$, and $5.315 < \Mmm < 5.339$~GeV/$c^2$).
This optimization reduces the expected limit by approximately 20\% compared to using a single bin with $\nn>0.7$. The final expected-upper-limit for the \bs\ (\bd ) search is \brbsmmeNF\ (\brbdmmeNF) at 95\% C.L. and 
$1.0\times 10^{-8}$ ($3.4\times 10^{-9}$) at 90\% C.L.

\section{Results}
\label{sec:results}
\begin{figure*}
  \centering
  \includegraphics[width=0.65\textwidth]{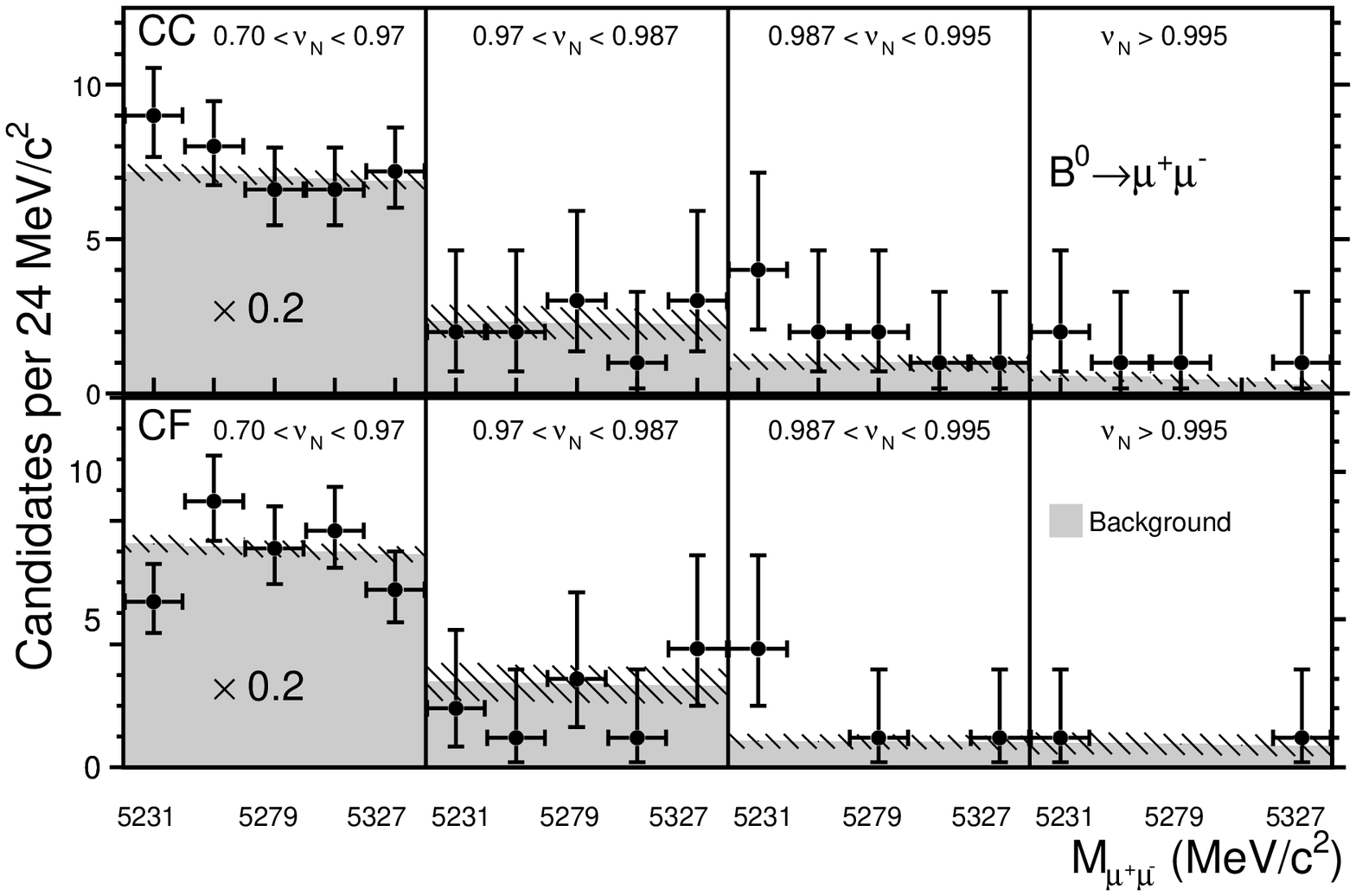}\\
  \includegraphics[width=0.65\textwidth]{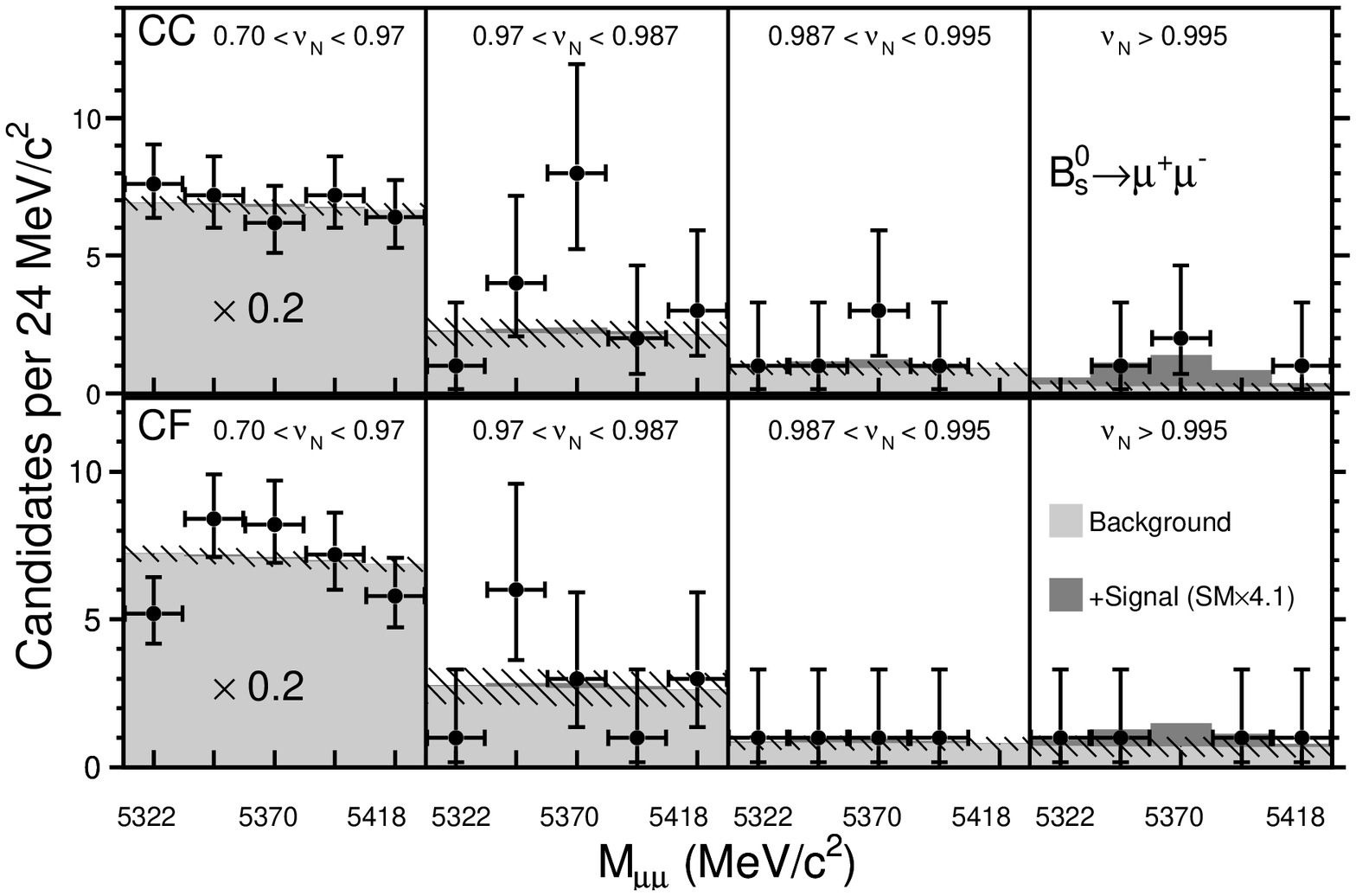}
  \caption{The background estimate (light gray) and its systematic 
    uncertainty (hashed area) is compared to the data (points), 
    and their Poisson uncertainties (error bars on points) for the CC and
    CF channels for the \bdmm\ (top) and \bsmm\ (bottom) searches.  
    Expectations that include signal at $4.1$ times the SM rate (dark 
    gray), corresponding to the fitted value from Fig.~\ref{fig:deltaChi}, 
    are also shown for the \bs\ results.  The lowest five NN 
    bins have been combined because the signal sensitivity is concentrated 
    in the highest three NN bins.
   }
   \label{fig:resultsPlots}
\end{figure*}

\begin{figure}
  \centering
  \includegraphics[width=0.45\textwidth]{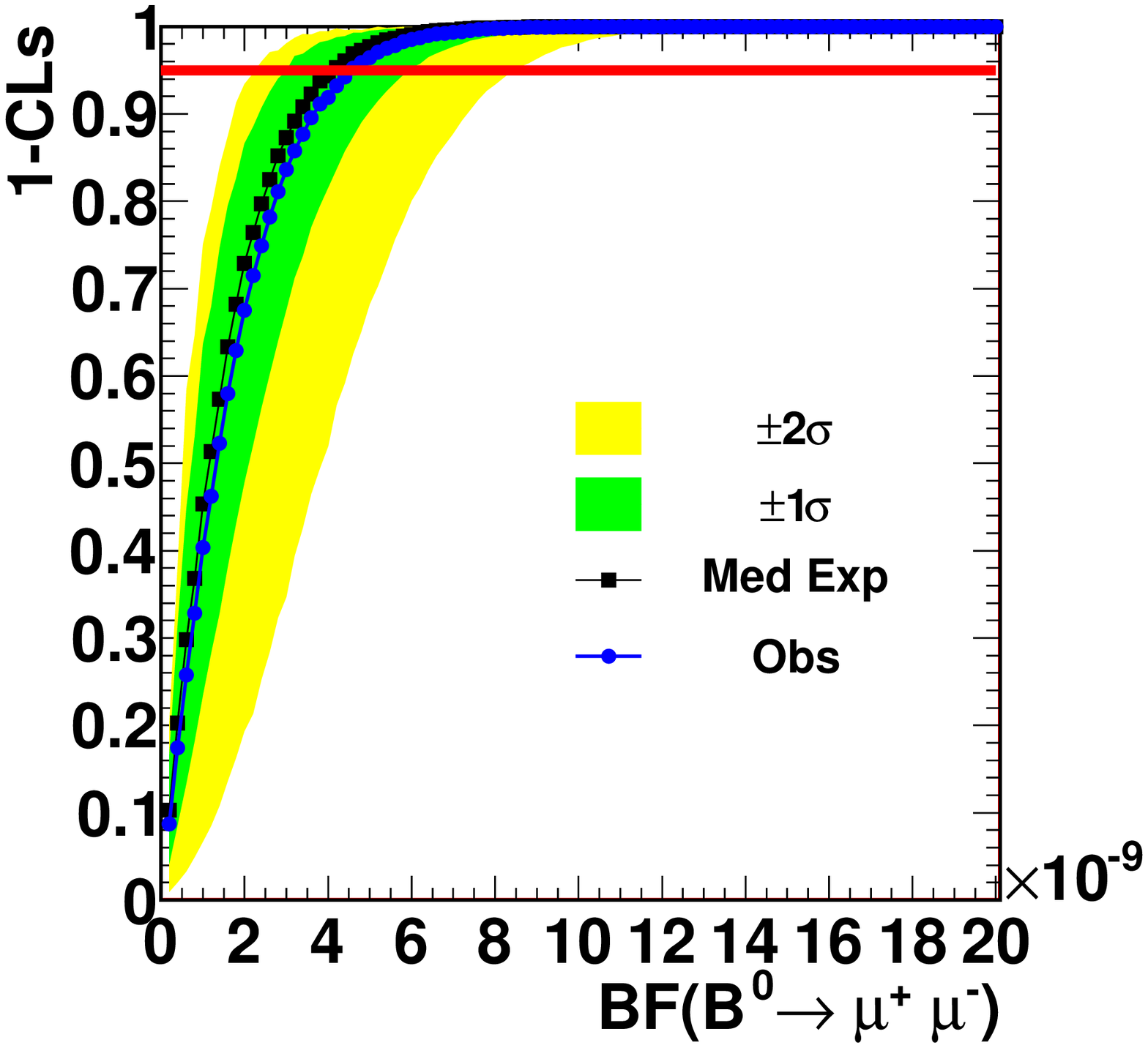}
  \caption{Distribution of 1--CLs as a function of \brbdmm. The expected 
   (observed) limit at 95\% C.L. is determined by the intersection of the 
   black (blue) points with the line at 1--CLs $=0.05$.  The shaded regions 
   indicate the spread of limits obtained from simulated experiments 
   due to fluctuations in the background alone.}
  \label{fig:bdCLs}
\end{figure}

The background estimates, systematic uncertainties on the background estimates, and the observed number of events for the \bdmm\ search are given in Table~\ref{tab:bdResults} and summarized in Fig.~\ref{fig:resultsPlots}. An ensemble of simulated experiments assuming the background-only hypothesis and including the effects of systematic uncertainties is used to estimate the probability that backgrounds alone could produce a $\rm{LLR}_{\rm{min}}$ value at least as small as the one observed in the data. The resulting \pval\ for the \bdmm\ search is 41\%, indicating that the observed
events are consistent with the background expectations.  The observed upper limits are shown in Fig.~\ref{fig:bdCLs} and are obtained using the CLs method to give \brbdmmoNF\ (\brbdmmoN) at 95\% (90\%) C.L.

\begin{table*}[htb]
  \begin{center}
    \caption{The results for the \bdmm\ search comparing the expected total 
      backgrounds and their uncertainty (Exp.) to the number of observed 
      events (Obs.) in each $( \nn , \Mmm )$ bin.  For each \nn\ bin, the 
      sum over mass bins is also shown.  The CC and CF channels 
      are given separately.}
    \label{tab:bdResults}
    \begin{tabular}{cccccccc}
      \hline\hline
      &&&&&&& \\
      \multicolumn{8}{l}{CC channel results:} \\ \hline
      & &\multicolumn{6}{c}{\Mmm\ bin (GeV/$c^2$)} \\
      \nn\ bin & & 5.310--5.334\: & 5.334--5.358\: & 5.358--5.382\:
      & 5.382--5.406\: & 5.406--5.430\:\:\: & Sum\\ \hline
      & Exp. & 10.80$\pm$0.74 & 10.70$\pm$0.74 & 10.61$\pm$0.73 & 10.51$\pm$0.72 & 10.41$\pm$0.72& 53.02\\ 
      0.700--0.760\:\: & Obs. & 15 & 14 & 10 & 7 & 11 & 57\\ \cline{2-8}
      & Exp. & 11.43$\pm$0.77 & 11.33$\pm$0.76 & 11.23$\pm$0.75 & 11.12$\pm$0.75 & 11.01$\pm$0.74& 56.12\\ 
      0.760--0.850\:\: & Obs. & 12 & 10 & 7 & 8 & 9 & 46\\ \cline{2-8}
      & Exp. & 4.88$\pm$0.47 & 4.83$\pm$0.47 & 4.78$\pm$0.46 & 4.73$\pm$0.46 & 4.68$\pm$0.46& 23.90\\ 
      0.850--0.900\:\: & Obs. & 10 & 3 & 6 & 6 & 5 & 30\\ \cline{2-8}
      & Exp. & 4.68$\pm$0.46 & 4.63$\pm$0.46 & 4.59$\pm$0.45 & 4.54$\pm$0.45 & 4.49$\pm$0.44& 22.92\\ 
      0.900--0.940\:\: & Obs. & 6 & 10 & 6 & 8 & 6 & 36\\ \cline{2-8}
      & Exp. & 4.03$\pm$0.42 & 3.99$\pm$0.42 & 3.94$\pm$0.42 & 3.89$\pm$0.41 & 3.85$\pm$0.41& 19.70\\ 
      0.940--0.970\:\: & Obs. & 2 & 3 & 4 & 4 & 5 & 18\\ \cline{2-8}
      & Exp. & 2.34$\pm$0.39 & 2.31$\pm$0.39 & 2.28$\pm$0.38 & 2.24$\pm$0.38 & 2.21$\pm$0.38& 11.39\\ 
      0.970--0.987\:\: & Obs. & 2 & 2 & 3 & 1 & 3 & 11\\ \cline{2-8}
      & Exp. & 1.04$\pm$0.27 & 1.03$\pm$0.27 & 1.00$\pm$0.27 & 0.96$\pm$0.26 & 0.93$\pm$0.26& 4.96\\ 
      0.987--0.995\:\: & Obs. & 4 & 2 & 2 & 1 & 1 & 10\\ \cline{2-8}
      & Exp. & 0.57$\pm$0.19 & 0.54$\pm$0.17 & 0.47$\pm$0.16 & 0.37$\pm$0.15 & 0.29$\pm$0.14& 2.24\\ 
      0.995--1.000\:\: & Obs. & 2 & 1 & 1 & 0 & 1 & 5\\ \hline
      &&&&&&& \\
      \multicolumn{8}{l}{CF channel results:} \\ \hline
      & &\multicolumn{6}{c}{\Mmm\ bin (GeV/$c^2$)} \\
      \nn\ bin & & 5.310--5.334& 5.334--5.358& 5.358--5.382& 5.382--5.406& 5.406--5.430& Sum\\ 
      \hline
      & Exp. & 10.65$\pm$0.75 & 10.53$\pm$0.74 & 10.40$\pm$0.73 & 10.28$\pm$0.73 & 10.15$\pm$0.72& 52.01\\ 
      0.700--0.760\:\: & Obs. & 8 & 13 & 12 & 16 & 10 & 59\\ \cline{2-8}
      & Exp. & 11.74$\pm$0.80 & 11.61$\pm$0.79 & 11.47$\pm$0.78 & 11.33$\pm$0.77 & 11.19$\pm$0.76& 57.33\\ 
      0.760--0.850\:\: & Obs. & 9 & 13 & 13 & 13 & 12 & 60\\ \cline{2-8}
      & Exp. & 6.40$\pm$0.56 & 6.32$\pm$0.55 & 6.24$\pm$0.55 & 6.17$\pm$0.54 & 6.09$\pm$0.53& 31.22\\ 
      0.850--0.900\:\: & Obs. & 3 & 4 & 3 & 2 & 1 & 13\\ \cline{2-8}
      & Exp. & 4.88$\pm$0.48 & 4.82$\pm$0.48 & 4.76$\pm$0.47 & 4.70$\pm$0.47 & 4.64$\pm$0.46& 23.80\\ 
      0.900--0.940\:\: & Obs. & 3 & 8 & 7 & 8 & 5 & 31\\ \cline{2-8}
      & Exp. & 4.14$\pm$0.44 & 4.09$\pm$0.44 & 4.04$\pm$0.43 & 3.99$\pm$0.43 & 3.94$\pm$0.42& 20.20\\ 
      0.940--0.970\:\: & Obs. & 5 & 7 & 2 & 1 & 2 & 17\\ \cline{2-8}
      & Exp. & 2.89$\pm$0.46 & 2.85$\pm$0.46 & 2.82$\pm$0.45 & 2.78$\pm$0.45 & 2.74$\pm$0.44& 14.07\\ 
      0.970--0.987\:\: & Obs. & 2 & 1 & 3 & 1 & 4 & 11\\ \cline{2-8}
      & Exp. & 0.88$\pm$0.26 & 0.87$\pm$0.25 & 0.86$\pm$0.25 & 0.85$\pm$0.25 & 0.83$\pm$0.24& 4.30\\ 
      0.987--0.995\:\: & Obs. & 4 & 0 & 1 & 0 & 1 & 6\\ \cline{2-8}
      & Exp. & 0.82$\pm$0.37 & 0.81$\pm$0.36 & 0.79$\pm$0.36 & 0.75$\pm$0.35 & 0.72$\pm$0.35& 3.89\\ 
      0.995--1.000\:\: & Obs. & 1 & 0 & 0 & 0 & 1 & 2\\ 
      \hline\hline
    \end{tabular}
  \end{center}
\end{table*}

\begin{figure}
  \centering
  \includegraphics[width=0.45\textwidth]{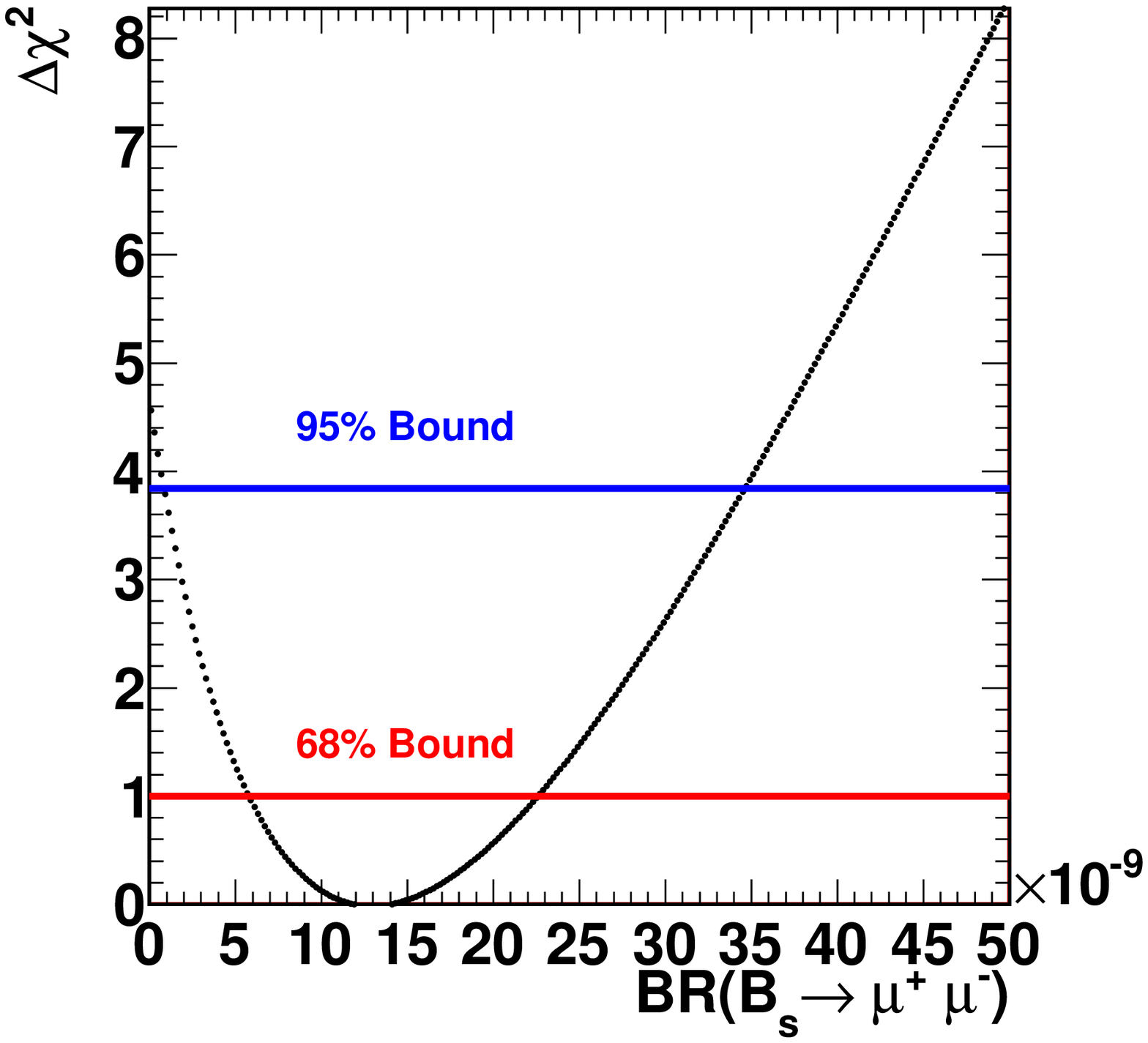}
  \caption{$\Delta\chi^2$ distribution as a function of \brbsmm.}
  \label{fig:deltaChi}
\end{figure}

\begin{figure}
  \centering
  \includegraphics[width=0.45\textwidth]{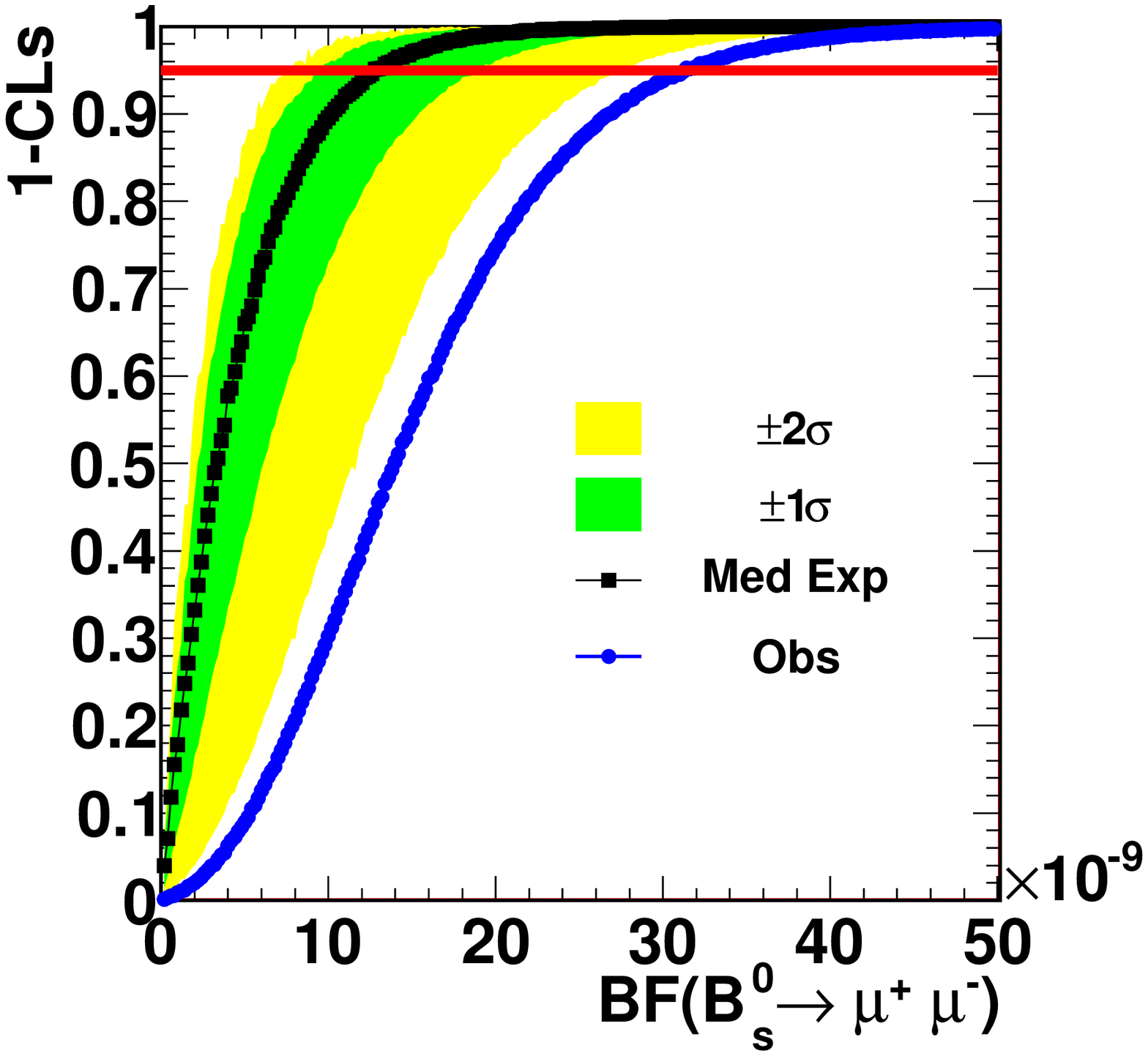}
  \caption{Distribution of 1--CLs as a function of \brbsmm. The expected 
   (observed) limit at 95\% C.L. is determined by the intersection of the 
   black (blue) points with the line at 1--CLs $=0.05$.  The shaded regions 
   indicate the spread of limits obtained from simulated experiments 
   due to fluctuations in the background alone.}
   \label{fig:bsCLs}
\end{figure}

The results of the \bsmm\ search are shown in Table~\ref{tab:bsResults} and
summarized in Fig.~\ref{fig:resultsPlots}.  A small excess of events is 
observed in the CC channel and populates the most sensitive \nn\ and \Mmm\ bins.  The probability that background alone could 
yield a $\rm{LLR}_{\rm{min}}$ value smaller than that observed is $0.94\%$, which corresponds to an excess greater than $2\sigma$.  
Using the expression $\Delta\chi^2 = \rm{LLR} - \rm{LLR}_{\rm{min}}$ we perform a fit to estimate \brbsmm.  The fit, shown in Fig.~\ref{fig:deltaChi}, includes systematic uncertainties and gives
$\brbsmm=1.3^{+0.9}_{-0.7}\times 10^{-8}$ at 68\% C.L. and 
$0.8\times 10^{-9} <\brbsmm<3.4\times10^{-8}$ at 95\% C.L.  A Bayesian method yields very similar results.  The probability that the SM, including
signal, could yield a value of $\rm{LLR}_{\rm{min}}$ smaller than that observed in the data is 6.8\% as determined using an ensemble of simulated experiments that include signal and background contributions, assuming the SM value for \brbsmm\ and including the effects of systematic uncertainties.  The observed upper limits from the CLs methodology are \brbsmmoNF\ (\brbsmmoN) at 95\% (90\%) C.L. and are shown in 
Fig.~\ref{fig:bsCLs}.   Relative to the previous analysis~\cite{CDFPRL2011}, which also reported a small excess using 30\% less data, the
significance of the excess has diminished and the estimate of \brbsmm\ is closer to the SM value.

\begin{table*}[htb]
  \begin{center}
    \caption{The results for the \bsmm\ search comparing the expected total 
      backgrounds and their uncertainty (Exp.) to the number of observed 
      events (Obs.) in each $( \nn , \Mmm )$ bin.  For each \nn\ bin, the 
      sum over mass bins is also shown.  The CC and CF channels 
      are given separately.}
    \label{tab:bsResults}
    \begin{tabular}{cccccccc}
      \hline\hline
      &&&&&&& \\
      \multicolumn{8}{l}{CC channel results:} \\ \hline
      & &\multicolumn{6}{c}{\Mmm\ bin (GeV/$c^2$)} \\
      \nn\ bin & & 5.310--5.334\: & 5.334--5.358\: & 5.358--5.382\:
      & 5.382--5.406\: & 5.406--5.430\:\:\: & Sum\\ \hline
      & Exp. & 10.43$\pm$0.72 & 10.33$\pm$0.71 & 10.23$\pm$0.71 & 10.14$\pm$0.70 & 10.04$\pm$0.69& 51.17\\ 
      0.700--0.760\:\: & Obs. & 13 & 8 & 7 & 6 & 7 & 41\\ \cline{2-8} 
      & Exp. & 11.03$\pm$0.74 & 10.93$\pm$0.74 & 10.82$\pm$0.73 & 10.72$\pm$0.72 & 10.62$\pm$0.72& 54.13\\ 
      0.760--0.850\:\: & Obs. & 9 & 8 & 12 & 15 & 8 & 52\\ \cline{2-8}
      & Exp. & 4.70$\pm$0.46 & 4.65$\pm$0.45 & 4.61$\pm$0.45 & 4.56$\pm$0.44 & 4.52$\pm$0.44& 23.03\\ 
      0.850--0.900\:\: & Obs. & 6 & 8 & 5 & 6 & 5 & 30\\ \cline{2-8}
      & Exp. & 4.50$\pm$0.45 & 4.45$\pm$0.44 & 4.41$\pm$0.44 & 4.37$\pm$0.43 & 4.33$\pm$0.43& 22.05\\ 
      0.900--0.940\:\: & Obs. & 5 & 5 & 5 & 6 & 8 & 29\\ \cline{2-8}
      & Exp. & 3.86$\pm$0.41 & 3.82$\pm$0.41 & 3.78$\pm$0.40 & 3.74$\pm$0.40 & 3.71$\pm$0.39& 18.91\\ 
      0.940--0.970\:\: & Obs. & 5 & 7 & 2 & 3 & 4 & 21\\ \cline{2-8}
      & Exp. & 2.22$\pm$0.38 & 2.19$\pm$0.37 & 2.17$\pm$0.37 & 2.15$\pm$0.37 & 2.12$\pm$0.36& 10.84\\ 
      0.970--0.987\:\: & Obs. & 1 & 4 & 8 & 2 & 3 & 18\\ \cline{2-8}
      & Exp. & 0.94$\pm$0.26 & 0.92$\pm$0.26 & 0.91$\pm$0.26 & 0.90$\pm$0.25 & 0.89$\pm$0.25& 4.56\\ 
      0.987--0.995\:\: & Obs. & 1 & 1 & 3 & 1 & 0 & 6\\ \cline{2-8}
      & Exp. & 0.31$\pm$0.14 & 0.26$\pm$0.14 & 0.25$\pm$0.14 & 0.24$\pm$0.14 & 0.23$\pm$0.14& 1.29\\ 
      0.995--1.000\:\: & Obs. & 0 & 1 & 2 & 0 & 1 & 4\\ \hline
      &&&&&&& \\
      \multicolumn{8}{l}{CF channel results:} \\ \hline
      & &\multicolumn{6}{c}{\Mmm\ bin (GeV/$c^2$)} \\
      \nn\ bin & & 5.310--5.334& 5.334--5.358& 5.358--5.382& 5.382--5.406& 5.406--5.430& Sum\\ 
      \hline
      & Exp. & 10.18$\pm$0.72 & 10.05$\pm$0.71 & 9.93$\pm$0.7 & 9.80$\pm$0.69 & 9.68$\pm$0.68& 49.64\\ 
      0.700--0.760\:\: & Obs. & 10 & 16 & 12 & 11 & 10 & 59\\ \cline{2-8}
      & Exp. & 11.22$\pm$0.76 & 11.08$\pm$0.75 & 10.94$\pm$0.74 & 10.8$\pm$0.73 & 10.67$\pm$0.72& 54.71\\ 
      0.760--0.850\:\: & Obs. & 8 & 13 & 9 & 13 & 4 & 47\\ \cline{2-8}
      & Exp. & 6.11$\pm$0.54 & 6.03$\pm$0.53 & 5.96$\pm$0.52 & 5.88$\pm$0.52 & 5.81$\pm$0.51& 29.79\\ 
      0.850--0.90\:\: & Obs. & 1 & 5 & 9 & 3 & 6 & 24\\ \cline{2-8}
      & Exp. & 4.65$\pm$0.46 & 4.60$\pm$0.46 & 4.54$\pm$0.45 & 4.48$\pm$0.44 & 4.42$\pm$0.44& 22.69\\ 
      0.900--0.940\:\: & Obs. & 6 & 2 & 8 & 5 & 4 & 25\\ \cline{2-8}
      & Exp. & 3.95$\pm$0.42 & 3.90$\pm$0.42 & 3.85$\pm$0.41 & 3.80$\pm$0.41 & 3.75$\pm$0.40& 19.25\\ 
      0.940--0.970\:\: & Obs. & 1 & 6 & 3 & 4 & 5 & 19\\ \cline{2-8}
      & Exp. & 2.75$\pm$0.44 & 2.71$\pm$0.44 & 2.68$\pm$0.43 & 2.64$\pm$0.43 & 2.61$\pm$0.42& 13.38\\ 
      0.970--0.987\:\: & Obs. & 1 & 6 & 3 & 1 & 3 & 14\\ \cline{2-8}
      & Exp. & 0.83$\pm$0.25 & 0.82$\pm$0.24 & 0.81$\pm$0.24 & 0.80$\pm$0.24 & 0.79$\pm$0.23& 4.06\\ 
      0.987--0.995\:\: & Obs. & 1 & 1 & 1 & 1 & 0 & 4\\ \cline{2-8}
      & Exp. & 0.73$\pm$0.35 & 0.71$\pm$0.34 & 0.69$\pm$0.34 & 0.68$\pm$0.34 & 0.67$\pm$0.33& 3.48\\ 
      0.995--1.000\:\: & Obs. & 1 & 1 & 0 & 1 & 1 & 4\\ 
      \hline\hline
    \end{tabular}
    
  \end{center}
\end{table*}

An excess is observed in the two highest NN bins of the CC channel, the most sensitive bins. The total background expectations for the $0.987<\nn<0.995$ and $0.995<\nn<1.000$ bins are 4.56 and 1.29 events while the SM expected signal yields are 0.75 and 0.20 events, respectively. We observe a total of 6 and 4 events, respectively, for these bins. As a check of consistency we redo the \brbsmm\ fit using only the two highest NN bins.  This yields a central value of $\brbsmm=1.0^{+0.8}_{-0.6}\times10^{-8}$, consistent with the full fit.
When considering only the two highest NN bins, the probability that 
background-only (background plus SM signal) could yield a log-likelihood ratio smaller than that observed in data is 2\% (22\%) including the effects of systematic uncertainties.

\begin{figure*}
  \centering
  \includegraphics[width=0.65\textwidth]{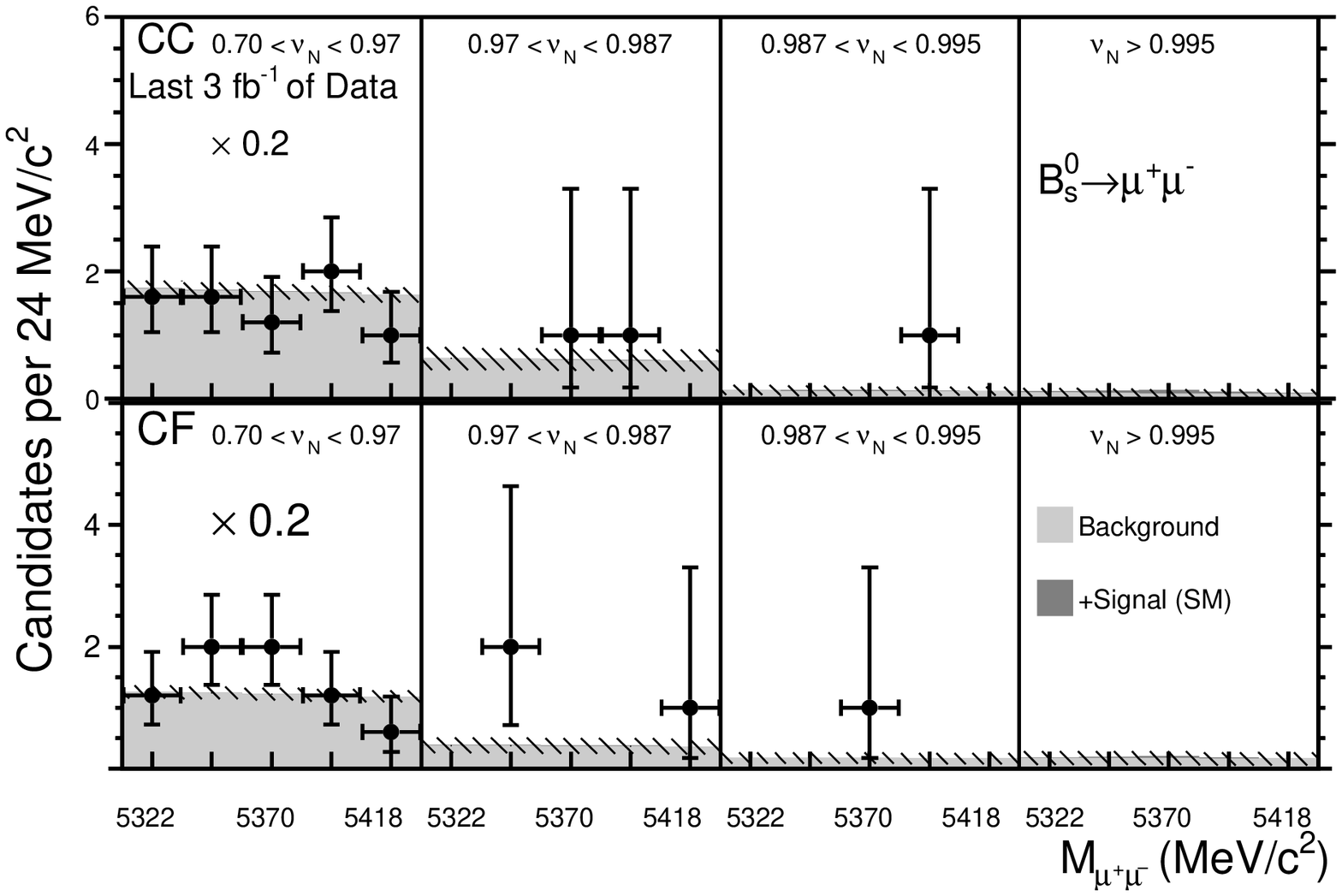}
  \caption{For the \bsmm\ search, the background estimate (light gray) and 
    its systematic uncertainty (hashed area) is compared to the data 
    (points), and its Poisson uncertainty (error bars on points) for the CC 
    and CF channels using only the last 3~\fb\ of data added for this
    analysis. In these plots the lowest five NN bins have been combined 
    because the signal sensitivity is concentrated in the highest three NN 
    bins.
  }
  \label{fig:newDataOnlyResults}
\end{figure*}

We also observe a data excess in the $0.970<\nn<0.987$ bin of the CC channel in the \bsmm\ search, where no significant signal contribution is expected.  Note that such an excess does not appear in the \bdmm\ search. This excess originated in the previous \bsmm\ analysis and was thoroughly investigated~\cite{CDFPRL2011}.  It was concluded that the excess in this bin was not caused by a problem with the background estimates, a NN bias, or any mis-modeling of the data and was likely due to a statistical upward fluctuation.  This conclusion is supported by 
Fig.~\ref{fig:newDataOnlyResults}, which compares the observed data to the background expectations for the \bsmm\ search for the 3~\fb\ of data
added for this analysis. No evidence of an excess in the
$0.970<\nn<0.987$ bin is found in the data added since the analysis of Ref.~\cite{CDFPRL2011}.

\begin{figure*}
  \centering
  \includegraphics[width=0.45\textwidth]{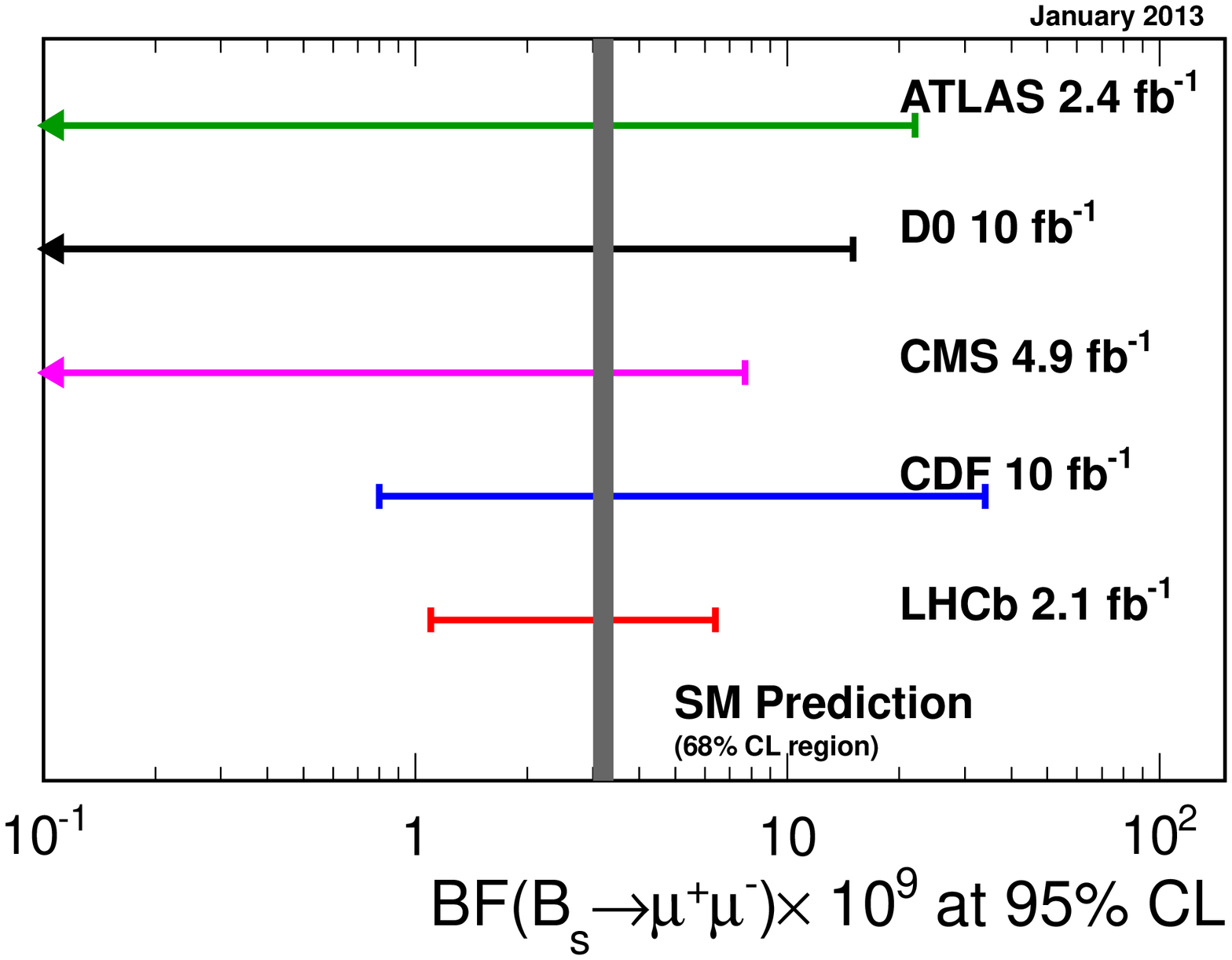}
  \caption{A comparison of current limits from ATLAS~\cite{ATLAS}, CMS~\cite{CMS}, D0~\cite{D0}, LHCb~\cite{LHCb}, and CDF (this paper) at 95\% C.L. The SM prediction~\cite{smbr}, at 68\% C.L., is shown as a thin grey vertical band.}
  \label{fig:limCompares}
\end{figure*}

Our \bdmm\ and \bsmm\ results are consistent with the bounds set by other experiments and with the SM expectations.  This is demonstrated in
Fig.~\ref{fig:limCompares} for the \bsmm\ result, where the small corrections ($< 10\%$) suggested by the recent work in Refs.~\cite{BMixEffects} and~\cite{BmmgammaEffects} have not been considered.

\section{Conclusion}
\label{sec:conclusion}
We report on the search for \bsmm\ and \bdmm\ decays using the full CDF 
Run II data set. These are the most sensitive searches for these decays at the Tevatron.  For the \bdmm\ search, the observed data are in agreement with background-only expectations and an upper limit of
$\brbdmm<\brbdmmoNF$ ($\brbdmmoN$) at 95\% (90\%) C.L. is set.
For the \bsmm\ search, a small excess of events is observed relative to 
expectations from background-only sources with a \pval\ of 0.94\% (6.8\%) assuming
only background (background plus SM signal).
Using a fit to the data we measure $\brbsmm=1.3^{+0.9}_{-0.7}\times 10^{-8}$ and the following bounds are set, 
$2.2\times 10^{-9}<\brbsmm<3.0\times 10^{-8}$ and 
$0.8\times 10^{-9}<\brbsmm<3.4\times 10^{-8}$ at 90\% and 95\% C.L., respectively.  These measurements are consistent with our previous result, the recent results from other experiments, and the SM expectations.

These are the final CDF results for searches for these rare FCNC decays and are the culmination of a program spanning nearly two decades.   
The sensitivity of the \bsmm\ analysis reported here is better than the pioneering measurement by CDF~\cite{CDF1996} by a factor of over 800, which exceeds by a factor of 35 the gain in sensitivity expected by just increasing the sample size.  The gains in search sensitivity originated from continual improvements to the analysis techniques employed.  Those techniques are described in detail to afford future experiments performing similar searches or measurements the opportunity to benefit from this research.  These results form the most sensitive search for \bsdmm\ decays performed previous to the LHC operational period and
remain competitive with the most recent LHC results.  The \bs\ and \bd\ results from all experiments are compatible with one another, indicate that there is no strong enhancement to the \bsmm\ decay rate, and strongly constrain new physics models that predict significant deviations from the standard model~\cite{recentReview}.

\section{\bf Acknowledgments}
We thank the Fermilab staff and the technical staffs of the participating institutions for their vital contributions. This work was supported by the U.S. Department of Energy and National Science Foundation; the Italian Istituto Nazionale di Fisica Nucleare; the Ministry of Education, Culture, Sports, Science and Technology of Japan; the Natural Sciences and Engineering Research Council of Canada; the National Science Council of the Republic of China; the Swiss National Science Foundation; the A.P. Sloan Foundation; the Bundesministerium f\"ur Bildung und Forschung, Germany; the Korean World Class University Program, the National Research Foundation of Korea; the Science and Technology Facilities Council and the Royal Society, UK; the Russian Foundation for Basic Research; the Ministerio de Ciencia e Innovaci\'{o}n, and Programa Consolider-Ingenio 2010, Spain; the Slovak R\&D Agency; the Academy of Finland; the Australian Research Council (ARC); and the EU community Marie Curie Fellowship contract 302103.

\pagebreak



\end{document}

%% file: November2012_Authors-withSperka.tex
\affiliation{Institute of Physics, Academia Sinica, Taipei, Taiwan 11529, Republic of China}
\affiliation{Argonne National Laboratory, Argonne, Illinois 60439, USA}
\affiliation{University of Athens, 157 71 Athens, Greece}
\affiliation{Institut de Fisica d'Altes Energies, ICREA, Universitat Autonoma de Barcelona, E-08193, Bellaterra (Barcelona), Spain}
\affiliation{Baylor University, Waco, Texas 76798, USA}
\affiliation{Istituto Nazionale di Fisica Nucleare Bologna, $^{ee}$University of Bologna, I-40127 Bologna, Italy}
\affiliation{University of California, Davis, Davis, California 95616, USA}
\affiliation{University of California, Los Angeles, Los Angeles, California 90024, USA}
\affiliation{Instituto de Fisica de Cantabria, CSIC-University of Cantabria, 39005 Santander, Spain}
\affiliation{Carnegie Mellon University, Pittsburgh, Pennsylvania 15213, USA}
\affiliation{Enrico Fermi Institute, University of Chicago, Chicago, Illinois 60637, USA}
\affiliation{Comenius University, 842 48 Bratislava, Slovakia; Institute of Experimental Physics, 040 01 Kosice, Slovakia}
\affiliation{Joint Institute for Nuclear Research, RU-141980 Dubna, Russia}
\affiliation{Duke University, Durham, North Carolina 27708, USA}
\affiliation{Fermi National Accelerator Laboratory, Batavia, Illinois 60510, USA}
\affiliation{University of Florida, Gainesville, Florida 32611, USA}
\affiliation{Laboratori Nazionali di Frascati, Istituto Nazionale di Fisica Nucleare, I-00044 Frascati, Italy}
\affiliation{University of Geneva, CH-1211 Geneva 4, Switzerland}
\affiliation{Glasgow University, Glasgow G12 8QQ, United Kingdom}
\affiliation{Harvard University, Cambridge, Massachusetts 02138, USA}
\affiliation{Division of High Energy Physics, Department of Physics, University of Helsinki and Helsinki Institute of Physics, FIN-00014, Helsinki, Finland}
\affiliation{University of Illinois, Urbana, Illinois 61801, USA}
\affiliation{The Johns Hopkins University, Baltimore, Maryland 21218, USA}
\affiliation{Institut f\"{u}r Experimentelle Kernphysik, Karlsruhe Institute of Technology, D-76131 Karlsruhe, Germany}
\affiliation{Center for High Energy Physics: Kyungpook National University, Daegu 702-701, Korea; Seoul National University, Seoul 151-742, Korea; Sungkyunkwan University, Suwon 440-746, Korea; Korea Institute of Science and Technology Information, Daejeon 305-806, Korea; Chonnam National University, Gwangju 500-757, Korea; Chonbuk National University, Jeonju 561-756, Korea; Ewha Womans University, Seoul, 120-750, Korea}
\affiliation{Ernest Orlando Lawrence Berkeley National Laboratory, Berkeley, California 94720, USA}
\affiliation{University of Liverpool, Liverpool L69 7ZE, United Kingdom}
\affiliation{University College London, London WC1E 6BT, United Kingdom}
\affiliation{Centro de Investigaciones Energeticas Medioambientales y Tecnologicas, E-28040 Madrid, Spain}
\affiliation{Massachusetts Institute of Technology, Cambridge, Massachusetts 02139, USA}
\affiliation{Institute of Particle Physics: McGill University, Montr\'{e}al, Qu\'{e}bec H3A~2T8, Canada; Simon Fraser University, Burnaby, British Columbia V5A~1S6, Canada; University of Toronto, Toronto, Ontario M5S~1A7, Canada; and TRIUMF, Vancouver, British Columbia V6T~2A3, Canada}
\affiliation{University of Michigan, Ann Arbor, Michigan 48109, USA}
\affiliation{Michigan State University, East Lansing, Michigan 48824, USA}
\affiliation{Institution for Theoretical and Experimental Physics, ITEP, Moscow 117259, Russia}
\affiliation{University of New Mexico, Albuquerque, New Mexico 87131, USA}
\affiliation{The Ohio State University, Columbus, Ohio 43210, USA}
\affiliation{Okayama University, Okayama 700-8530, Japan}
\affiliation{Osaka City University, Osaka 588, Japan}
\affiliation{University of Oxford, Oxford OX1 3RH, United Kingdom}
\affiliation{Istituto Nazionale di Fisica Nucleare, Sezione di Padova-Trento, $^{ff}$University of Padova, I-35131 Padova, Italy}
\affiliation{University of Pennsylvania, Philadelphia, Pennsylvania 19104, USA}
\affiliation{Istituto Nazionale di Fisica Nucleare Pisa, $^{gg}$University of Pisa, $^{hh}$University of Siena and $^{ii}$Scuola Normale Superiore, I-56127 Pisa, Italy, $^{mm}$INFN Pavia and University of Pavia, I-27100 Pavia, Italy}
\affiliation{University of Pittsburgh, Pittsburgh, Pennsylvania 15260, USA}
\affiliation{Purdue University, West Lafayette, Indiana 47907, USA}
\affiliation{University of Rochester, Rochester, New York 14627, USA}
\affiliation{The Rockefeller University, New York, New York 10065, USA}
\affiliation{Istituto Nazionale di Fisica Nucleare, Sezione di Roma 1, $^{jj}$Sapienza Universit\`{a} di Roma, I-00185 Roma, Italy}
\affiliation{Texas A\&M University, College Station, Texas 77843, USA}
\affiliation{Istituto Nazionale di Fisica Nucleare Trieste/Udine; $^{nn}$University of Trieste, I-34127 Trieste, Italy; $^{kk}$University of Udine, I-33100 Udine, Italy}
\affiliation{University of Tsukuba, Tsukuba, Ibaraki 305, Japan}
\affiliation{Tufts University, Medford, Massachusetts 02155, USA}
\affiliation{University of Virginia, Charlottesville, Virginia 22906, USA}
\affiliation{Waseda University, Tokyo 169, Japan}
\affiliation{Wayne State University, Detroit, Michigan 48201, USA}
\affiliation{University of Wisconsin, Madison, Wisconsin 53706, USA}
\affiliation{Yale University, New Haven, Connecticut 06520, USA}

\author{T.~Aaltonen}
\affiliation{Division of High Energy Physics, Department of Physics, University of Helsinki and Helsinki Institute of Physics, FIN-00014, Helsinki, Finland}
\author{S.~Amerio}
\affiliation{Istituto Nazionale di Fisica Nucleare, Sezione di Padova-Trento, $^{ff}$University of Padova, I-35131 Padova, Italy}
\author{D.~Amidei}
\affiliation{University of Michigan, Ann Arbor, Michigan 48109, USA}
\author{A.~Anastassov$^x$}
\affiliation{Fermi National Accelerator Laboratory, Batavia, Illinois 60510, USA}
\author{A.~Annovi}
\affiliation{Laboratori Nazionali di Frascati, Istituto Nazionale di Fisica Nucleare, I-00044 Frascati, Italy}
\author{J.~Antos}
\affiliation{Comenius University, 842 48 Bratislava, Slovakia; Institute of Experimental Physics, 040 01 Kosice, Slovakia}
\author{G.~Apollinari}
\affiliation{Fermi National Accelerator Laboratory, Batavia, Illinois 60510, USA}
\author{J.A.~Appel}
\affiliation{Fermi National Accelerator Laboratory, Batavia, Illinois 60510, USA}
\author{T.~Arisawa}
\affiliation{Waseda University, Tokyo 169, Japan}
\author{A.~Artikov}
\affiliation{Joint Institute for Nuclear Research, RU-141980 Dubna, Russia}
% ...

\collaboration{CDF Collaboration\footnote{With visitors from
$^a$Istituto Nazionale di Fisica Nucleare, Sezione di Cagliari, 09042 Monserrato (Cagliari), Italy,
$^b$University of California Irvine, Irvine, CA 92697, USA,
$^c$University of California Santa Barbara, Santa Barbara, CA 93106, USA,
$^d$University of California Santa Cruz, Santa Cruz, CA 95064, USA,
$^e$Institute of Physics, Academy of Sciences of the Czech Republic, 182~21, Czech Republic,
$^f$CERN, CH-1211 Geneva, Switzerland,
$^g$Cornell University, Ithaca, NY 14853, USA,
$^h$University of Cyprus, Nicosia CY-1678, Cyprus,
$^i$Office of Science, U.S. Department of Energy, Washington, DC 20585, USA,
$^j$University College Dublin, Dublin 4, Ireland,
$^k$ETH, 8092 Z\"{u}rich, Switzerland,
$^l$University of Fukui, Fukui City, Fukui Prefecture, Japan 910-0017,
$^m$Universidad Iberoamericana, Lomas de Santa Fe, M\'{e}xico, C.P. 01219, Distrito Federal,
$^n$University of Iowa, Iowa City, IA 52242, USA,
$^o$Kinki University, Higashi-Osaka City, Japan 577-8502,
$^p$Kansas State University, Manhattan, KS 66506, USA,
$^q$Brookhaven National Laboratory, Upton, NY 11973, USA,
$^r$University of Manchester, Manchester M13 9PL, United Kingdom,
$^s$Queen Mary, University of London, London, E1 4NS, United Kingdom,
$^t$University of Melbourne, Victoria 3010, Australia,
$^u$Muons, Inc., Batavia, IL 60510, USA,
$^v$Nagasaki Institute of Applied Science, Nagasaki 851-0193, Japan,
$^w$National Research Nuclear University, Moscow 115409, Russia,
$^x$Northwestern University, Evanston, IL 60208, USA,
$^y$University of Notre Dame, Notre Dame, IN 46556, USA,
$^z$Universidad de Oviedo, E-33007 Oviedo, Spain,
$^{aa}$CNRS-IN2P3, Paris, F-75205 France,
$^{bb}$Texas Tech University, Lubbock, TX 79609, USA,
$^{cc}$Universidad Tecnica Federico Santa Maria, 110v Valparaiso, Chile,
$^{dd}$Yarmouk University, Irbid 211-63, Jordan,
$^{ll}$Universite catholique de Louvain, 1348 Louvain-La-Neuve, Belgium,
$^{oo}$University of Z\"{u}rich, 8006 Z\"{u}rich, Switzerland,
$^{pp}$Massachusetts General Hospital and Harvard Medical School, Boston, MA 02114 USA,
$^{qq}$Hampton University, Hampton, VA 23668, USA,
$^{rr}$Los Alamos National Laboratory, Los Alamos, NM 87544, USA
}}
\noaffiliation